\documentclass[twocolumn,aps,pra,footinbib,floatfix]{revtex4-2} 
\usepackage[utf8]{inputenc}

\usepackage{amsmath,amssymb,bm}
\usepackage{hyperref,graphicx,siunitx,xfrac}
\usepackage[dvipsnames]{xcolor}

\usepackage{tikz} 
\usetikzlibrary{arrows,decorations.pathmorphing,backgrounds,positioning,fit,matrix,calc}
\tikzset{snake it/.style={decorate, decoration=snake,segment length=0.1cm}}

\newcommand{\diag}{\mathrm{diag}}
\newcommand{\ket}[1]{|#1\rangle}

\newcommand{\braOket}[3]{\langle #1|#2|#3\rangle}

\begin{document}
\title{Attractive Haldane bilayers for trapping non-Abelian anyons}

\author{Valentin Cr\'epel}
\affiliation{Center for Computational Quantum Physics, Flatiron Institute, New York, New York 10010, USA}
\author{Nicolas Regnault}
\affiliation{Laboratoire de Physique de l’Ecole normale sup\'erieure,ENS, Universit\'e PSL, CNRS, Sorbonne Universit\'e}
\affiliation{Department of Physics, Princeton University, Princeton, NJ 08544, USA}

\begin{abstract}
We study the interplay between intrinsic topological order and superconductivity in a two-component Haldane bilayer, where the two layers are coupled by an attractive force. We obtain the phase diagram of the model with exact diagonalization in finite-size, and develop arguments to assess the stability of the observed phases in the thermodynamic limit. Our main result is that a finite critical attraction strength is needed to pair fermions forming a fractional topological order. This behavior can be harnessed to create clean interfaces between a fractional topological insulator and a superconductor by gating, wherein non-Abelian parafermionic modes are trapped. We discuss realization of such interfaces in the bulk of double bilayers of transition metal dichalcogenides by inhomogenous electrostatic gating, which should mitigate the spurious effects of disorder and crystalline defects present on physical edges.
\end{abstract}

\maketitle

\section{Introduction} \label{sec_introduction}

\subsection{Motivation and scope}

\begin{figure}[h!]
\centering
\includegraphics[width=\columnwidth]{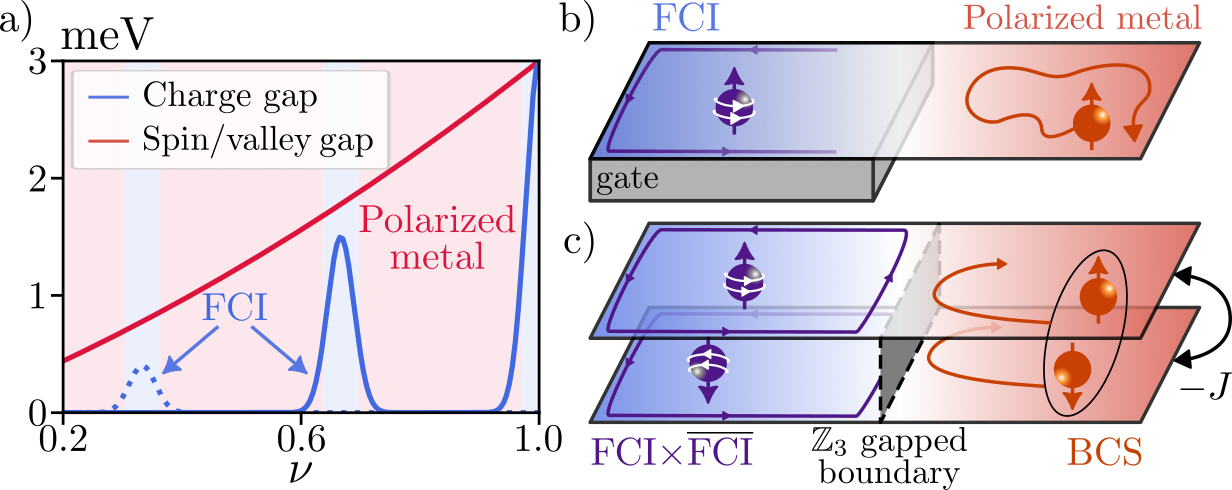}
\caption{a) Schematic phase diagram of moir\'e semiconductors featuring a weakly dispersing topological band. 
Strong local interactions yields a finite flavor gap in a wide range of moir\'e filling $\nu\leq 1$ (light red), resulting in a flavor-polarized metal that is destabilized in favor of FCIs at certain rational fillings. 
The series of fractions observed and the exact values of the gaps are material dependent and provided here as orders of magnitude only. For instance, $\nu=1/3$ is not seen in $3.7^\circ$-twisted MoTe$_2$ (dotted line), whereas $\nu=2/3$ is.
b) The application of gate potentials can tune between the compressible and insulating phases. 
c) We study a model describing two superimposed copies of a material described by (a) coupled by a weak local attractive interaction $J$, with a stacking favoring polarization of the copies into opposite spin and chirality. 
When both copies are in the polarized metallic phase, the weak attraction drives a Cooper instability yielding superconductivity (BCS). 
In contrast, we will show that the fractional topological insulator made of two FCI with opposite chirality (${\rm FCI}\times \overline{\rm FCI}$) is perturbatively immune to such attractive term: its topological properties remain intact for small $J$. 
The interface between these phases hosts $\mathbb{Z}_3$ non-abelian anyons.}
\label{fig_artview}
\end{figure}

The recent observation of fractional Chern insulators in 3.7$^\circ$-twisted MoTe2 homobilayers~\cite{cai2023signatures,xu2023observation,zeng2023thermodynamic,park2023observation} and in a hBN/pentalayer graphene heterostructure~\cite{lu2023fractional} has renewed theoretical interests in anyonic phases of matter at zero magnetic field. 
For carrier densities below one charge per moir\'e unit cell, the phase diagram of these two materials (schematically depicted in Fig.~\ref{fig_artview}a) is dominated by an extended chiral flavor-polarized metallic phase, whose spontaneous polarization in a single spin/valley component highlights the enhanced role of local interactions in the flat moir\'e band~\cite{crepel2023anomalous}, while its chirality comes from the non-zero spin-Chern number of the partially occupied band~\cite{wu2019topological,devakul2021magic} and is manifest, \textit{e.g.}, in finite values of the transverse conductivity or circular dichroism~\cite{cai2023signatures,xu2023observation,zeng2023thermodynamic,park2023observation,lu2023fractional}. 
At certain commensurate fillings, and at temperature below the interaction-induced flavor gap, additional charge gaps open and fractional Chern insulators (FCIs) are stabilized~\cite{regnault2011fractional,sheng2011fractional,neupert2011fractional}, as experimentally revealed by a fractionally quantized Hall conductance~\cite{zeng2023thermodynamic,park2023observation,lu2023fractional}. 
For twisted transition metal dichalcogenide (TMD) homobilayers, the interplay and hierarchy between flavor and charge gaps can be grasped using an extended Kane-Mele-like model capturing the energetics and topology of the topmost valence bands~\cite{devakul2021magic,wu2019topological}, though the latter cannot describe some continuum features such as anomalies~\cite{sheffer2021chiral,estienne2023ideal,parhizkar2023generic} and only indirectly accounts for microscopic details that can change the landscape of observed phase (\textit{e.g.} lattice relaxation, or FCI charge gaps)~\cite{zhang2024polarization,wang2024fractional,yu2024fractional,abouelkomsan2023band,jia2024moire,redekop2024direct,reddy2023fractional}.
Tuning between metallic and FCI phases is in practice achieved by electrostatic gating (Fig.~\ref{fig_artview}b).

At the moment, all experimentally observed FCIs phases are Abelian and correspond to the topological order of a 1/3-Laughlin fractional quantum Hall (FQH) state~\cite{laughlin1983anomalous} or to Jain's principal sequence~\cite{jain1989composite}. 
A natural question opening up is the realization of non-Abelian anyons in these moir\'e platforms~\cite{kang2024observation}. 
Following proposals originally focused on FQH bilayers~\cite{mong2014universal,vaezi2014superconducting}, the polarized-metallic phase hosting FCIs sketched in Fig.~\ref{fig_artview}a can be envisioned as building blocks to construct more complicated setups in which non-Abelian anyons may be trapped and manipulated. 
More precisely, the edge of quantum Hall bilayers in which each layer feels an opposite magnetic field and stabilizes a Laughlin 1/3 topological order is predicted to be gapped and host $\mathbb{Z}_3$ parafermions when put in proximity with a superconductor~\cite{mong2014universal,vaezi2014superconducting,repellin2018numerical,barkeshli2016charge,barkeshli2011bilayer,barkeshli2012topological,katzir2020superconducting,liu2019fractional} (analogous to Fig.~\ref{fig_artview}c). 
Beyond the fundamental impact of realizing such non-abelian anyons, these parafermionic gapped boundaries were proved to be sufficient resources for universal topological quantum computation~\cite{cong2017universal}, provided they may be moved in space to perform braiding operations. 
Numerical studies in finite-size hinted that weak proximity-induced attractive interactions in the bulk of the Hall bilayer does not spoil the double Laughlin topological order~\cite{chen2012interaction,furukawa2017quantum}. 
Instead, the attraction strength needed to exceed a finite threshold --- larger than the Laughlin charge gap but comparable to the bare coulomb interaction scale stabilizing the latter --- to drive the system into a superfluid phase.

Our goal is to study a similar interplay between topological order and superconductivity in lattice systems, and describe pathways to realize non-abelian boundaries using two copies of moir\'e materials described by Fig.~\ref{fig_artview}a.
For this, we will introduce a tight-binding toy model known to capture the energetics and topology of the topmost band of spin/valley polarized twisted TMD homobilayers~\cite{crepel2024bridging}. 
While inspired by specific moir\'e heterostuctures, it will be clear from the methods used, the free parameters of our theory, and the interpretation of our results, that our objective is not a quantitative microscopic description of specific materials. 
Our goal is rather the description of the competing forces at play using a representative model containing the essential features of the problem. 
For this, we will include a weak, local, and phenomenological attractive interaction between two superimposed copies of the above model, together with a direct repulsive interaction between them. 
We shall also focus on a single commensurate fraction where FCI appear in this model, and choose a total filling of $\nu=2/3$ (\textit{i.e.} 1/3 in each copy). 
While this fraction is not observed in the experiments of Refs.~\cite{cai2023signatures,xu2023observation,zeng2023thermodynamic,park2023observation,lu2023fractional} due to either a small/vanishing flavor gap or a competing charge density wave~\cite{li2021spontaneous,song2023phase,wilhelm2021interplay,reddy2023toward}, it remains perfectly valid in our context where spin/valley polarization is assumed in each copy and an approximate particle-hole symmetry holds~\cite{lauchli2013hierarchy}. 
This filling has the advantage to be less sensitive to band mixing effects that are important at filling 2/3 in twisted MoTe$_2$~\cite{yu2024fractional}, but would unnecessarily obscure the physics described here.

\subsection{Expectations and related works}

Let us quickly brush out the different phases expected to arise in our toy model. 
To describe the assumed layer-Chern and layer-spin locking, we introduce a ``layer'' pseudo-index $\ell = \pm$, which determines both the spin of the free charge carriers and Chern number of the band they occupy: $\sigma_\ell = C_\ell = \ell$.

As discussed previously, for generic fillings of the two layers $\nu_\ell$, the system in absence of attraction simply consists of two spin polarized metals. 
When the two layers have similar fillings $\nu_\ell$ such that their Fermi surfaces are not too dissimilar, the system undergoes a Cooper instability resulting in a $s$-wave interlayer paired superconducting state in presence of an \emph{arbitrarily weak} attractive potential $J$~\cite{bardeen1957microscopic}. 
Due to the non-retarded nature of interactions, the gap of the superconducting state is proportional to $\Delta_{0} \propto \sqrt{E_F W} e^{-\alpha W/J}$ with $E_F$ the Fermi energy, $W$ the overall bandwidth, and $\alpha$ an order one constant~\cite{randeria1989bound,crepel2021new}.

This Fermi-liquid instability at infinitesimal values of $J$  contrasts with the physics when the density of both layers reach commensurate fractions and realize FCIs. 
Here, we shall primarily consider the layer filling $\nu_\ell = 1/3$, for which the stabilized FCIs fall in the same universality class as the 1/3-Laughlin state~\cite{laughlin1983anomalous}, and can be understood as band insulators of composite fermions, which are made of one charge carrier attached to two fluxes~\cite{jain1989composite,sohal2018chern}. 
Due to their different $C_\ell$, particles in both layers bind to opposite fluxes, leading to a fractional topological insulator (FTI) described by the $\bm{K}$-matrix $\bm{K} = \diag(3,-3)$~\cite{levin2009fractional,neupert2015fractional,stern2016fractional}. 
Note that the FTI is protected by an effective time-reversal symmetry that exchanges the two layers.
In some TMD hetero-stacks, the FTI can be furthermore protected by U(1) spin symmetry arising from the strong Ising spin-orbit coupling~\cite{kang2024observation}.
The time-reversal invariant FQH state described by the same $\bm{K}$-matrix has been shown to be robust against an attraction of strength smaller than its gap by exact diagonalization on small clusters~\cite{chen2012interaction,furukawa2017quantum}. 
In the composite fermion language, these results can be intuitively understood as follows: to bind fermions together through the attraction $J$, one first need to undress them from their fluxes which requires a attraction strength greater than $J_c > 0$. 
This argument has been qualitatively corroborated by a renormalization group (RG) analysis in the context of gapless composite Fermi liquid, where attractive interactions need to overcome a critical value to drive a pairing transition~\cite{metlitski2015cooper}.

\subsection{Outline and main results}

Our work is directly connected to the quantum Hall literature summarized above, but contains new insights on the interplay between pairing and topological order applied to moir\'e materials. In more details, after presenting our model in Sec.~\ref{sec_MicroscopicModel}, we
\begin{itemize}
\item Show that the essential physics observed in quantum Hall systems can also occur in lattice models, \textit{i.e.} in absence of magnetic field, and that the expected phases described above are all present in the phase diagram that we obtain on a finite-cluster using exact diagonalization (Sec.~\ref{sec_phasediagram}). 
\item Provide arguments identifying which of our finite-size observations persist in the thermodynamic limit (Sec.~\ref{sec_thermolimit}). For instance, the need of a finite attraction $J_c$ to pair composite fermions is proved irrespective of the microscopic model and beyond finite-size using a coupled-wire construction capturing the universal behavior of the topological order. Our analysis of thermodynamic properties also conveys that the residual three-fold degenerate phases observed in our numerics, and in other similar calculations~\cite{neupert2015fractional}, are finite-size effects. 
\item Discuss direct application of our theory to moir\'e material (Sec.~\ref{sec_physicalrealization}), and provide prescriptions for material realization of our model (\textit{e.g.} stacking of the two copies favoring polarization with layer-chirality locking). We also propose a new method to realize the long sought-after FTI-to-superconductor interface in the bulk of the sample by spatially changing either the strength of attractive interaction or by mere inhomogeneous electrostatic gating of the sample (Fig.~\ref{fig_artview}c). In both cases, the FTI-to-superconductor interface is realized in the bulk of the heterostructure, thereby bypassing the need of proximitized coupling and the strong disorder effects induced by crystalline defects at the edge of the system. 
\end{itemize}

\section{Microscopic model} \label{sec_MicroscopicModel}

We consider two superimposed copies of Haldane's model~\cite{haldane1988model,kane2005z2} distinguished by a ``layer'' pseudo-index $\ell = \pm 1$, which we assume decoupled at the single particle level:
\begin{subequations} \label{eq_fullmodel}
\begin{align} \label{eq_Haldanebilayer}
\mathcal{H}_0 & = - \sum_{\ell = \pm 1} \sum_{n=1}^3 \sum_{\langle r , r' \rangle_n} \left( t_n e^{i \ell \phi_n} c_{r,\ell}^\dagger c_{r',\ell} + hc \right) , \\ 
\phi_1 & = \phi_3 = 0, \quad \phi_2 = \frac{2 \pi}{3} \equiv \phi , \quad 8 t_3 = 3 t_2 = - t_1  \equiv t , \notag
\end{align}
with $\langle r, r' \rangle_n$ running over $n$-th nearest neighbor pairs of the honeycomb lattice with the convention that $\langle r, r' \rangle_2$ turns right. Our choice for the phase $\phi$ and for the tunneling amplitudes $t_{1,2,3}$ are motivated by a recent analysis of magic-angle twisted TMDs~\cite{devakul2021magic,crepel2024bridging}; it yields a bandwidth-to-gap ratio of about $1/7$ and a Berry curvature standard deviation equal to $\simeq 0.8$ times its mean. Other parameters could have been chosen as long as the resulting bands have similar features~\cite{wu2012zoology,regnault2011fractional}. Each Haldane layer describes spin-polarized electrons, which is a valid description of systems represented by Fig.~\ref{fig_artview}a at energies below the flavor-gap induced by intra-layer on-site interactions (see Sec.~\ref{sec_introduction}). We have also assumed that the Chern number of the lowest band was locked with the layer index $C_\ell = \ell$, the microscopic reasons for this are discussed in Sec.~\ref{sec_physicalrealization}.

To study the interplay between pairing and topological order, we introduce a phenomenological local attractive interaction of strength $J$ between the layers, and account for intra- and inter-layer Coulomb repulsion through positive nearest-neighbor interaction coefficients $V_\parallel$ and $V_\perp$, respectively:
\begin{align} \label{eq_intralayerrepulsion}
\mathcal{H}_\parallel & = V_\parallel \sum_{\ell=\pm 1, \langle r, r' \rangle_1} n_{r,\ell} \, n_{r',\ell} \\
\mathcal{H}_\perp & = V_\perp \sum_{\langle r, r' \rangle_1} n_{r,+1} n_{r',-1}  - J \sum_{r} n_{r,+1} n_{r,-1} ,  \label{eq_interlayerinteraction}
\end{align}
with $n_{r,\ell} = c_{r,\ell}^\dagger c_{r,\ell}$ and $V_\parallel, V_\perp, J>0$. Putative material realization of this model will likely feature ($i$) a weak effective attraction coefficient as any local pairing needs to first overcome the bare on-site electronic Coulomb repulsion between the layers, and ($ii$) inter-layer nearest neighbor interactions weaker than intra-layer ones due to the finite separation between the two layers. We shall henceforth focus on the regime $J, V_\perp \leq V_\parallel$. 
The total Hamiltonian 
\begin{equation}
\mathcal{H} = \mathcal{H}_0 + \mathcal{H}_\parallel + \mathcal{H}_\perp
\end{equation}
\end{subequations}
independently conserves the number of particles in the top and bottom layers. We can therefore diagonalize it after fixing the carrier density $\nu_\ell$ in both layers, which greatly simplifies the numerical simulations and allows to access larger system sizes.

\section{Phase diagram at 2/3 filling} \label{sec_phasediagram}

In this Section, we demonstrate the stability of the bilayer fractional topological insulator, stabilized for $J=V_\perp=0$, with respect to inter-layer interactions. Of particular interest for applications (see Sec.~\ref{sec_physicalrealization}), we show that the FTI retains its topological order for finite attraction strength $J<J_c$, with $J_c$ estimated from our finite-size numerics (dashed line in Fig.~\ref{fig_GSdegeneracy}). 
As explained in Sec.~\ref{sec_introduction}, we focus on the  total filling $\nu = \nu_+ + \nu_- = 2/3$, where we obtain the phase diagram of the model using band projected exact diagonalization in the flat band limit with up to $N_{\rm cell} = 18$ unit cells. This phase diagram hosts both an FTI and superconducting phases, respectively anticipated at small and large $J$, separated by an intermediate phases featuring three nearly degenerate ground states (see Fig.~\ref{fig_GSdegeneracy}). Checking which of these phases persist in the thermodynamic limit is the focus of Sec.~\ref{sec_thermolimit}.

\subsection{Methods}

To determine the phase diagram of the Hamiltonian Eq.~\ref{eq_fullmodel} at filling $\nu = 2/3$ in presence of interactions that exceeds the small bandwidth of the lowest bands, we perform band projected exact diagonalization (ED) in the flat-band limit using the free open-source software \href{https://nick-ux.org/diagham/index.php/Main_Page}{``DiagHam''}. To mitigate finite-size effects, we use tilted finite size clusters with aspect ratios as close to the thermodynamic value $\sqrt{3}/2$, which corresponds to having the same number of unit cell along the two primitive vectors $(a_1, a_2)$ of the triangular Bravais lattice. The total density $\nu$ and our numerical resources limit the accessible system sizes to a total of unit cells $N_{\rm cell} \leq 18$, and we will only present data obtained for $N_{\rm cell} \geq 12$ as no FTIs were observed on smaller clusters. This leaves three different system sizes, for which we choose the tilting vectors $T_{i=1,2} = n_{i,1} a_1 + n_{i,2} a_2$, as defined in Ref.~\cite{repellin2014z}, and the number of unit cell in those directions $N_{i=1,2}$ to be 
\begin{equation} \label{eq_tiltedclusters}
\begin{array}{c||c|c|c|c|c|c}
N_{\rm cell} & N_1 & N_2 & n_{1,1} & n_{1,2} & n_{2,1} & n_{2,2} \\ \hline 
12 & 6 & 2 & 2 &-4 & 2 & 2  \\ 
15 & 3 & 5 & 1 &-5 & 3 & 0  \\ 
18 & 9 & 2 & 3 &-6 & 2 & 2 
\end{array} \, .
\end{equation}
Only the $N_{\rm cell}=15$ aspect ratio significantly differs from $\sqrt{3}/2$, which leads to a larger spread of the FTI ground state manifold in our numerics. The largest Hilbert spaces considered correspond to a balanced bilayer with $12$ particles, \textit{i.e.} $6$ in each layer, populating $2 N_{\rm cell}=36$ different one-body orbitals, and have dimension $\simeq 1.9\cdot 10^7$ in each of the $N_{\rm cell}=18$ possible many-body center of mass momentum sectors. 

\subsection{Phase diagram} \label{subsec_phasediag}

To investigate the phase diagram, we first determine the layer polarization $P_z = \frac{\nu_+ - \nu_-}{\nu_+ + \nu_-}$ of the ground state by diagonalization of Eq.~\ref{eq_fullmodel} in all the $(\nu_+, \nu_-)$ sectors compatible with $\nu=\nu_+ + \nu_-=2/3$. Because the model is invariant under an effective time-reversal symmetry $\Tilde{\mathcal{T}}$ consisting of layer inversion followed by complex conjugation, we restricted our attention to the $\nu_+ \geq \nu_-$ sectors, and found the balanced bilayer $P_z = 0$ to be energetically favored for all the parameters and system sizes that considered in our study (see App.~\ref{app_layerpolarization}). We thus set $\nu_+ = \nu_-$ from now on. 

\begin{figure}
\centering
\includegraphics[width=\columnwidth]{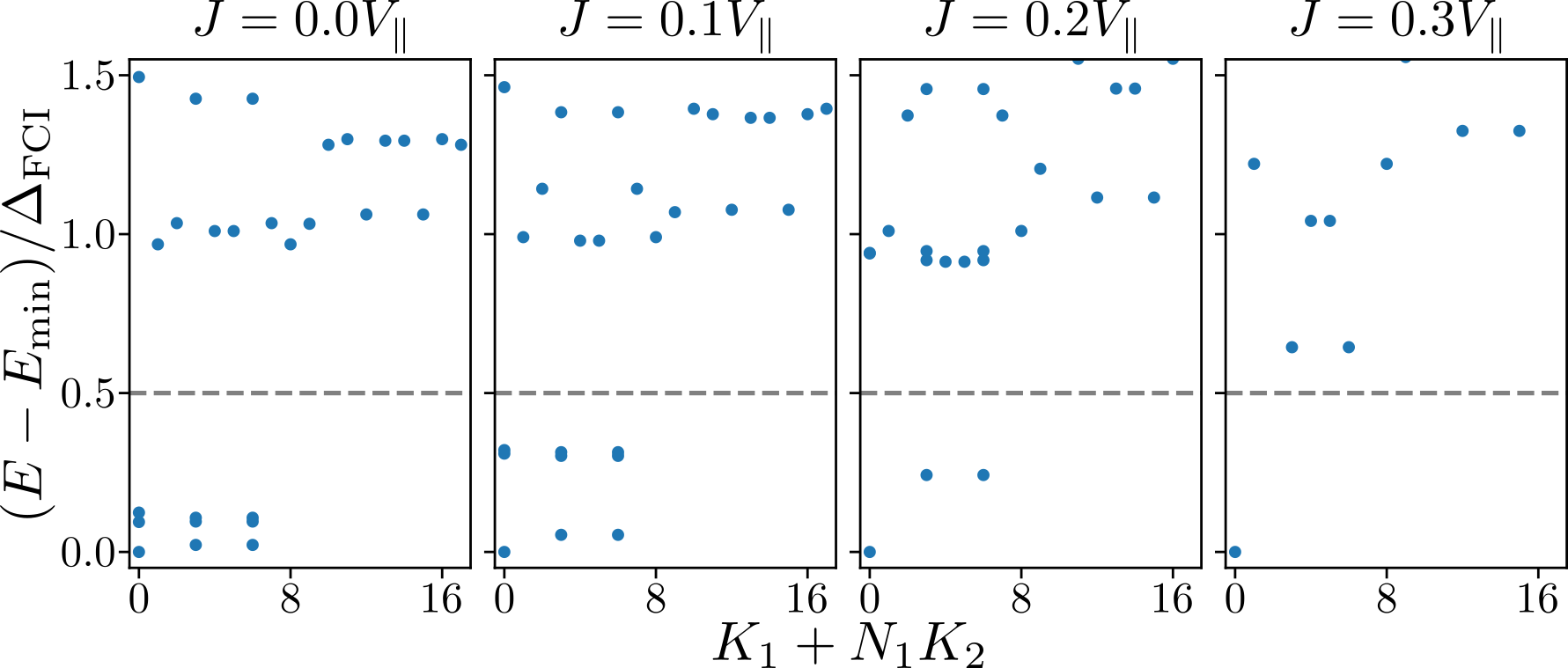}
\caption{Representative examples of momentum resolved ED spectra, obtained for $N_{\rm cell}=18$ and $V_\perp = 0.1 V_\parallel$ for different values of $J$. $(K_1,K_2)$ denote the many-body momenta along the $T_{1/2}$ directions. Here, energies are rescaled with respect to the gap $\Delta_{\rm FCI}$ of the FCI phase from a single layer at $\nu=1/3$. 
We also show how the cutoff $\varepsilon$ allows to differentiate the 9-fold, 3-fold and singly degenerate phases.}
\label{fig_edexamples}
\end{figure}

\begin{figure}
\centering
\includegraphics[width=\columnwidth]{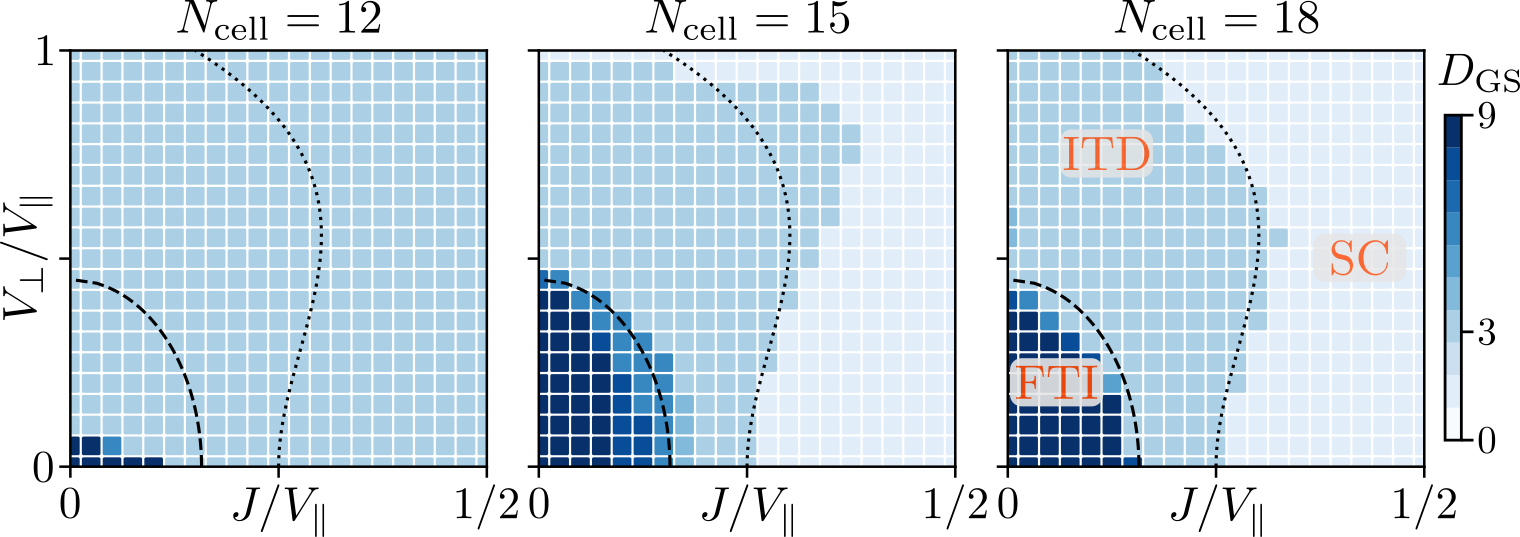}
\caption{Ground state degeneracy $D_{\rm GS}$ of the ground state manifold of Eq.~\ref{eq_fullmodel} for different system sizes $N_{\rm cell}$. 
The anticipated fractional topological insulator (FTI) and superconductor (SC) phases are separated by an intermediate three-fold degenerate (ITD) region. 
The dashed and dotted lines represents the numerically estimated boundaries between these phases for $N_{\rm cell} = 18$ and should only be used as guides to the eye.}
\label{fig_GSdegeneracy}
\end{figure}

To quickly identify the phases appearing in our model, we estimate the ground state degeneracy $D_{\rm GS}$ of the balanced bilayer. For this, we choose a coarse energy cutoff $\varepsilon$ and count the number of many-body states whose energy difference with the ground state is smaller than $\varepsilon$, as represented graphically in Fig.~\ref{fig_edexamples}. To determine a consistent cutoff for all system sizes, we diagonalize the model in the layer-polarized regime $\nu_+ = 1/3$ and $\nu_-=0$, where the system behaves as an extended Haldane model which is known to host an FCI ground state (see App.~\ref{app_proofFCI}), we extract its gap many-body gap $\Delta_{\rm FCI}$ (independent of $V_\perp,J$, and proportional to $V_\parallel$ in the flat band limit that we are considering here) and arbitrarily set $\varepsilon = \Delta_{\rm FCI}/2$. With this choice, we obtain the ground state degeneracy $D_{\rm GS}$ depicted in Fig.~\ref{fig_GSdegeneracy}, where three regions can be clearly identified: a nine-fold degenerate phase adiabatically connected to the decoupled layer limit $J=V_\perp=0$; a singly-degenerate ground state obtained for large attractive interactions, which emerges for $J \gtrsim 0.25 V_\parallel$ in our numerical simulations; and an intermediate three-fold degenerate (ITD) region separating the two previous phases. While these three phases appear for any choice of $\varepsilon$, their exact boundaries depend on the cutoff $\varepsilon$. 
We show in App.~\ref{app_phaseboundaries} that the phase boundaries obtained in Fig.~\ref{fig_GSdegeneracy} provides good estimates of those determined by direct inspection of the numerical spread to gap ratios, which guided our choice for $\varepsilon$. 

\begin{figure}
\centering
\includegraphics[width=\columnwidth]{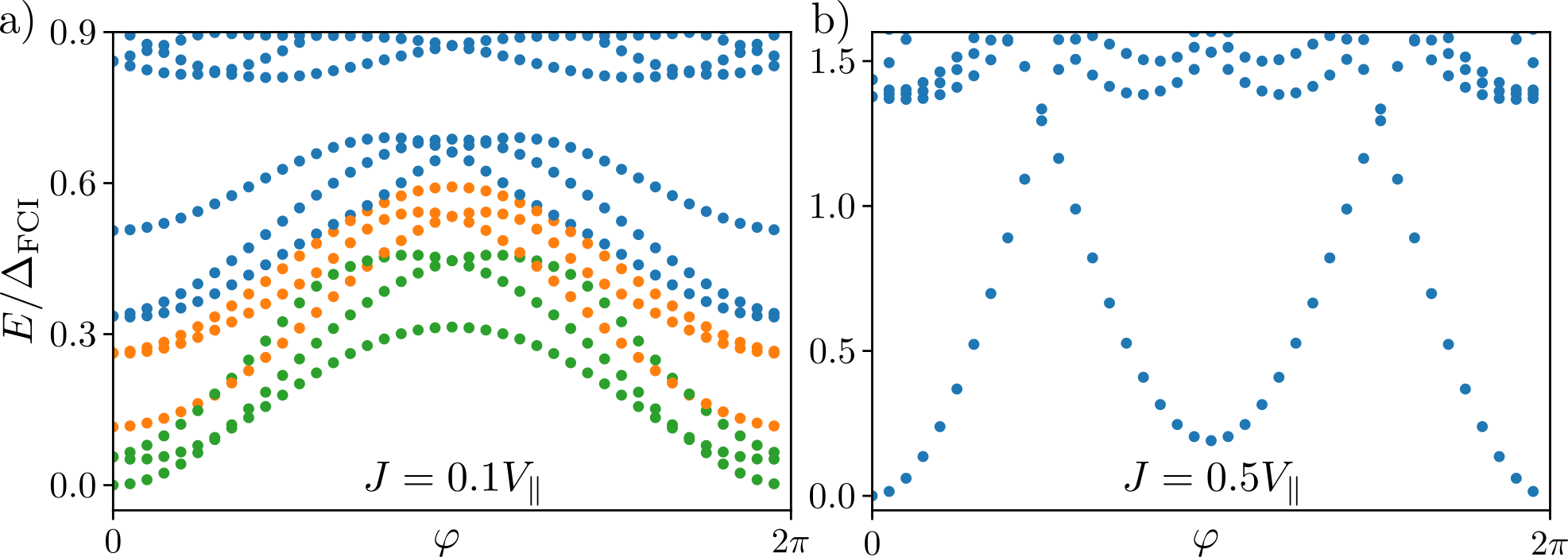}
\caption{Flow of the low-energy state of upon $\varphi = 0 \to 2\pi$ flux insertion in the direction of the tilted lattice vector $T_1$ for $N_{\rm cell} = 15$, $V_\perp = 0$ in FTI phase $J=0.1V_\parallel$ (a), and the SC phase $J=0.5V_\parallel$ (b). The colors in (a) are guides to the eye highlighting the $6\pi$-periodicity of the 9-fold degenerate ground state manifold under flux insertion, they carry no physical meaning.
}
\label{fig_chargeflux}
\end{figure}

The nature of the two first phases is clear from the specific limit they are connected to: the nine-fold degenerate phase is a FTI connected to the tensor product of two layer-polarized FCIs related by the effective time-reversal symmetry $\Tilde{\mathcal{T}}$; the singly degenerate state appears in the limit $J>0$ and $V_{\perp}, V_\parallel = 0$ where the system becomes unstable to inter-layer pairing and becomes a superconductor (SC). This is confirmed by the spectral flows of the ground state manifold under a $2\pi$-flux insertion in the direction of $T_1$, which are shown in Fig.~\ref{fig_chargeflux}. There, we see that the two phases respectively have a periodicity of $3\times 2\pi=6\pi$ (FTI) and $2\pi/2=\pi$ (SC), highlighting the charge $e/3$ and $2e$ carried by elementary excitations in each of these phases~\cite{scalapino1993insulator,loder2008crossover}. 

The intermediate three-fold degenerate region (ITD) is investigated further in Sec.~\ref{subsec_threefoldisfinitesize}, where we argue that it is an artefact of the momentum discretization of our finite-size clusters (Eq.~\ref{eq_tiltedclusters}). In App.~\ref{app_phaseseparation}, we provide additional data for values of $V_\perp > V_\parallel$ beyond the physical regime identified in Sec.~\ref{sec_MicroscopicModel}, where the system shows signs of phase separation. Note that a similar tendency to phase separation was observed in Ref.~\cite{chen2012interaction} for a continuum model based on Landau levels with opposite magnetic to realize the time-reversal symmetry.

\section{Thermodynamic properties} \label{sec_thermolimit}

This section is devoted to checking which of the phases observed on finite-size clusters persist in the thermodynamic limit. Sec.~\ref{subsec_coupledwire} shows that the FTI ground state manifold keeps its topological order in the thermodynamic limit for non-zero attractive interactions in spite of the increasing spread observed in our finite-size numerical calculations when $J$ or $V_\perp$ increase. Sec.~\ref{subsec_threefoldisfinitesize} argues that the intermediate phase with three nearly degenerate ground states observed in our finite-size numerics is an artefact of the momentum-space discretization involved in our numerics and disappears in the thermodynamic limit.

\subsection{Stability of the topological order}  \label{subsec_coupledwire}

Our numerical simulations on finite clusters corroborate earlier studies~\cite{chen2012interaction,furukawa2017quantum} suggesting that the FTI order at $\nu = 1/3+1/3$ is stable against perturbative attractive interactions, which confirms our intuitive picture that a finite coupling strength $J_c$ is needed to undress the composite fermions from their fluxes and pair the original fermionic degrees of freedom. To demonstrate the universal character of this observation, and prove that it also holds in the thermodynamic limit, we now use a coupled-wire construction for the fractional topological order and show that it is robust to attractive interactions.

At $J = V_\perp = 0$, the FTI is the product of two time-reversal copies of the Laughlin 1/3 state. This state is the prototypical example of intrinsic topological order captured by a coupled wire construction~\cite{kane2002fractional,teo2014luttinger,crepel2020microscopic}. As a result, we can describe the universal features of the FTI using two decoupled copies of the Laughlin's $1/3$ wire construction feeling opposite fluxes (one for each layer)~\cite{neupert2014wire,klinovaja2014quantum,santos2015fractional}, as described by the low-energy action~\footnote{see Refs.~\cite{kane2002fractional,teo2014luttinger,neupert2014wire,klinovaja2014quantum,santos2015fractional,mukhopadhyay2001sliding} for details and App.~\ref{app_exactpoint} for a short derivation of this effective action.}
\begin{align} \label{eq_coupledwirebeforeattration}
& S = \sum_{\substack{ \ell = \pm \\ j \in {\rm links} }} \int {\rm d}x {\rm d}\tau \left[ \frac{i \ell}{3 \pi} (\partial_\tau \varphi_{j,\ell})(\partial_x \theta_{j,\ell}) - \mathcal{H}^{(j,\ell)} \right] \\
& \left[ \partial_x \theta_{j,\ell} (x), \varphi_{j',\ell'}(x') \right] = 3 i \pi \ell \delta_{j,j'} \delta_{\ell,\ell'} \delta (x-x') ,
\end{align} 
where $\partial_x \theta_{j,\ell}$ and $\varphi_{j,\ell}$ are low-energy density and phase variables, respectively, for layer $\ell$ that are located on the link between wire $j$ and $j+1$. In the present context, the Hamiltonian is split in three parts $\mathcal{H} = \mathcal{H}_{\rm LL} + \mathcal{H}_{\rm SG} + \mathcal{H}_{J}$ representing \begin{itemize}
\item the low-energy Luttinger liquid on each wire $\mathcal{H}_{\rm LL}^{(j,\ell)} = K_0^{-1} (\partial_x \theta_{j,\ell})^2 + K_0 (\partial_x \varphi_{j,\ell})^2$ where $K_0>0$.
\item the sine-Gordon mass term $\mathcal{H}_{\rm SG} = g \cos ( 2 \theta_{j,\ell})$ coming from allowed inter-wire tunneling.
\item the local density-density attractive interactions of our model that primarily couple to the low-energy degrees of freedom as an on-wire perturbation $\mathcal{H}_{J} = - 2 J^{\rm eff} (\partial_x \theta_{j,+}) (\partial_x \theta_{j,-})$.
\end{itemize} 
In absence of this last term, namely $J^{\rm eff}=0$, the flow of $g$ toward strong coupling together with the commutation relations between the density and phase variables in Eq.~\ref{eq_coupledwirebeforeattration} reproduce the desired topological order and captures its topological quasiparticle excitations~\cite{kane2002fractional,teo2014luttinger,neupert2014wire,klinovaja2014quantum,santos2015fractional}. In fact, when $J^{\rm eff} > 0$ is small enough, the bosonic theory for density operators obtained after integrating out the phase variable $\varphi$ remains unchanged when expressed in terms of charge $\theta_c = (\theta_+ + \theta_-)/\sqrt{2}$ and pseudo-spin $\theta_s = (\theta_+ - \theta_-)/\sqrt{2}$ variables, except for renormalized Luttinger parameters $(K_c/K_0)^{-1} = 3 \sqrt{1-K_0 J^{\rm eff}}$ and $(K_s/K_0)^{-1} = 3 \sqrt{1+K_0 J^{\rm eff}}$ (see App.~\ref{app_exactpoint}). Therefore, as long as $J^{\rm eff} < J_c^{\rm eff} = K_0^{-1}$, the low energy content of the coupled-wire construction is unchanged and the system remains in the same topological phase.

While $J^{\rm eff}$ may differ from the physical attraction strength $J$ due to the low energy projection and the phenomenological nature of the coupled-wire construction, our analysis proves that the universal topological content of the FTI is robust against perturbative attractive interaction. Together with the evidence of a robust many-body gap provided by our finite-size numerical calculations, this demonstrate the stability of the FTI against small attractive interactions. Note that a similar argument holds true for the repulsive inter-layer density-density interactions (change $-J^{\rm eff}$ to a positive coefficient $V_\perp^{\rm eff}$ above).

\subsection{Finite-size intermediate 3-fold degeneracy} \label{subsec_threefoldisfinitesize}

We now analyze in more detail the three-fold degenerate intermediate phase appearing in our finite size calculations, the ITD phase in Fig.~\ref{fig_GSdegeneracy}. We need to say that a similar three-fold degenerate phase was also observed in a repulsive model~\cite{neupert2011fractional,neupert2011fractional2}. 
If topologically ordered, the ground state degeneracy of this phase would only be consistent with a non-Abelian fractional topological order~\cite{neupert2015fractional}. 
Here, we show evidence that this intermediate region only exists for certain system sizes and momentum discretizations. 
This observation hinders the interpretation of this region as a robust phase with topological order, but is consistent and easily explained by finite-size level repulsion on the small momentum-space discretized grids involved in the numerical simulations.

\begin{figure}
\centering
\includegraphics[width=\columnwidth]{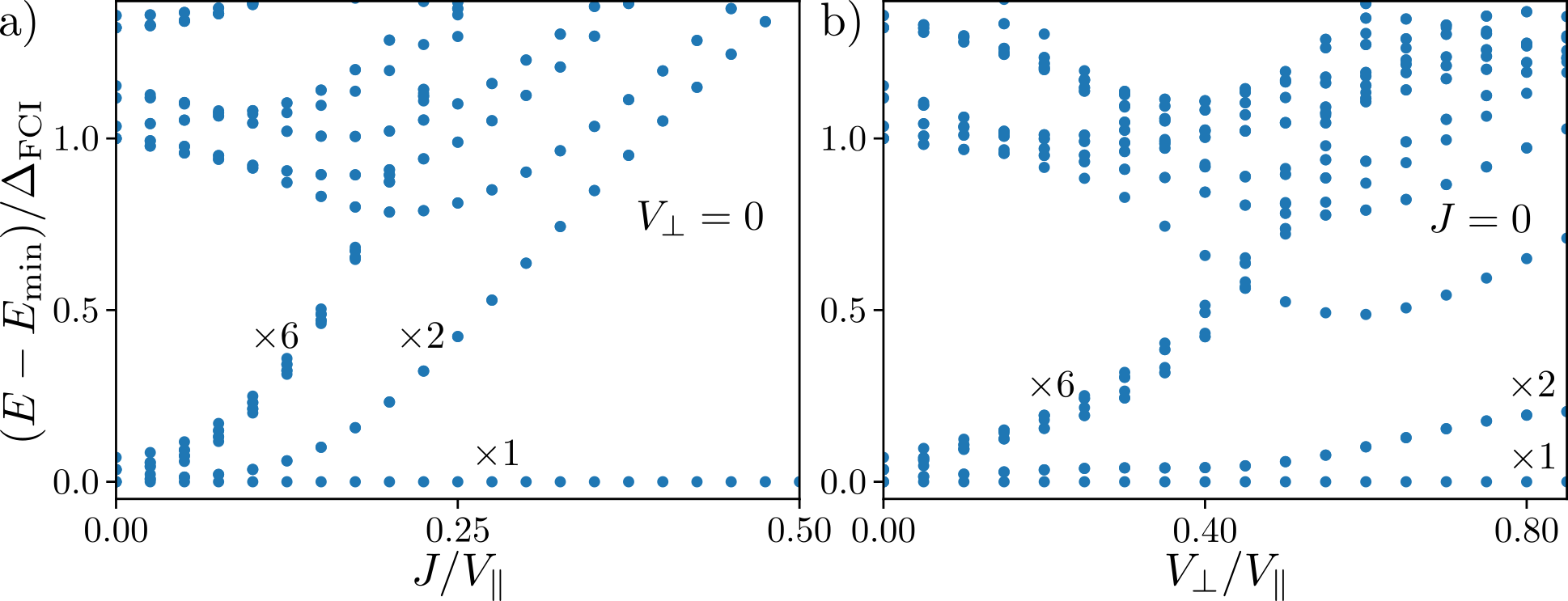}
\caption{Many-body spectra, with energy measured relative to the ground state, as a function of $J$ at $V_\perp=0$ (a) and of $V_\perp$ at $J=0$ (b), obtained for $N_{\rm cell}=18$. 
}
\label{fig_threestates}
\end{figure}

To better understand the finite size structure of the lattice considered, we display the full many-body spectrum along the $V_\perp =0$ and $J=0$ lines in Fig.~\ref{fig_threestates}. When the inter-layer interaction coefficients $V_\perp$ and $J$ are small, we observe that the nine-fold degenerate FTI ground state manifold split into two degenerate groups. The group with lowest energy contains three states, the one with higher energy has six and keeps increasing in energy until it merges with the many-body continuum. For larger values of the inter-layer couplings, the low energy group splits again to yield a non-degenerate ground state and two excited states. The resulting singly degenerate state corresponds to the superconductor labeled SC in Fig.~\ref{fig_GSdegeneracy}. The three-fold degenerate intermediate phase is therefore rooted into the splitting of the FTI manifold as $9=3+6$ that occurs for small inter-layer interaction.

We attribute this behavior to finite-size effects and argue that the intermediate phase does not appear in the thermodynamic limit. More precisely, for all finite clusters considered, the FTI ground states appears in triplets in three distinct many-body momentum sectors (see Fig.~\ref{fig_proofFCI} in App.~\ref{app_proofFCI}). In finite-size, nearly degenerate states within the same momentum sector experience level repulsion and split more rapidly than degenerate states originally lying in different momentum sectors. In each of the FTI triplets, the first effect of inter-layer interactions is therefore to lower the energy of a single state through level repulsion, leading to a three-fold degenerate phase whose states occupy different momentum sectors -- which we checked is the case in the ITD. 

Following the same reasoning, we expect that any lattice where all FTI states lie in the same many-body momentum sector, which includes the system in the thermodynamic limit~\cite{bernevig2012emergent}, will not feature any ITD region but directly transition from the FTI to the SC. Indeed, the interlayer interaction would single out only one state from the ground state manifold (see also additional details in App.~\ref{app_rankoneperturbation}). 
To check this hypothesis, we performed exact diagonalization of the model of Ref.~\cite{crepel2024bridging} using a different set of parameter than Eq.~\ref{eq_fullmodel} for which an FTI with all nine degenerate ground states appear in the same momentum sector is stabilized on a $6 \times 3$ cluster for $J=V_\perp=0$ -- see details in App.~\ref{app_rankoneperturbation}. 
Increasing $J$ as in Fig.~\ref{fig_threestates}a, we observe no intermediate regions with an approximate three-fold degeneracy.
The small region between the FTI and SC phases shows a splitting of the nine original FTI ground states as 1+8, with the lowest energy states adiabatically connected to the SC and the other eight others merging into the many-body continuum as $J$ increases. 
This additional piece of evidence supports our interpretation of the ITD region as a finite-size artifact.

\section{Physical realization with stacked twisted bilayers} \label{sec_physicalrealization}

In this section, we identify two-dimensional heterostructures that may realize the model Eq.~\ref{eq_fullmodel}. We then envision practical ways to create FTI-to-superconductor interfaces within the bulk of these heterostructures, paving the way for the clean trapping and manipulation of non-Abelian anyons (as discussed in Sec.~\ref{sec_introduction}). In this context, the emergence of parafermions at the interface has been theoretically demonstrated several times by studying the low-energy properties of the FTI edge modes coupled to the superconductor~\cite{mong2014universal,vaezi2014superconducting,repellin2018numerical,barkeshli2016charge,barkeshli2011bilayer,barkeshli2012topological,katzir2020superconducting}. Our goal here is not to repeat those calculations, but rather to identify physical platform where this physics could be realized.

\subsection{TMD double bilayers}

\begin{figure}
\centering
\includegraphics[width=\columnwidth]{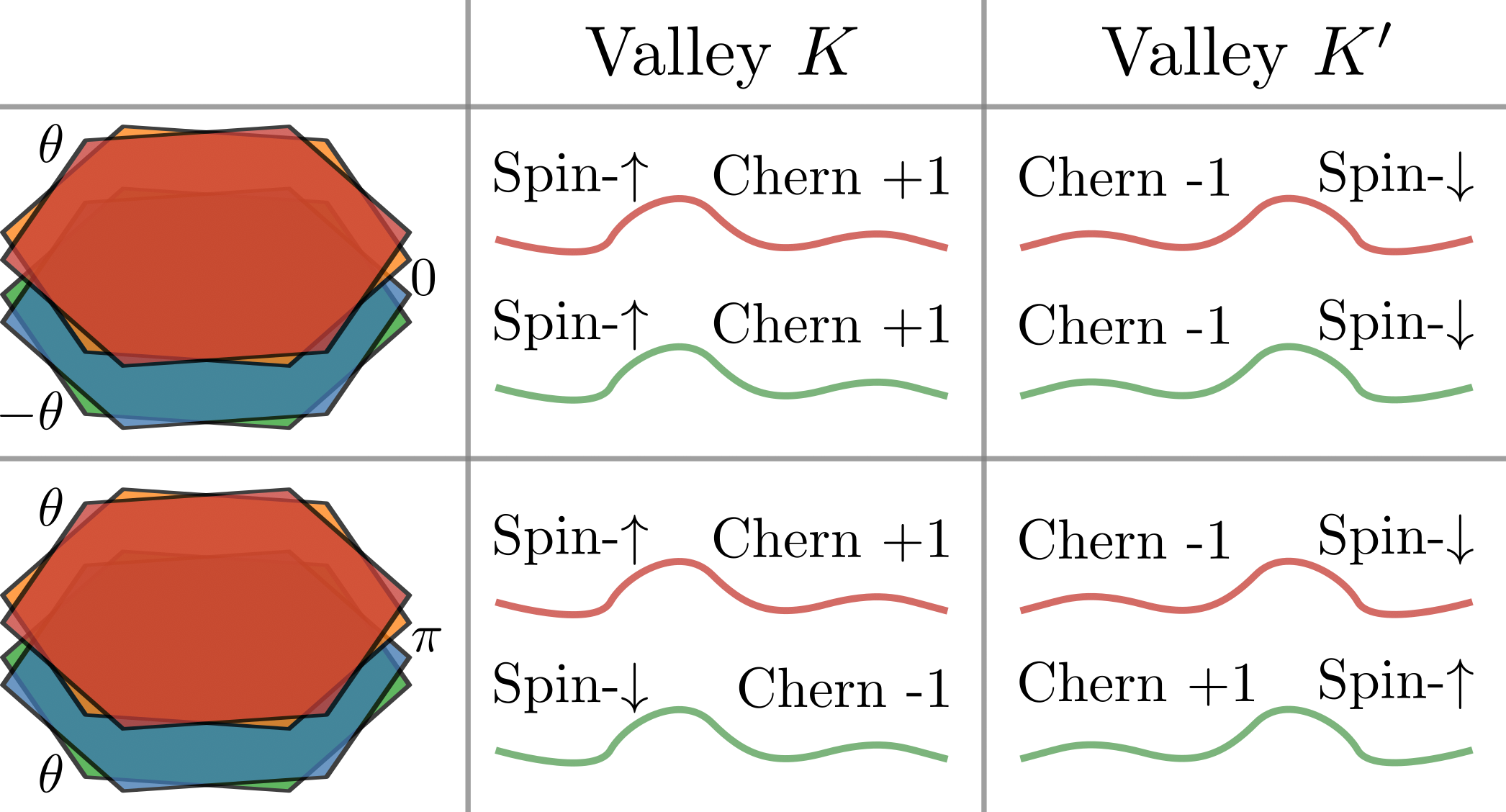}
\caption{Chern number and spin-polarization of the topmost valence band of the two stacked tTMD (in absence of inter-tTMD coupling) in valley $K$ and $K'$ for the two sequences of twist angles $(\theta , 0 , -\theta)$ and $(\theta , \pi , \theta)$ for which the stacked tTMD have overlapping moir\'e honeycomb lattice. 
}
\label{fig_sketchdoublebilayer}
\end{figure}

Recalling that twisted transition metal dichalcogenide homobilayers (tTMD) spontaneously polarize and realize~\cite{crepel2023anomalous} a single copy of an extended Haldane model at temperatures $T < \Delta_{\rm spin}$ for moir\'e filling below unity~\cite{devakul2021magic}, we consider two such tTMD stacked on top of one another in order to realize the two ``Haldane layers'' of Eq.~\ref{eq_Haldanebilayer}. There are only two choices of twist angles and stacking for which the moir\'e honeycomb lattice of the top and bottom tTMD match, corresponding to the sequence of twist angles $(\theta , 0 , -\theta)$ and $(\theta , \pi , \theta)$ between consecutive layers. These two different stacking configurations, sketched in Fig.~\ref{fig_sketchdoublebilayer}, differ in the chirality and spin carried by the moir\'e bands from the two stacked tTMD in each valley. 

In the following paragraphs, we examine two mechanisms for achieving spin-polarization with opposite chiralities in the two tTMD, and inducing a weak pairing potential between them. The first and most straightforward one consist in placing an $s$-wave superconductor in proximity with the double bilayer, thereby inducing antiferromagnetic and attractive interactions into the latter. The second mechanism is \textit{intrinsic} to the double bilayer, \textit{i.e.} it does not rely on externally imposed potentials or proximitization effects. Indeed, we will see that the perturbative effect of the weak inter-tTMD tunneling is precisely to energetically favor states of opposite chirality and to induce a weak inter-tTMD attraction. Both mechanisms only apply to the $(\theta , 0 , -\theta)$ configuration, making it a more promising platform to materialize the physics displayed in this work. 

In realistic settings, the second situation is less likely to occur as the effective attraction induced by the weak inter-tTMD tunneling should overcome the bare Coulomb repulsion. The precise balance in this competition including all forces involved in mediating attractive interactions (\textit{e.g.} phonons) goes beyond the scope of the present work. Here, we do not provide a detailed microscopic description of the double bilayer and assume that the band structure of the individual tTMDs is unaffected by the stacking. Even if the second and third layer are commensurately stacked (without twist), lattice relaxations can still renormalize the parameters of our theory and play an important role~\cite{carr2018relaxation}. We leave a proper \textit{ab-initio} study to future work. Nonetheless, at the phenomenological level, the intrinsic pairing potential existing due to the weak inter-tTMD tunneling opens a promiseful route towards the spontaneous emergence of superconductivity in TMD heterostructures (see also Refs.~\cite{crepel2022spin, crepel2022unconventional,crepel2023topological}), and the trapping of non-Abelian anyon by mere application of electrostatic potentials.

\subsection{Proximity induced superconductivity} 

\subsubsection{Superconducting potential}

We first investigate the effect of a $s$-wave superconducting pairing potential induced on the heterostructure by proximity effect. An externally imposed $s$-wave superconducting pairing potential will couple to pairs of electrons having ($i$) opposite spin, ($ii$) zero center of mass and hence carrying opposite valley index, and ($iii$) no orbital angular momentum and therefore filling bands with opposite Chern number. Given these rules, we observe that the superconducting potential does not couple the two stacked tTMD in the $(\theta,\pi,\theta)$ stacking configuration (see Fig.~\ref{fig_sketchdoublebilayer}), and we therefore discard this situation from now on. 
In $(\theta,0,-\theta)$ double bilayers, the intra-layer ferromagnetism and superconducting coupling are compatible and combine to yield states where opposite layers are filled with electrons with opposite spin, opposite valley and opposite chirality (see Fig.~\ref{fig_sketchdoublebilayer}). This situation is described by the bilayer Haldane model Eq.~\ref{eq_Haldanebilayer}, where the residual pairing induced by the superconducting potential can be phenomenologically captured by the attractive interaction of Eq.~\ref{eq_interlayerinteraction}.

\subsubsection{Gate-defined superconductor-to-FTI interface}

Stacked tTMD with $(\theta,0,-\theta)$ stacking in proximity to a $s$-wave superconductor provides a first realization of our model Eq.~\ref{eq_fullmodel}. We now discuss how to engineer a superconductor-to-FTI interface within the bulk of this double bilayer using this proximity effect. For this, we realize that the coherence length of conventional superconductor can be of order or larger than the typical size of two-dimensional stacked heterostructures. As an example, supercurrents induced by a $\sim \SI{250}{\nano\meter}$-distant superconducting electrode have been measured in graphene~\cite{heersche2007bipolar}; a distance that should be compared with typical moir\'e lattice constants in the range $\sim \SI{5}{\nano\meter}$. 

\begin{figure}
\centering
\includegraphics[width=\columnwidth]{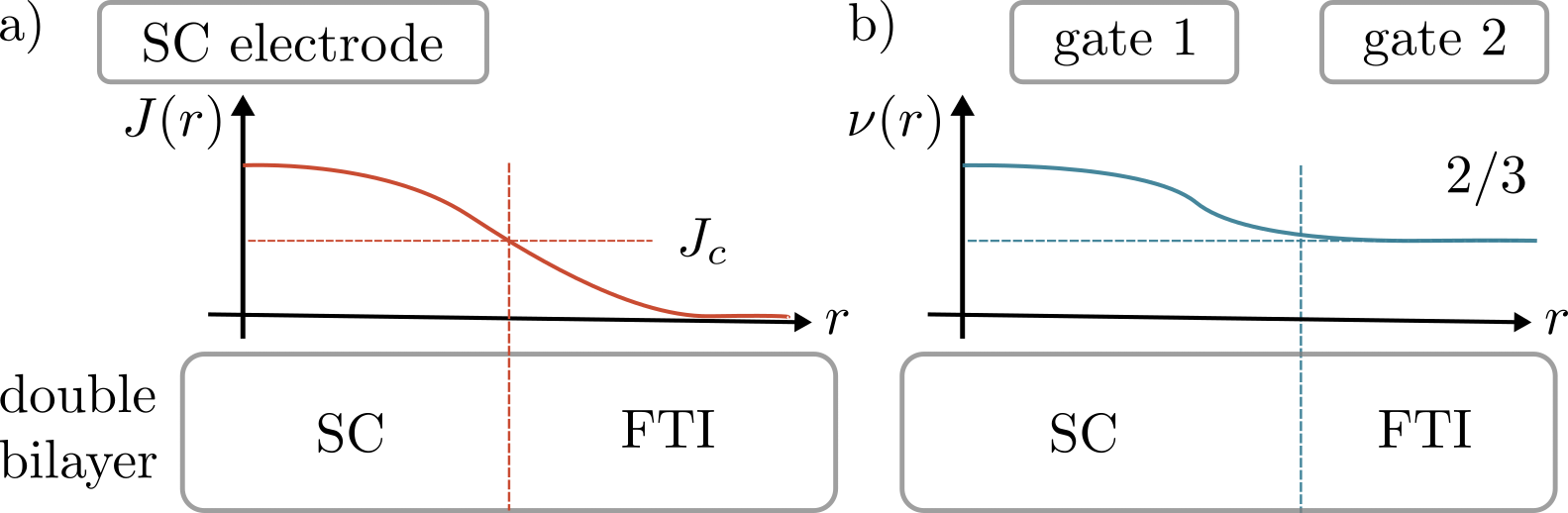}
\caption{Realization of a superconductor-to-FTI interface in the bulk of the $(\theta,0,-\theta)$ tTMD bilayer using ($a$) proximity induced superconductivity, and ($b$) electrostatic gating to tune the density in presence of an intrinsic pairing in the double bilayer. 
}
\label{fig_sketchinterface}
\end{figure}

As sketched in Fig.~\ref{fig_sketchinterface}a, a superconducting electrode localized on one side of the heterostructure will therefore induce a spatially varying pairing potential in the heterostructure. In Eq.~\ref{eq_fullmodel}, this would correspond to a non-uniform $J(r)$ coefficient decreasing with the $r$ distance to the electrode. Setting the total filling of the heterostructure to $\nu=2/3$ through electrostatic gating, the system will stabilize a FTI far away from the electrode where $J(r\to\infty)\simeq 0$ [labeled FTI in Fig.~\ref{fig_GSdegeneracy}] and a superconductor in the region where $J(r\to 0)$ exceeds $J_c$ close to the electrode [labeled SC in Fig.~\ref{fig_GSdegeneracy}].

\subsection{Spontaneous emergence due to pertubative interlayer tunneling} 

We now argue that the $(\theta,0,-\theta)$ double bilayers may not necessarily need the application of an external superconducting potential for the realization of Haldane copies with opposite chirality, nor for the emergence of attractive interactions. Indeed, both of these effects can spontaneously emerge when we account for the action of the weak inter-tTMD tunneling. Although perturbatively weak, these effects are sufficient to drive, for instance, superconducting instabilities in Fermi liquid states. As stated above, this weak attraction competes with the bare Coulomb repulsion between the two stacked bilayers. Our aim here is only to uncover this potential pairing mechanism, the precise balance between all attractive and repulsive contributions to the inter-tTMD interaction kernel going beyond the scope of this work.

Let us first highlight that the tunneling amplitude between aligned TMD monolayers typically range between $5-\SI{10}{\milli\electronvolt}$~\cite{tong2017topological}, which is smaller than the inter-layer tunneling of tTMD. The latter can be estimated by fitting continuum models to large-scale \textit{ab-initio} calculations of the twisted bilayer, which has lead to a maximum tunneling amplitude of $3 \times\SI{23.8}{\milli\electronvolt} = \SI{71.4}{\milli\electronvolt}$ in the moir\'e unit cell for twisted MoTe$_2$~\cite{wang2024fractional}. This stronger hybridization is usually explained by the important lattice relaxation occurring in the twisted heterostructures~\cite{wang2024fractional,yu2024fractional,devakul2021magic}. For the present discussion, this difference between tunneling parameters justifies including the tunneling $t_\perp$ between the two stacked tTMD perturbatively.

\subsubsection{Pauli-blocking induced antiferromagnetism}

In absence of any inter-tTMD couplings and at filling $\nu_+, \nu_- < 1$, the two stacked bilayers spontaneously develop a spin-valley gap $\Delta_{\rm spin}$ and valley-polarize (Fig.~\ref{fig_artview}a). In a simplified picture, the spin gap can be seen as an energy shift applied to all single particle state of the unpopulated valleys. The inclusion of $t_\perp$ differentiates the cases where polarization of the two stacked tTMD occurs in the same or in opposite valleys. Indeed, since $t_\perp$ is local, it carries no momentum and only provides intra-valley hoppings. If both tTMD are polarized in the same valley, this hopping is Pauli-blocked and has almost no effect on the total energy of the system. On the other hand, when the two tTMD are polarized in opposite valleys, the virtual hopping to states with energy $\Delta_{\rm spin}$ yields an energy gain $\delta E \simeq - (\nu_+ + \nu_-) t_\perp^2 / \Delta_{\rm spin}$. This favors opposite valley polarization in the two stacked tTMD, leading to the realization of two Haldane copies with opposite chirality when the tTMD are $(\theta,0,-\theta)$-stacked (see Fig.~\ref{fig_sketchdoublebilayer}).

The previous argument becomes analytically exact in the chiral limit of tTMD~\cite{crepel2023chiral,crepel2024topologically} at filling $\nu_+ = \nu_- = 1$, the argument proceeding in a manner exactly analogous to the description of ferromagnetic states appearing at integer filling in the chiral limit of twisted bilayer graphene~\cite{bultinck2020ground}. The phenomenological description given above can also be seen as an adiabatic evolution away from this exactly solvable point.

\subsubsection{Short-range attractive potential}

In addition to this valley-antiferromagnetic coupling between the two tTMD, second order processes in $t_\perp$ also induce an effective interaction between the two stacked bilayers. To estimate this contribution, we perform a diagrammatic expansion within the random phase approximation, and evaluate all diagrams to lowest order in $t_\perp$ (second order) using the decoupled tTMD case as zero-th order in the expansion, which simply corresponds to the intra-tTMD interaction kernel $v_\parallel(q)$. The relevant terms can be represented by the following diagrams
\begin{align}
&
\begin{tikzpicture}[baseline=0cm,scale=0.8]
\draw[color=BrickRed,line width=1.25] (0.25,0) arc(0:180:0.25);
\draw[color=OliveGreen,line width=1.25] (-0.25,0) arc(180:360:0.25);
\draw[fill] (0.25,0) circle (0.075); \draw[fill] (-0.25,0) circle (0.075);
\draw[color=BrickRed,line width=1.25] (-0.5,0.75) -- (0.5,0.75); 
\draw[color=OliveGreen,line width=1.25] (-0.5,-0.75) -- (0.5,-0.75); 
\draw[snake it] (0.,0.75) -- (0.,0.25); \draw[snake it] (0.,-0.75) -- (0,-0.25);
\end{tikzpicture} , \quad
\begin{tikzpicture}[baseline=0cm,scale=0.8]
\begin{scope}[rotate=45]
\draw[color=BrickRed,line width=1.25] (0.25,0) arc(0:180:0.25);
\draw[color=OliveGreen,line width=1.25] (-0.25,0) arc(180:360:0.25);
\draw[fill] (0.25,0) circle (0.075); \draw[fill] (-0.25,0) circle (0.075);
\end{scope}
\draw[color=OliveGreen,line width=1.25] (1.,0.) circle (0.25);
\draw[color=BrickRed,line width=1.25] (-0.5,0.75) -- (1.5,0.75); 
\draw[color=OliveGreen,line width=1.25] (-0.5,-0.75) -- (1.5,-0.75); 
\draw[snake it] (0.,0.75) -- (0.,0.25); \draw[snake it] (0.25,0.) -- (0.75,0.); \draw[snake it] (1.,-0.75) -- (1,-0.25);
\end{tikzpicture} , \quad
\begin{tikzpicture}[baseline=0cm,scale=0.8]
\begin{scope}[shift={(1.,0.)},rotate=45]
\draw[color=BrickRed,line width=1.25] (0.25,0) arc(0:180:0.25);
\draw[color=OliveGreen,line width=1.25] (-0.25,0) arc(180:360:0.25);
\draw[fill] (0.25,0) circle (0.075); \draw[fill] (-0.25,0) circle (0.075);
\end{scope}
\draw[color=BrickRed,line width=1.25] (0.,0.) circle (0.25);
\draw[color=BrickRed,line width=1.25] (-0.5,0.75) -- (1.5,0.75); 
\draw[color=OliveGreen,line width=1.25] (-0.5,-0.75) -- (1.5,-0.75); 
\draw[snake it] (0.,0.75) -- (0.,0.25); \draw[snake it] (0.25,0.) -- (0.75,0.); \draw[snake it] (1.,-0.75) -- (1,-0.25);
\end{tikzpicture} , \quad
\begin{tikzpicture}[baseline=0cm,scale=0.8]
\begin{scope}[rotate=45]
\draw[color=BrickRed,line width=1.25] (0.25,0) arc(0:180:0.25);
\draw[color=OliveGreen,line width=1.25] (-0.25,0) arc(180:360:0.25);
\draw[fill] (0.25,0) circle (0.075); \draw[fill] (-0.25,0) circle (0.075);
\end{scope}
\draw[color=OliveGreen,line width=1.25] (1.,0.) circle (0.25);
\draw[color=OliveGreen,line width=1.25] (2.,0.) circle (0.25);
\draw[color=BrickRed,line width=1.25] (-0.5,0.75) -- (2.5,0.75); 
\draw[color=OliveGreen,line width=1.25] (-0.5,-0.75) -- (2.5,-0.75); 
\draw[snake it] (0.,0.75) -- (0.,0.25); \draw[snake it] (0.25,0.) -- (0.75,0.); \draw[snake it] (1.25,0.) -- (1.75,0.); \draw[snake it] (2.,-0.75) -- (2,-0.25);
\end{tikzpicture} , \notag \\
&
\begin{tikzpicture}[baseline=0cm,scale=0.8]
\begin{scope}[shift={(1,0)},rotate=90]
\draw[color=BrickRed,line width=1.25] (0.25,0) arc(0:180:0.25);
\draw[color=OliveGreen,line width=1.25] (-0.25,0) arc(180:360:0.25);
\draw[fill] (0.25,0) circle (0.075); \draw[fill] (-0.25,0) circle (0.075);
\end{scope}
\draw[color=BrickRed,line width=1.25] (0.,0.) circle (0.25);
\draw[color=OliveGreen,line width=1.25] (2.,0.) circle (0.25);
\draw[color=BrickRed,line width=1.25] (-0.5,0.75) -- (2.5,0.75); 
\draw[color=OliveGreen,line width=1.25] (-0.5,-0.75) -- (2.5,-0.75); 
\draw[snake it] (0.,0.75) -- (0.,0.25); \draw[snake it] (0.25,0.) -- (0.75,0.); \draw[snake it] (1.25,0.) -- (1.75,0.); \draw[snake it] (2.,-0.75) -- (2,-0.25);
\end{tikzpicture} , \quad 
\begin{tikzpicture}[baseline=0cm,scale=0.8]
\begin{scope}[shift={(2,0)},rotate=45]
\draw[color=BrickRed,line width=1.25] (0.25,0) arc(0:180:0.25);
\draw[color=OliveGreen,line width=1.25] (-0.25,0) arc(180:360:0.25);
\draw[fill] (0.25,0) circle (0.075); \draw[fill] (-0.25,0) circle (0.075);
\end{scope}
\draw[color=BrickRed,line width=1.25] (0.,0.) circle (0.25);
\draw[color=BrickRed,line width=1.25] (1.,0.) circle (0.25);
\draw[color=BrickRed,line width=1.25] (-0.5,0.75) -- (2.5,0.75); 
\draw[color=OliveGreen,line width=1.25] (-0.5,-0.75) -- (2.5,-0.75); 
\draw[snake it] (0.,0.75) -- (0.,0.25); \draw[snake it] (0.25,0.) -- (0.75,0.); \draw[snake it] (1.25,0.) -- (1.75,0.); \draw[snake it] (2.,-0.75) -- (2,-0.25);
\end{tikzpicture} , \cdots \label{eq_RPadiagrams}
\end{align}
where red and green lines respectively correspond to the bottom and top tTMD, wiggly lines represent intra-tTMD interaction, and thick dots stand for inter-tTMD tunneling $t_\perp$ events. If we write the intra-tTMD susceptibility as $\chi_\parallel(q)$ (red and green bubble in Eq.~\ref{eq_RPadiagrams}, equal due to $\Tilde{\mathcal{T}}$ symmetry) and the inter-tTMD one as $\chi_\perp(q)$ (dotted bubble in  Eq.~\ref{eq_RPadiagrams}) with $q$ denoting the exchanged momentum; the above series can be summed analytically and yields the inter-tTMD interaction kernel
\begin{equation} \label{eq_perturbativeattraction}
v_\perp (q) = \frac{\chi_\perp(q) v_\parallel^2 (q)}{[1-\chi_\parallel(q) v_\parallel(q)]^2} ,
\end{equation}
whose dependence on and perturbative order in $t_\perp$ is hidden in $\chi_\perp (q)$. To identify the nature of this perturbatively induced inter-tTMD interaction, we set $q=0$ and use the leading order approximation~\cite{crepel2022unconventional}
\begin{equation}
\chi_\perp(q) \simeq - \frac{t_\perp^2}{\Delta_{\rm spin}^3} , 
\end{equation}
which sets the sign of $v_\perp (q\to 0) < 0$ (all other factors are perfect squares in Eq.~\ref{eq_perturbativeattraction}). This shows that the perturbative account of $t_\perp$ gives rise to weak attractive interactions between the two stacked tTMD, accounted for by the phenomenological $J$ in Eq.~\ref{eq_interlayerinteraction}.

\subsubsection{Gate-defined superconductor-to-FTI interface}

The advantage of this intrinsic and weak pairing potential is that it dramatically simplifies the experimental requirement for the realization of a clean superconductor-to-FTI interface in the bulk of the heterostructure. Indeed, the FTI stabilized at $\nu=1/3+1/3$ is stable against such weak attractive interactions (as shown in Sec.~\ref{sec_phasediagram}). On the other hand, the Fermi liquids appearing in the tTMD at filling away from $1/3$ (Fig.~\ref{fig_artview}a) are unstable against arbitrarily weak attractive interactions. Thus, changing the total density of the bilayer from $\nu = 2/3$ to $\nu = 2/3 + x$ with $x>0$ by application of different gate potentials in two adjacent regions of the heterostructure will create the desired interface, as sketched in Fig.~\ref{fig_sketchinterface}b.

\section{Conclusion}

In this article, we studied the interplay between time-reversal symmetric fractional topological order obtained by flux attachment and superconductivity in a lattice model representative of certain moir\'e heterostructures. Our finite size numerical calculations shows that, when two composite fermion can be coupled by a pairing potential --- \textit{i.e.} when they carry opposite spin and opposite flux/chirality --- they remain robust against a finite strength of attraction. This matches earlier numerical observations made in the continuum and in presence of large and opposite magnetic fields in two proximitized quantum Hall layers~\cite{chen2012interaction,furukawa2017quantum}. 

Contrary to other studies however, we have kept some direct repulsion between layers $V_\perp$, and show that the results remain true in the thermodynamic limit using a coupled wire construction capturing the universal low-energy features of the topological phase. In a different physical language, this robustness could be interpreted as the energy needed to unbind the fluxes from the composite fermions and pair the original fermions of the theory. A similar result was obtained for gapless states of composite fermions using field theory methods. 

Our analysis of thermodynamic properties also conveys that residual three-fold degenerate phase observed in our numerics, and in other similar calculations, are finite-size effects. They come from the finite size of the cluster considered, which send the degenerate topological ground state at different many-body momenta, and finite-size level repulsion within each of these momentum sectors. 

Finally, we discussed possible realization of this physics in double TMD bilayers. We identified the stacking between bilayer necessary for the emergence of pairing between the two stacked copies of the bilayer, and highlighted a possible route toward spontaneously generated pairing from the inter-bilayer tunneling. Tuning either the attraction strength or density across the bulk of these heterostructures allows to create interfaces between the fractional topological order and a superconducting phase. We hope to study such interfaces using the methods developed in Refs.~\cite{crepel2019microscopic,crepel2019model,crepel2019variational,jaworowski2020model,jaworowski2021model,zhu2020topological,li2021dynamics} in the future.

As explained in the introduction, non-Abelian parafermion are predicted to be trapped at these interfaces. It is important to note that the transition does not necessarily occurs at the edge of the system where many stacking fault and strong disorder exist, but can rather be defined electrostatically in the bulk of the heterostructure where we expected smoother behaviors. We also note that, in the quantum Hall context, non-Abelian phases were predicted to appear for sufficiently screened on-site interactions~\cite{haldane1988spin,crepel2019matrix,repellin2017creating,furukawa2014global,chen2012interaction} and may be realized in similar heterostructures as the double bilayers considered here. 

As an outlook, let us highlight that our model and the realization of a clean superconductor-to-FTI interface through proximitization with a superconductor (Fig.~\ref{fig_sketchinterface}a) may find its realization in simpler system than the double twisted TMD homobilayers studied in Sec.~\ref{sec_physicalrealization}. For instance, the two Haldane layers of Eq.~\ref{eq_fullmodel} could correspond to the two valleys of a single TMD homobilayers, which would exhibit a phase diagram similar to the one presented in Sec.~\ref{sec_phasediagram} for a commensurate total density $\nu$ where both valleys are equally populated. This could describe the phenomenology of twisted MoTe$_2$ bilayers at moir\'e filling fraction 4/3, although evidence for FTI behaviors at this density have for the moment been lacking. At different filling of the moir\'e bands, the interplay of superconductivity and repulsive interactions could also drive a single tTMDh into a non-abelian spin-unpolarized phase where the multicomponent nature of the system could lead to potentially richer physics~\cite{hansson2017quantum,crepel2018matrix}.

\section{Acknowledgement}  

The Flatiron Institute is a division of
the Simons Foundation. We acknowledge Titus Neupert for fruitful discussions. VC thanks Junkai Dong for insightful questions on the model. N.R. thanks Jiabin Yu, Jonah Herzog-Arbeitman, Yves Kwan and Andrei Bernevig for collaborations on related topics. N.R. acknowledges support from the QuantERA II Programme which has received funding from the European Union’s Horizon 2020 research and innovation programme under Grant Agreement No 101017733.

\bibliography{AFMFCI}

\begin{thebibliography}{90}%
\makeatletter
\providecommand \@ifxundefined [1]{%
 \@ifx{#1\undefined}
}%
\providecommand \@ifnum [1]{%
 \ifnum #1\expandafter \@firstoftwo
 \else \expandafter \@secondoftwo
 \fi
}%
\providecommand \@ifx [1]{%
 \ifx #1\expandafter \@firstoftwo
 \else \expandafter \@secondoftwo
 \fi
}%
\providecommand \natexlab [1]{#1}%
\providecommand \enquote  [1]{``#1''}%
\providecommand \bibnamefont  [1]{#1}%
\providecommand \bibfnamefont [1]{#1}%
\providecommand \citenamefont [1]{#1}%
\providecommand \href@noop [0]{\@secondoftwo}%
\providecommand \href [0]{\begingroup \@sanitize@url \@href}%
\providecommand \@href[1]{\@@startlink{#1}\@@href}%
\providecommand \@@href[1]{\endgroup#1\@@endlink}%
\providecommand \@sanitize@url [0]{\catcode `\\12\catcode `\$12\catcode `\&12\catcode `\#12\catcode `\^12\catcode `\_12\catcode `\%12\relax}%
\providecommand \@@startlink[1]{}%
\providecommand \@@endlink[0]{}%
\providecommand \url  [0]{\begingroup\@sanitize@url \@url }%
\providecommand \@url [1]{\endgroup\@href {#1}{\urlprefix }}%
\providecommand \urlprefix  [0]{URL }%
\providecommand \Eprint [0]{\href }%
\providecommand \doibase [0]{https://doi.org/}%
\providecommand \selectlanguage [0]{\@gobble}%
\providecommand \bibinfo  [0]{\@secondoftwo}%
\providecommand \bibfield  [0]{\@secondoftwo}%
\providecommand \translation [1]{[#1]}%
\providecommand \BibitemOpen [0]{}%
\providecommand \bibitemStop [0]{}%
\providecommand \bibitemNoStop [0]{.\EOS\space}%
\providecommand \EOS [0]{\spacefactor3000\relax}%
\providecommand \BibitemShut  [1]{\csname bibitem#1\endcsname}%
\let\auto@bib@innerbib\@empty
\bibitem [{\citenamefont {Cai}\ \emph {et~al.}(2023)\citenamefont {Cai}, \citenamefont {Anderson}, \citenamefont {Wang}, \citenamefont {Zhang}, \citenamefont {Liu}, \citenamefont {Holtzmann}, \citenamefont {Zhang}, \citenamefont {Fan}, \citenamefont {Taniguchi}, \citenamefont {Watanabe} \emph {et~al.}}]{cai2023signatures}%
  \BibitemOpen
  \bibfield  {author} {\bibinfo {author} {\bibfnamefont {J.}~\bibnamefont {Cai}}, \bibinfo {author} {\bibfnamefont {E.}~\bibnamefont {Anderson}}, \bibinfo {author} {\bibfnamefont {C.}~\bibnamefont {Wang}}, \bibinfo {author} {\bibfnamefont {X.}~\bibnamefont {Zhang}}, \bibinfo {author} {\bibfnamefont {X.}~\bibnamefont {Liu}}, \bibinfo {author} {\bibfnamefont {W.}~\bibnamefont {Holtzmann}}, \bibinfo {author} {\bibfnamefont {Y.}~\bibnamefont {Zhang}}, \bibinfo {author} {\bibfnamefont {F.}~\bibnamefont {Fan}}, \bibinfo {author} {\bibfnamefont {T.}~\bibnamefont {Taniguchi}}, \bibinfo {author} {\bibfnamefont {K.}~\bibnamefont {Watanabe}}, \emph {et~al.},\ }\bibfield  {title} {\bibinfo {title} {Signatures of fractional quantum anomalous hall states in twisted mote2},\ }\href@noop {} {\bibfield  {journal} {\bibinfo  {journal} {Nature}\ }\textbf {\bibinfo {volume} {622}},\ \bibinfo {pages} {63} (\bibinfo {year} {2023})}\BibitemShut {NoStop}%
\bibitem [{\citenamefont {Xu}\ \emph {et~al.}(2023)\citenamefont {Xu}, \citenamefont {Sun}, \citenamefont {Jia}, \citenamefont {Liu}, \citenamefont {Xu}, \citenamefont {Li}, \citenamefont {Gu}, \citenamefont {Watanabe}, \citenamefont {Taniguchi}, \citenamefont {Tong} \emph {et~al.}}]{xu2023observation}%
  \BibitemOpen
  \bibfield  {author} {\bibinfo {author} {\bibfnamefont {F.}~\bibnamefont {Xu}}, \bibinfo {author} {\bibfnamefont {Z.}~\bibnamefont {Sun}}, \bibinfo {author} {\bibfnamefont {T.}~\bibnamefont {Jia}}, \bibinfo {author} {\bibfnamefont {C.}~\bibnamefont {Liu}}, \bibinfo {author} {\bibfnamefont {C.}~\bibnamefont {Xu}}, \bibinfo {author} {\bibfnamefont {C.}~\bibnamefont {Li}}, \bibinfo {author} {\bibfnamefont {Y.}~\bibnamefont {Gu}}, \bibinfo {author} {\bibfnamefont {K.}~\bibnamefont {Watanabe}}, \bibinfo {author} {\bibfnamefont {T.}~\bibnamefont {Taniguchi}}, \bibinfo {author} {\bibfnamefont {B.}~\bibnamefont {Tong}}, \emph {et~al.},\ }\bibfield  {title} {\bibinfo {title} {Observation of integer and fractional quantum anomalous hall states in twisted bilayer mote2},\ }\href@noop {} {\bibfield  {journal} {\bibinfo  {journal} {arXiv preprint arXiv:2308.06177}\ } (\bibinfo {year} {2023})}\BibitemShut {NoStop}%
\bibitem [{\citenamefont {Zeng}\ \emph {et~al.}(2023)\citenamefont {Zeng}, \citenamefont {Xia}, \citenamefont {Kang}, \citenamefont {Zhu}, \citenamefont {Kn{\"u}ppel}, \citenamefont {Vaswani}, \citenamefont {Watanabe}, \citenamefont {Taniguchi}, \citenamefont {Mak},\ and\ \citenamefont {Shan}}]{zeng2023thermodynamic}%
  \BibitemOpen
  \bibfield  {author} {\bibinfo {author} {\bibfnamefont {Y.}~\bibnamefont {Zeng}}, \bibinfo {author} {\bibfnamefont {Z.}~\bibnamefont {Xia}}, \bibinfo {author} {\bibfnamefont {K.}~\bibnamefont {Kang}}, \bibinfo {author} {\bibfnamefont {J.}~\bibnamefont {Zhu}}, \bibinfo {author} {\bibfnamefont {P.}~\bibnamefont {Kn{\"u}ppel}}, \bibinfo {author} {\bibfnamefont {C.}~\bibnamefont {Vaswani}}, \bibinfo {author} {\bibfnamefont {K.}~\bibnamefont {Watanabe}}, \bibinfo {author} {\bibfnamefont {T.}~\bibnamefont {Taniguchi}}, \bibinfo {author} {\bibfnamefont {K.~F.}\ \bibnamefont {Mak}},\ and\ \bibinfo {author} {\bibfnamefont {J.}~\bibnamefont {Shan}},\ }\bibfield  {title} {\bibinfo {title} {Thermodynamic evidence of fractional chern insulator in moir{\'e} mote2},\ }\href@noop {} {\bibfield  {journal} {\bibinfo  {journal} {Nature}\ }\textbf {\bibinfo {volume} {622}},\ \bibinfo {pages} {69} (\bibinfo {year} {2023})}\BibitemShut {NoStop}%
\bibitem [{\citenamefont {Park}\ \emph {et~al.}(2023)\citenamefont {Park}, \citenamefont {Cai}, \citenamefont {Anderson}, \citenamefont {Zhang}, \citenamefont {Zhu}, \citenamefont {Liu}, \citenamefont {Wang}, \citenamefont {Holtzmann}, \citenamefont {Hu}, \citenamefont {Liu} \emph {et~al.}}]{park2023observation}%
  \BibitemOpen
  \bibfield  {author} {\bibinfo {author} {\bibfnamefont {H.}~\bibnamefont {Park}}, \bibinfo {author} {\bibfnamefont {J.}~\bibnamefont {Cai}}, \bibinfo {author} {\bibfnamefont {E.}~\bibnamefont {Anderson}}, \bibinfo {author} {\bibfnamefont {Y.}~\bibnamefont {Zhang}}, \bibinfo {author} {\bibfnamefont {J.}~\bibnamefont {Zhu}}, \bibinfo {author} {\bibfnamefont {X.}~\bibnamefont {Liu}}, \bibinfo {author} {\bibfnamefont {C.}~\bibnamefont {Wang}}, \bibinfo {author} {\bibfnamefont {W.}~\bibnamefont {Holtzmann}}, \bibinfo {author} {\bibfnamefont {C.}~\bibnamefont {Hu}}, \bibinfo {author} {\bibfnamefont {Z.}~\bibnamefont {Liu}}, \emph {et~al.},\ }\bibfield  {title} {\bibinfo {title} {Observation of fractionally quantized anomalous hall effect},\ }\href@noop {} {\bibfield  {journal} {\bibinfo  {journal} {Nature}\ }\textbf {\bibinfo {volume} {622}},\ \bibinfo {pages} {74} (\bibinfo {year} {2023})}\BibitemShut {NoStop}%
\bibitem [{\citenamefont {Lu}\ \emph {et~al.}(2023)\citenamefont {Lu}, \citenamefont {Han}, \citenamefont {Yao}, \citenamefont {Reddy}, \citenamefont {Yang}, \citenamefont {Seo}, \citenamefont {Watanabe}, \citenamefont {Taniguchi}, \citenamefont {Fu},\ and\ \citenamefont {Ju}}]{lu2023fractional}%
  \BibitemOpen
  \bibfield  {author} {\bibinfo {author} {\bibfnamefont {Z.}~\bibnamefont {Lu}}, \bibinfo {author} {\bibfnamefont {T.}~\bibnamefont {Han}}, \bibinfo {author} {\bibfnamefont {Y.}~\bibnamefont {Yao}}, \bibinfo {author} {\bibfnamefont {A.~P.}\ \bibnamefont {Reddy}}, \bibinfo {author} {\bibfnamefont {J.}~\bibnamefont {Yang}}, \bibinfo {author} {\bibfnamefont {J.}~\bibnamefont {Seo}}, \bibinfo {author} {\bibfnamefont {K.}~\bibnamefont {Watanabe}}, \bibinfo {author} {\bibfnamefont {T.}~\bibnamefont {Taniguchi}}, \bibinfo {author} {\bibfnamefont {L.}~\bibnamefont {Fu}},\ and\ \bibinfo {author} {\bibfnamefont {L.}~\bibnamefont {Ju}},\ }\bibfield  {title} {\bibinfo {title} {Fractional quantum anomalous hall effect in a graphene moire superlattice},\ }\href@noop {} {\bibfield  {journal} {\bibinfo  {journal} {arXiv preprint arXiv:2309.17436}\ } (\bibinfo {year} {2023})}\BibitemShut {NoStop}%
\bibitem [{\citenamefont {Cr{\'e}pel}\ and\ \citenamefont {Fu}(2023)}]{crepel2023anomalous}%
  \BibitemOpen
  \bibfield  {author} {\bibinfo {author} {\bibfnamefont {V.}~\bibnamefont {Cr{\'e}pel}}\ and\ \bibinfo {author} {\bibfnamefont {L.}~\bibnamefont {Fu}},\ }\bibfield  {title} {\bibinfo {title} {Anomalous hall metal and fractional chern insulator in twisted transition metal dichalcogenides},\ }\href@noop {} {\bibfield  {journal} {\bibinfo  {journal} {Physical Review B}\ }\textbf {\bibinfo {volume} {107}},\ \bibinfo {pages} {L201109} (\bibinfo {year} {2023})}\BibitemShut {NoStop}%
\bibitem [{\citenamefont {Wu}\ \emph {et~al.}(2019)\citenamefont {Wu}, \citenamefont {Lovorn}, \citenamefont {Tutuc}, \citenamefont {Martin},\ and\ \citenamefont {MacDonald}}]{wu2019topological}%
  \BibitemOpen
  \bibfield  {author} {\bibinfo {author} {\bibfnamefont {F.}~\bibnamefont {Wu}}, \bibinfo {author} {\bibfnamefont {T.}~\bibnamefont {Lovorn}}, \bibinfo {author} {\bibfnamefont {E.}~\bibnamefont {Tutuc}}, \bibinfo {author} {\bibfnamefont {I.}~\bibnamefont {Martin}},\ and\ \bibinfo {author} {\bibfnamefont {A.}~\bibnamefont {MacDonald}},\ }\bibfield  {title} {\bibinfo {title} {Topological insulators in twisted transition metal dichalcogenide homobilayers},\ }\href@noop {} {\bibfield  {journal} {\bibinfo  {journal} {Physical review letters}\ }\textbf {\bibinfo {volume} {122}},\ \bibinfo {pages} {086402} (\bibinfo {year} {2019})}\BibitemShut {NoStop}%
\bibitem [{\citenamefont {Devakul}\ \emph {et~al.}(2021)\citenamefont {Devakul}, \citenamefont {Cr{\'e}pel}, \citenamefont {Zhang},\ and\ \citenamefont {Fu}}]{devakul2021magic}%
  \BibitemOpen
  \bibfield  {author} {\bibinfo {author} {\bibfnamefont {T.}~\bibnamefont {Devakul}}, \bibinfo {author} {\bibfnamefont {V.}~\bibnamefont {Cr{\'e}pel}}, \bibinfo {author} {\bibfnamefont {Y.}~\bibnamefont {Zhang}},\ and\ \bibinfo {author} {\bibfnamefont {L.}~\bibnamefont {Fu}},\ }\bibfield  {title} {\bibinfo {title} {Magic in twisted transition metal dichalcogenide bilayers},\ }\href@noop {} {\bibfield  {journal} {\bibinfo  {journal} {Nature communications}\ }\textbf {\bibinfo {volume} {12}},\ \bibinfo {pages} {6730} (\bibinfo {year} {2021})}\BibitemShut {NoStop}%
\bibitem [{\citenamefont {Regnault}\ and\ \citenamefont {Bernevig}(2011)}]{regnault2011fractional}%
  \BibitemOpen
  \bibfield  {author} {\bibinfo {author} {\bibfnamefont {N.}~\bibnamefont {Regnault}}\ and\ \bibinfo {author} {\bibfnamefont {B.~A.}\ \bibnamefont {Bernevig}},\ }\bibfield  {title} {\bibinfo {title} {Fractional chern insulator},\ }\href@noop {} {\bibfield  {journal} {\bibinfo  {journal} {Physical Review X}\ }\textbf {\bibinfo {volume} {1}},\ \bibinfo {pages} {021014} (\bibinfo {year} {2011})}\BibitemShut {NoStop}%
\bibitem [{\citenamefont {Sheng}\ \emph {et~al.}(2011)\citenamefont {Sheng}, \citenamefont {Gu}, \citenamefont {Sun},\ and\ \citenamefont {Sheng}}]{sheng2011fractional}%
  \BibitemOpen
  \bibfield  {author} {\bibinfo {author} {\bibfnamefont {D.}~\bibnamefont {Sheng}}, \bibinfo {author} {\bibfnamefont {Z.-C.}\ \bibnamefont {Gu}}, \bibinfo {author} {\bibfnamefont {K.}~\bibnamefont {Sun}},\ and\ \bibinfo {author} {\bibfnamefont {L.}~\bibnamefont {Sheng}},\ }\bibfield  {title} {\bibinfo {title} {Fractional quantum hall effect in the absence of landau levels},\ }\href@noop {} {\bibfield  {journal} {\bibinfo  {journal} {Nature communications}\ }\textbf {\bibinfo {volume} {2}},\ \bibinfo {pages} {389} (\bibinfo {year} {2011})}\BibitemShut {NoStop}%
\bibitem [{\citenamefont {Neupert}\ \emph {et~al.}(2011{\natexlab{a}})\citenamefont {Neupert}, \citenamefont {Santos}, \citenamefont {Ryu}, \citenamefont {Chamon},\ and\ \citenamefont {Mudry}}]{neupert2011fractional}%
  \BibitemOpen
  \bibfield  {author} {\bibinfo {author} {\bibfnamefont {T.}~\bibnamefont {Neupert}}, \bibinfo {author} {\bibfnamefont {L.}~\bibnamefont {Santos}}, \bibinfo {author} {\bibfnamefont {S.}~\bibnamefont {Ryu}}, \bibinfo {author} {\bibfnamefont {C.}~\bibnamefont {Chamon}},\ and\ \bibinfo {author} {\bibfnamefont {C.}~\bibnamefont {Mudry}},\ }\bibfield  {title} {\bibinfo {title} {Fractional topological liquids with time-reversal symmetry and their lattice realization},\ }\href@noop {} {\bibfield  {journal} {\bibinfo  {journal} {Physical Review B—Condensed Matter and Materials Physics}\ }\textbf {\bibinfo {volume} {84}},\ \bibinfo {pages} {165107} (\bibinfo {year} {2011}{\natexlab{a}})}\BibitemShut {NoStop}%
\bibitem [{\citenamefont {Sheffer}\ and\ \citenamefont {Stern}(2021)}]{sheffer2021chiral}%
  \BibitemOpen
  \bibfield  {author} {\bibinfo {author} {\bibfnamefont {Y.}~\bibnamefont {Sheffer}}\ and\ \bibinfo {author} {\bibfnamefont {A.}~\bibnamefont {Stern}},\ }\bibfield  {title} {\bibinfo {title} {Chiral magic-angle twisted bilayer graphene in a magnetic field: Landau level correspondence, exact wave functions, and fractional chern insulators},\ }\href@noop {} {\bibfield  {journal} {\bibinfo  {journal} {Physical Review B}\ }\textbf {\bibinfo {volume} {104}},\ \bibinfo {pages} {L121405} (\bibinfo {year} {2021})}\BibitemShut {NoStop}%
\bibitem [{\citenamefont {Estienne}\ \emph {et~al.}(2023)\citenamefont {Estienne}, \citenamefont {Regnault},\ and\ \citenamefont {Cr\'epel}}]{estienne2023ideal}%
  \BibitemOpen
  \bibfield  {author} {\bibinfo {author} {\bibfnamefont {B.}~\bibnamefont {Estienne}}, \bibinfo {author} {\bibfnamefont {N.}~\bibnamefont {Regnault}},\ and\ \bibinfo {author} {\bibfnamefont {V.}~\bibnamefont {Cr\'epel}},\ }\bibfield  {title} {\bibinfo {title} {Ideal chern bands as landau levels in curved space},\ }\href {https://doi.org/10.1103/PhysRevResearch.5.L032048} {\bibfield  {journal} {\bibinfo  {journal} {Phys. Rev. Res.}\ }\textbf {\bibinfo {volume} {5}},\ \bibinfo {pages} {L032048} (\bibinfo {year} {2023})}\BibitemShut {NoStop}%
\bibitem [{\citenamefont {Parhizkar}\ and\ \citenamefont {Galitski}(2023)}]{parhizkar2023generic}%
  \BibitemOpen
  \bibfield  {author} {\bibinfo {author} {\bibfnamefont {A.}~\bibnamefont {Parhizkar}}\ and\ \bibinfo {author} {\bibfnamefont {V.}~\bibnamefont {Galitski}},\ }\bibfield  {title} {\bibinfo {title} {A generic topological criterion for flat bands in two dimensions},\ }\href@noop {} {\bibfield  {journal} {\bibinfo  {journal} {arXiv preprint arXiv:2301.00824}\ } (\bibinfo {year} {2023})}\BibitemShut {NoStop}%
\bibitem [{\citenamefont {Zhang}\ \emph {et~al.}(2024)\citenamefont {Zhang}, \citenamefont {Wang}, \citenamefont {Liu}, \citenamefont {Fan}, \citenamefont {Cao},\ and\ \citenamefont {Xiao}}]{zhang2024polarization}%
  \BibitemOpen
  \bibfield  {author} {\bibinfo {author} {\bibfnamefont {X.-W.}\ \bibnamefont {Zhang}}, \bibinfo {author} {\bibfnamefont {C.}~\bibnamefont {Wang}}, \bibinfo {author} {\bibfnamefont {X.}~\bibnamefont {Liu}}, \bibinfo {author} {\bibfnamefont {Y.}~\bibnamefont {Fan}}, \bibinfo {author} {\bibfnamefont {T.}~\bibnamefont {Cao}},\ and\ \bibinfo {author} {\bibfnamefont {D.}~\bibnamefont {Xiao}},\ }\bibfield  {title} {\bibinfo {title} {Polarization-driven band topology evolution in twisted mote2 and wse2},\ }\href@noop {} {\bibfield  {journal} {\bibinfo  {journal} {Nature Communications}\ }\textbf {\bibinfo {volume} {15}},\ \bibinfo {pages} {4223} (\bibinfo {year} {2024})}\BibitemShut {NoStop}%
\bibitem [{\citenamefont {Wang}\ \emph {et~al.}(2024)\citenamefont {Wang}, \citenamefont {Zhang}, \citenamefont {Liu}, \citenamefont {He}, \citenamefont {Xu}, \citenamefont {Ran}, \citenamefont {Cao},\ and\ \citenamefont {Xiao}}]{wang2024fractional}%
  \BibitemOpen
  \bibfield  {author} {\bibinfo {author} {\bibfnamefont {C.}~\bibnamefont {Wang}}, \bibinfo {author} {\bibfnamefont {X.-W.}\ \bibnamefont {Zhang}}, \bibinfo {author} {\bibfnamefont {X.}~\bibnamefont {Liu}}, \bibinfo {author} {\bibfnamefont {Y.}~\bibnamefont {He}}, \bibinfo {author} {\bibfnamefont {X.}~\bibnamefont {Xu}}, \bibinfo {author} {\bibfnamefont {Y.}~\bibnamefont {Ran}}, \bibinfo {author} {\bibfnamefont {T.}~\bibnamefont {Cao}},\ and\ \bibinfo {author} {\bibfnamefont {D.}~\bibnamefont {Xiao}},\ }\bibfield  {title} {\bibinfo {title} {Fractional chern insulator in twisted bilayer mote$_2$},\ }\href@noop {} {\bibfield  {journal} {\bibinfo  {journal} {Physical Review Letters}\ }\textbf {\bibinfo {volume} {132}},\ \bibinfo {pages} {036501} (\bibinfo {year} {2024})}\BibitemShut {NoStop}%
\bibitem [{\citenamefont {Yu}\ \emph {et~al.}(2024)\citenamefont {Yu}, \citenamefont {Herzog-Arbeitman}, \citenamefont {Wang}, \citenamefont {Vafek}, \citenamefont {Bernevig},\ and\ \citenamefont {Regnault}}]{yu2024fractional}%
  \BibitemOpen
  \bibfield  {author} {\bibinfo {author} {\bibfnamefont {J.}~\bibnamefont {Yu}}, \bibinfo {author} {\bibfnamefont {J.}~\bibnamefont {Herzog-Arbeitman}}, \bibinfo {author} {\bibfnamefont {M.}~\bibnamefont {Wang}}, \bibinfo {author} {\bibfnamefont {O.}~\bibnamefont {Vafek}}, \bibinfo {author} {\bibfnamefont {B.~A.}\ \bibnamefont {Bernevig}},\ and\ \bibinfo {author} {\bibfnamefont {N.}~\bibnamefont {Regnault}},\ }\bibfield  {title} {\bibinfo {title} {Fractional chern insulators versus nonmagnetic states in twisted bilayer mote 2},\ }\href@noop {} {\bibfield  {journal} {\bibinfo  {journal} {Physical Review B}\ }\textbf {\bibinfo {volume} {109}},\ \bibinfo {pages} {045147} (\bibinfo {year} {2024})}\BibitemShut {NoStop}%
\bibitem [{\citenamefont {Abouelkomsan}\ \emph {et~al.}(2023)\citenamefont {Abouelkomsan}, \citenamefont {Reddy}, \citenamefont {Fu},\ and\ \citenamefont {Bergholtz}}]{abouelkomsan2023band}%
  \BibitemOpen
  \bibfield  {author} {\bibinfo {author} {\bibfnamefont {A.}~\bibnamefont {Abouelkomsan}}, \bibinfo {author} {\bibfnamefont {A.~P.}\ \bibnamefont {Reddy}}, \bibinfo {author} {\bibfnamefont {L.}~\bibnamefont {Fu}},\ and\ \bibinfo {author} {\bibfnamefont {E.~J.}\ \bibnamefont {Bergholtz}},\ }\bibfield  {title} {\bibinfo {title} {Band mixing in the quantum anomalous hall regime of twisted semiconductor bilayers},\ }\href@noop {} {\bibfield  {journal} {\bibinfo  {journal} {arXiv preprint arXiv:2309.16548}\ } (\bibinfo {year} {2023})}\BibitemShut {NoStop}%
\bibitem [{\citenamefont {Jia}\ \emph {et~al.}(2024)\citenamefont {Jia}, \citenamefont {Yu}, \citenamefont {Liu}, \citenamefont {Herzog-Arbeitman}, \citenamefont {Qi}, \citenamefont {Pi}, \citenamefont {Regnault}, \citenamefont {Weng}, \citenamefont {Bernevig},\ and\ \citenamefont {Wu}}]{jia2024moire}%
  \BibitemOpen
  \bibfield  {author} {\bibinfo {author} {\bibfnamefont {Y.}~\bibnamefont {Jia}}, \bibinfo {author} {\bibfnamefont {J.}~\bibnamefont {Yu}}, \bibinfo {author} {\bibfnamefont {J.}~\bibnamefont {Liu}}, \bibinfo {author} {\bibfnamefont {J.}~\bibnamefont {Herzog-Arbeitman}}, \bibinfo {author} {\bibfnamefont {Z.}~\bibnamefont {Qi}}, \bibinfo {author} {\bibfnamefont {H.}~\bibnamefont {Pi}}, \bibinfo {author} {\bibfnamefont {N.}~\bibnamefont {Regnault}}, \bibinfo {author} {\bibfnamefont {H.}~\bibnamefont {Weng}}, \bibinfo {author} {\bibfnamefont {B.~A.}\ \bibnamefont {Bernevig}},\ and\ \bibinfo {author} {\bibfnamefont {Q.}~\bibnamefont {Wu}},\ }\bibfield  {title} {\bibinfo {title} {Moir{\'e} fractional chern insulators. i. first-principles calculations and continuum models of twisted bilayer mote 2},\ }\href@noop {} {\bibfield  {journal} {\bibinfo  {journal} {Physical Review B}\ }\textbf {\bibinfo {volume} {109}},\ \bibinfo {pages} {205121} (\bibinfo {year} {2024})}\BibitemShut {NoStop}%
\bibitem [{\citenamefont {Redekop}\ \emph {et~al.}(2024)\citenamefont {Redekop}, \citenamefont {Zhang}, \citenamefont {Park}, \citenamefont {Cai}, \citenamefont {Anderson}, \citenamefont {Sheekey}, \citenamefont {Arp}, \citenamefont {Babikyan}, \citenamefont {Salters}, \citenamefont {Watanabe} \emph {et~al.}}]{redekop2024direct}%
  \BibitemOpen
  \bibfield  {author} {\bibinfo {author} {\bibfnamefont {E.}~\bibnamefont {Redekop}}, \bibinfo {author} {\bibfnamefont {C.}~\bibnamefont {Zhang}}, \bibinfo {author} {\bibfnamefont {H.}~\bibnamefont {Park}}, \bibinfo {author} {\bibfnamefont {J.}~\bibnamefont {Cai}}, \bibinfo {author} {\bibfnamefont {E.}~\bibnamefont {Anderson}}, \bibinfo {author} {\bibfnamefont {O.}~\bibnamefont {Sheekey}}, \bibinfo {author} {\bibfnamefont {T.}~\bibnamefont {Arp}}, \bibinfo {author} {\bibfnamefont {G.}~\bibnamefont {Babikyan}}, \bibinfo {author} {\bibfnamefont {S.}~\bibnamefont {Salters}}, \bibinfo {author} {\bibfnamefont {K.}~\bibnamefont {Watanabe}}, \emph {et~al.},\ }\bibfield  {title} {\bibinfo {title} {Direct magnetic imaging of fractional chern insulators in twisted mote $ \_2 $ with a superconducting sensor},\ }\href@noop {} {\bibfield  {journal} {\bibinfo  {journal} {arXiv preprint arXiv:2405.10269}\ } (\bibinfo {year} {2024})}\BibitemShut {NoStop}%
\bibitem [{\citenamefont {Reddy}\ \emph {et~al.}(2023)\citenamefont {Reddy}, \citenamefont {Alsallom}, \citenamefont {Zhang}, \citenamefont {Devakul},\ and\ \citenamefont {Fu}}]{reddy2023fractional}%
  \BibitemOpen
  \bibfield  {author} {\bibinfo {author} {\bibfnamefont {A.~P.}\ \bibnamefont {Reddy}}, \bibinfo {author} {\bibfnamefont {F.}~\bibnamefont {Alsallom}}, \bibinfo {author} {\bibfnamefont {Y.}~\bibnamefont {Zhang}}, \bibinfo {author} {\bibfnamefont {T.}~\bibnamefont {Devakul}},\ and\ \bibinfo {author} {\bibfnamefont {L.}~\bibnamefont {Fu}},\ }\bibfield  {title} {\bibinfo {title} {Fractional quantum anomalous hall states in twisted bilayer mote 2 and wse 2},\ }\href@noop {} {\bibfield  {journal} {\bibinfo  {journal} {Physical Review B}\ }\textbf {\bibinfo {volume} {108}},\ \bibinfo {pages} {085117} (\bibinfo {year} {2023})}\BibitemShut {NoStop}%
\bibitem [{\citenamefont {Laughlin}(1983)}]{laughlin1983anomalous}%
  \BibitemOpen
  \bibfield  {author} {\bibinfo {author} {\bibfnamefont {R.~B.}\ \bibnamefont {Laughlin}},\ }\bibfield  {title} {\bibinfo {title} {Anomalous quantum hall effect: an incompressible quantum fluid with fractionally charged excitations},\ }\href@noop {} {\bibfield  {journal} {\bibinfo  {journal} {Physical Review Letters}\ }\textbf {\bibinfo {volume} {50}},\ \bibinfo {pages} {1395} (\bibinfo {year} {1983})}\BibitemShut {NoStop}%
\bibitem [{\citenamefont {Jain}(1989)}]{jain1989composite}%
  \BibitemOpen
  \bibfield  {author} {\bibinfo {author} {\bibfnamefont {J.~K.}\ \bibnamefont {Jain}},\ }\bibfield  {title} {\bibinfo {title} {Composite-fermion approach for the fractional quantum hall effect},\ }\href@noop {} {\bibfield  {journal} {\bibinfo  {journal} {Physical review letters}\ }\textbf {\bibinfo {volume} {63}},\ \bibinfo {pages} {199} (\bibinfo {year} {1989})}\BibitemShut {NoStop}%
\bibitem [{\citenamefont {Kang}\ \emph {et~al.}(2024)\citenamefont {Kang}, \citenamefont {Shen}, \citenamefont {Qiu}, \citenamefont {Watanabe}, \citenamefont {Taniguchi}, \citenamefont {Shan},\ and\ \citenamefont {Mak}}]{kang2024observation}%
  \BibitemOpen
  \bibfield  {author} {\bibinfo {author} {\bibfnamefont {K.}~\bibnamefont {Kang}}, \bibinfo {author} {\bibfnamefont {B.}~\bibnamefont {Shen}}, \bibinfo {author} {\bibfnamefont {Y.}~\bibnamefont {Qiu}}, \bibinfo {author} {\bibfnamefont {K.}~\bibnamefont {Watanabe}}, \bibinfo {author} {\bibfnamefont {T.}~\bibnamefont {Taniguchi}}, \bibinfo {author} {\bibfnamefont {J.}~\bibnamefont {Shan}},\ and\ \bibinfo {author} {\bibfnamefont {K.~F.}\ \bibnamefont {Mak}},\ }\bibfield  {title} {\bibinfo {title} {Observation of the fractional quantum spin hall effect in moir$\backslash$'e mote2},\ }\href@noop {} {\bibfield  {journal} {\bibinfo  {journal} {arXiv preprint arXiv:2402.03294}\ } (\bibinfo {year} {2024})}\BibitemShut {NoStop}%
\bibitem [{\citenamefont {Mong}\ \emph {et~al.}(2014)\citenamefont {Mong}, \citenamefont {Clarke}, \citenamefont {Alicea}, \citenamefont {Lindner}, \citenamefont {Fendley}, \citenamefont {Nayak}, \citenamefont {Oreg}, \citenamefont {Stern}, \citenamefont {Berg}, \citenamefont {Shtengel} \emph {et~al.}}]{mong2014universal}%
  \BibitemOpen
  \bibfield  {author} {\bibinfo {author} {\bibfnamefont {R.~S.}\ \bibnamefont {Mong}}, \bibinfo {author} {\bibfnamefont {D.~J.}\ \bibnamefont {Clarke}}, \bibinfo {author} {\bibfnamefont {J.}~\bibnamefont {Alicea}}, \bibinfo {author} {\bibfnamefont {N.~H.}\ \bibnamefont {Lindner}}, \bibinfo {author} {\bibfnamefont {P.}~\bibnamefont {Fendley}}, \bibinfo {author} {\bibfnamefont {C.}~\bibnamefont {Nayak}}, \bibinfo {author} {\bibfnamefont {Y.}~\bibnamefont {Oreg}}, \bibinfo {author} {\bibfnamefont {A.}~\bibnamefont {Stern}}, \bibinfo {author} {\bibfnamefont {E.}~\bibnamefont {Berg}}, \bibinfo {author} {\bibfnamefont {K.}~\bibnamefont {Shtengel}}, \emph {et~al.},\ }\bibfield  {title} {\bibinfo {title} {Universal topological quantum computation from a superconductor-abelian quantum hall heterostructure},\ }\href@noop {} {\bibfield  {journal} {\bibinfo  {journal} {Physical Review X}\ }\textbf {\bibinfo {volume} {4}},\ \bibinfo {pages} {011036} (\bibinfo {year} {2014})}\BibitemShut {NoStop}%
\bibitem [{\citenamefont {Vaezi}(2014)}]{vaezi2014superconducting}%
  \BibitemOpen
  \bibfield  {author} {\bibinfo {author} {\bibfnamefont {A.}~\bibnamefont {Vaezi}},\ }\bibfield  {title} {\bibinfo {title} {Superconducting analogue of the parafermion fractional quantum hall states},\ }\href@noop {} {\bibfield  {journal} {\bibinfo  {journal} {Physical Review X}\ }\textbf {\bibinfo {volume} {4}},\ \bibinfo {pages} {031009} (\bibinfo {year} {2014})}\BibitemShut {NoStop}%
\bibitem [{\citenamefont {Repellin}\ \emph {et~al.}(2018)\citenamefont {Repellin}, \citenamefont {Cook}, \citenamefont {Neupert},\ and\ \citenamefont {Regnault}}]{repellin2018numerical}%
  \BibitemOpen
  \bibfield  {author} {\bibinfo {author} {\bibfnamefont {C.}~\bibnamefont {Repellin}}, \bibinfo {author} {\bibfnamefont {A.~M.}\ \bibnamefont {Cook}}, \bibinfo {author} {\bibfnamefont {T.}~\bibnamefont {Neupert}},\ and\ \bibinfo {author} {\bibfnamefont {N.}~\bibnamefont {Regnault}},\ }\bibfield  {title} {\bibinfo {title} {Numerical investigation of gapped edge states in fractional quantum hall-superconductor heterostructures},\ }\href@noop {} {\bibfield  {journal} {\bibinfo  {journal} {npj Quantum Materials}\ }\textbf {\bibinfo {volume} {3}},\ \bibinfo {pages} {14} (\bibinfo {year} {2018})}\BibitemShut {NoStop}%
\bibitem [{\citenamefont {Barkeshli}(2016)}]{barkeshli2016charge}%
  \BibitemOpen
  \bibfield  {author} {\bibinfo {author} {\bibfnamefont {M.}~\bibnamefont {Barkeshli}},\ }\bibfield  {title} {\bibinfo {title} {Charge 2 e/3 superconductivity and topological degeneracies without localized zero modes in bilayer fractional quantum hall states},\ }\href@noop {} {\bibfield  {journal} {\bibinfo  {journal} {Physical review letters}\ }\textbf {\bibinfo {volume} {117}},\ \bibinfo {pages} {096803} (\bibinfo {year} {2016})}\BibitemShut {NoStop}%
\bibitem [{\citenamefont {Barkeshli}\ and\ \citenamefont {Wen}(2011)}]{barkeshli2011bilayer}%
  \BibitemOpen
  \bibfield  {author} {\bibinfo {author} {\bibfnamefont {M.}~\bibnamefont {Barkeshli}}\ and\ \bibinfo {author} {\bibfnamefont {X.-G.}\ \bibnamefont {Wen}},\ }\bibfield  {title} {\bibinfo {title} {Bilayer quantum hall phase transitions and the orbifold non-abelian fractional quantum hall states},\ }\bibfield  {journal} {\bibinfo  {journal} {Physical Review B}\ }\textbf {\bibinfo {volume} {84}},\ \href {https://doi.org/10.1103/physrevb.84.115121} {10.1103/physrevb.84.115121} (\bibinfo {year} {2011})\BibitemShut {NoStop}%
\bibitem [{\citenamefont {Barkeshli}\ and\ \citenamefont {Qi}(2012)}]{barkeshli2012topological}%
  \BibitemOpen
  \bibfield  {author} {\bibinfo {author} {\bibfnamefont {M.}~\bibnamefont {Barkeshli}}\ and\ \bibinfo {author} {\bibfnamefont {X.-L.}\ \bibnamefont {Qi}},\ }\bibfield  {title} {\bibinfo {title} {Topological nematic states and non-abelian lattice dislocations},\ }\href {https://doi.org/10.1103/PhysRevX.2.031013} {\bibfield  {journal} {\bibinfo  {journal} {Phys. Rev. X}\ }\textbf {\bibinfo {volume} {2}},\ \bibinfo {pages} {031013} (\bibinfo {year} {2012})}\BibitemShut {NoStop}%
\bibitem [{\citenamefont {Katzir}\ \emph {et~al.}(2020)\citenamefont {Katzir}, \citenamefont {Stern}, \citenamefont {Berg},\ and\ \citenamefont {Lindner}}]{katzir2020superconducting}%
  \BibitemOpen
  \bibfield  {author} {\bibinfo {author} {\bibfnamefont {B.~A.}\ \bibnamefont {Katzir}}, \bibinfo {author} {\bibfnamefont {A.}~\bibnamefont {Stern}}, \bibinfo {author} {\bibfnamefont {E.}~\bibnamefont {Berg}},\ and\ \bibinfo {author} {\bibfnamefont {N.~H.}\ \bibnamefont {Lindner}},\ }\bibfield  {title} {\bibinfo {title} {Superconducting fractional quantum hall edges via repulsive interactions},\ }\href@noop {} {\bibfield  {journal} {\bibinfo  {journal} {arXiv preprint arXiv:2011.13950}\ } (\bibinfo {year} {2020})}\BibitemShut {NoStop}%
\bibitem [{\citenamefont {Liu}\ and\ \citenamefont {Bergholtz}(2019)}]{liu2019fractional}%
  \BibitemOpen
  \bibfield  {author} {\bibinfo {author} {\bibfnamefont {Z.}~\bibnamefont {Liu}}\ and\ \bibinfo {author} {\bibfnamefont {E.~J.}\ \bibnamefont {Bergholtz}},\ }\bibfield  {title} {\bibinfo {title} {Fractional quantum hall states with gapped boundaries in an extreme lattice limit},\ }\href@noop {} {\bibfield  {journal} {\bibinfo  {journal} {Physical Review B}\ }\textbf {\bibinfo {volume} {99}},\ \bibinfo {pages} {195122} (\bibinfo {year} {2019})}\BibitemShut {NoStop}%
\bibitem [{\citenamefont {Cong}\ \emph {et~al.}(2017)\citenamefont {Cong}, \citenamefont {Cheng},\ and\ \citenamefont {Wang}}]{cong2017universal}%
  \BibitemOpen
  \bibfield  {author} {\bibinfo {author} {\bibfnamefont {I.}~\bibnamefont {Cong}}, \bibinfo {author} {\bibfnamefont {M.}~\bibnamefont {Cheng}},\ and\ \bibinfo {author} {\bibfnamefont {Z.}~\bibnamefont {Wang}},\ }\bibfield  {title} {\bibinfo {title} {Universal quantum computation with gapped boundaries},\ }\href@noop {} {\bibfield  {journal} {\bibinfo  {journal} {Physical Review Letters}\ }\textbf {\bibinfo {volume} {119}},\ \bibinfo {pages} {170504} (\bibinfo {year} {2017})}\BibitemShut {NoStop}%
\bibitem [{\citenamefont {Chen}\ and\ \citenamefont {Yang}(2012)}]{chen2012interaction}%
  \BibitemOpen
  \bibfield  {author} {\bibinfo {author} {\bibfnamefont {H.}~\bibnamefont {Chen}}\ and\ \bibinfo {author} {\bibfnamefont {K.}~\bibnamefont {Yang}},\ }\bibfield  {title} {\bibinfo {title} {Interaction-driven quantum phase transitions in fractional topological insulators},\ }\href@noop {} {\bibfield  {journal} {\bibinfo  {journal} {Physical Review B}\ }\textbf {\bibinfo {volume} {85}},\ \bibinfo {pages} {195113} (\bibinfo {year} {2012})}\BibitemShut {NoStop}%
\bibitem [{\citenamefont {Furukawa}\ and\ \citenamefont {Ueda}(2017)}]{furukawa2017quantum}%
  \BibitemOpen
  \bibfield  {author} {\bibinfo {author} {\bibfnamefont {S.}~\bibnamefont {Furukawa}}\ and\ \bibinfo {author} {\bibfnamefont {M.}~\bibnamefont {Ueda}},\ }\bibfield  {title} {\bibinfo {title} {Quantum hall phase diagram of two-component bose gases: Intercomponent entanglement and pseudopotentials},\ }\href@noop {} {\bibfield  {journal} {\bibinfo  {journal} {Physical Review A}\ }\textbf {\bibinfo {volume} {96}},\ \bibinfo {pages} {053626} (\bibinfo {year} {2017})}\BibitemShut {NoStop}%
\bibitem [{\citenamefont {Crépel}\ and\ \citenamefont {Millis}(2024)}]{crepel2024bridging}%
  \BibitemOpen
  \bibfield  {author} {\bibinfo {author} {\bibfnamefont {V.}~\bibnamefont {Crépel}}\ and\ \bibinfo {author} {\bibfnamefont {A.}~\bibnamefont {Millis}},\ }\href@noop {} {\bibinfo {title} {Bridging the small and large in twisted transition metal dicalcogenide homobilayers: a tight binding model capturing orbital interference and topology across a wide range of twist angles}} (\bibinfo {year} {2024}),\ \Eprint {https://arxiv.org/abs/2403.15546} {arXiv:2403.15546 [cond-mat.str-el]} \BibitemShut {NoStop}%
\bibitem [{\citenamefont {Li}\ \emph {et~al.}(2021{\natexlab{a}})\citenamefont {Li}, \citenamefont {Kumar}, \citenamefont {Sun},\ and\ \citenamefont {Lin}}]{li2021spontaneous}%
  \BibitemOpen
  \bibfield  {author} {\bibinfo {author} {\bibfnamefont {H.}~\bibnamefont {Li}}, \bibinfo {author} {\bibfnamefont {U.}~\bibnamefont {Kumar}}, \bibinfo {author} {\bibfnamefont {K.}~\bibnamefont {Sun}},\ and\ \bibinfo {author} {\bibfnamefont {S.-Z.}\ \bibnamefont {Lin}},\ }\bibfield  {title} {\bibinfo {title} {Spontaneous fractional chern insulators in transition metal dichalcogenide moir{\'e} superlattices},\ }\href@noop {} {\bibfield  {journal} {\bibinfo  {journal} {Physical Review Research}\ }\textbf {\bibinfo {volume} {3}},\ \bibinfo {pages} {L032070} (\bibinfo {year} {2021}{\natexlab{a}})}\BibitemShut {NoStop}%
\bibitem [{\citenamefont {Song}\ \emph {et~al.}(2023)\citenamefont {Song}, \citenamefont {Zhang},\ and\ \citenamefont {Senthil}}]{song2023phase}%
  \BibitemOpen
  \bibfield  {author} {\bibinfo {author} {\bibfnamefont {X.-Y.}\ \bibnamefont {Song}}, \bibinfo {author} {\bibfnamefont {Y.-H.}\ \bibnamefont {Zhang}},\ and\ \bibinfo {author} {\bibfnamefont {T.}~\bibnamefont {Senthil}},\ }\bibfield  {title} {\bibinfo {title} {Phase transitions out of quantum hall states in moir\'e tmd bilayers},\ }\href@noop {} {\bibfield  {journal} {\bibinfo  {journal} {arXiv preprint arXiv:2308.10903}\ } (\bibinfo {year} {2023})}\BibitemShut {NoStop}%
\bibitem [{\citenamefont {Wilhelm}\ \emph {et~al.}(2021)\citenamefont {Wilhelm}, \citenamefont {Lang},\ and\ \citenamefont {L{\"a}uchli}}]{wilhelm2021interplay}%
  \BibitemOpen
  \bibfield  {author} {\bibinfo {author} {\bibfnamefont {P.}~\bibnamefont {Wilhelm}}, \bibinfo {author} {\bibfnamefont {T.~C.}\ \bibnamefont {Lang}},\ and\ \bibinfo {author} {\bibfnamefont {A.~M.}\ \bibnamefont {L{\"a}uchli}},\ }\bibfield  {title} {\bibinfo {title} {Interplay of fractional chern insulator and charge density wave phases in twisted bilayer graphene},\ }\href@noop {} {\bibfield  {journal} {\bibinfo  {journal} {Physical Review B}\ }\textbf {\bibinfo {volume} {103}},\ \bibinfo {pages} {125406} (\bibinfo {year} {2021})}\BibitemShut {NoStop}%
\bibitem [{\citenamefont {Reddy}\ and\ \citenamefont {Fu}(2023)}]{reddy2023toward}%
  \BibitemOpen
  \bibfield  {author} {\bibinfo {author} {\bibfnamefont {A.~P.}\ \bibnamefont {Reddy}}\ and\ \bibinfo {author} {\bibfnamefont {L.}~\bibnamefont {Fu}},\ }\bibfield  {title} {\bibinfo {title} {Toward a global phase diagram of the fractional quantum anomalous hall effect},\ }\href@noop {} {\bibfield  {journal} {\bibinfo  {journal} {Physical Review B}\ }\textbf {\bibinfo {volume} {108}},\ \bibinfo {pages} {245159} (\bibinfo {year} {2023})}\BibitemShut {NoStop}%
\bibitem [{\citenamefont {L{\"a}uchli}\ \emph {et~al.}(2013)\citenamefont {L{\"a}uchli}, \citenamefont {Liu}, \citenamefont {Bergholtz},\ and\ \citenamefont {Moessner}}]{lauchli2013hierarchy}%
  \BibitemOpen
  \bibfield  {author} {\bibinfo {author} {\bibfnamefont {A.~M.}\ \bibnamefont {L{\"a}uchli}}, \bibinfo {author} {\bibfnamefont {Z.}~\bibnamefont {Liu}}, \bibinfo {author} {\bibfnamefont {E.~J.}\ \bibnamefont {Bergholtz}},\ and\ \bibinfo {author} {\bibfnamefont {R.}~\bibnamefont {Moessner}},\ }\bibfield  {title} {\bibinfo {title} {Hierarchy of fractional chern insulators and competing compressible states},\ }\href@noop {} {\bibfield  {journal} {\bibinfo  {journal} {Physical Review Letters}\ }\textbf {\bibinfo {volume} {111}},\ \bibinfo {pages} {126802} (\bibinfo {year} {2013})}\BibitemShut {NoStop}%
\bibitem [{\citenamefont {Bardeen}\ \emph {et~al.}(1957)\citenamefont {Bardeen}, \citenamefont {Cooper},\ and\ \citenamefont {Schrieffer}}]{bardeen1957microscopic}%
  \BibitemOpen
  \bibfield  {author} {\bibinfo {author} {\bibfnamefont {J.}~\bibnamefont {Bardeen}}, \bibinfo {author} {\bibfnamefont {L.~N.}\ \bibnamefont {Cooper}},\ and\ \bibinfo {author} {\bibfnamefont {J.~R.}\ \bibnamefont {Schrieffer}},\ }\bibfield  {title} {\bibinfo {title} {Microscopic theory of superconductivity},\ }\href@noop {} {\bibfield  {journal} {\bibinfo  {journal} {Physical Review}\ }\textbf {\bibinfo {volume} {106}},\ \bibinfo {pages} {162} (\bibinfo {year} {1957})}\BibitemShut {NoStop}%
\bibitem [{\citenamefont {Randeria}\ \emph {et~al.}(1989)\citenamefont {Randeria}, \citenamefont {Duan},\ and\ \citenamefont {Shieh}}]{randeria1989bound}%
  \BibitemOpen
  \bibfield  {author} {\bibinfo {author} {\bibfnamefont {M.}~\bibnamefont {Randeria}}, \bibinfo {author} {\bibfnamefont {J.-M.}\ \bibnamefont {Duan}},\ and\ \bibinfo {author} {\bibfnamefont {L.-Y.}\ \bibnamefont {Shieh}},\ }\bibfield  {title} {\bibinfo {title} {Bound states, cooper pairing, and bose condensation in two dimensions},\ }\href@noop {} {\bibfield  {journal} {\bibinfo  {journal} {Physical review letters}\ }\textbf {\bibinfo {volume} {62}},\ \bibinfo {pages} {981} (\bibinfo {year} {1989})}\BibitemShut {NoStop}%
\bibitem [{\citenamefont {Cr{\'e}pel}\ and\ \citenamefont {Fu}(2021)}]{crepel2021new}%
  \BibitemOpen
  \bibfield  {author} {\bibinfo {author} {\bibfnamefont {V.}~\bibnamefont {Cr{\'e}pel}}\ and\ \bibinfo {author} {\bibfnamefont {L.}~\bibnamefont {Fu}},\ }\bibfield  {title} {\bibinfo {title} {New mechanism and exact theory of superconductivity from strong repulsive interaction},\ }\href@noop {} {\bibfield  {journal} {\bibinfo  {journal} {Science Advances}\ }\textbf {\bibinfo {volume} {7}},\ \bibinfo {pages} {eabh2233} (\bibinfo {year} {2021})}\BibitemShut {NoStop}%
\bibitem [{\citenamefont {Sohal}\ \emph {et~al.}(2018)\citenamefont {Sohal}, \citenamefont {Santos},\ and\ \citenamefont {Fradkin}}]{sohal2018chern}%
  \BibitemOpen
  \bibfield  {author} {\bibinfo {author} {\bibfnamefont {R.}~\bibnamefont {Sohal}}, \bibinfo {author} {\bibfnamefont {L.~H.}\ \bibnamefont {Santos}},\ and\ \bibinfo {author} {\bibfnamefont {E.}~\bibnamefont {Fradkin}},\ }\bibfield  {title} {\bibinfo {title} {Chern-simons composite fermion theory of fractional chern insulators},\ }\href@noop {} {\bibfield  {journal} {\bibinfo  {journal} {Physical Review B}\ }\textbf {\bibinfo {volume} {97}},\ \bibinfo {pages} {125131} (\bibinfo {year} {2018})}\BibitemShut {NoStop}%
\bibitem [{\citenamefont {Levin}\ and\ \citenamefont {Stern}(2009)}]{levin2009fractional}%
  \BibitemOpen
  \bibfield  {author} {\bibinfo {author} {\bibfnamefont {M.}~\bibnamefont {Levin}}\ and\ \bibinfo {author} {\bibfnamefont {A.}~\bibnamefont {Stern}},\ }\bibfield  {title} {\bibinfo {title} {Fractional topological insulators},\ }\href@noop {} {\bibfield  {journal} {\bibinfo  {journal} {Physical review letters}\ }\textbf {\bibinfo {volume} {103}},\ \bibinfo {pages} {196803} (\bibinfo {year} {2009})}\BibitemShut {NoStop}%
\bibitem [{\citenamefont {Neupert}\ \emph {et~al.}(2015)\citenamefont {Neupert}, \citenamefont {Chamon}, \citenamefont {Iadecola}, \citenamefont {Santos},\ and\ \citenamefont {Mudry}}]{neupert2015fractional}%
  \BibitemOpen
  \bibfield  {author} {\bibinfo {author} {\bibfnamefont {T.}~\bibnamefont {Neupert}}, \bibinfo {author} {\bibfnamefont {C.}~\bibnamefont {Chamon}}, \bibinfo {author} {\bibfnamefont {T.}~\bibnamefont {Iadecola}}, \bibinfo {author} {\bibfnamefont {L.~H.}\ \bibnamefont {Santos}},\ and\ \bibinfo {author} {\bibfnamefont {C.}~\bibnamefont {Mudry}},\ }\bibfield  {title} {\bibinfo {title} {Fractional (chern and topological) insulators},\ }\href {https://doi.org/10.1088/0031-8949/2015/t164/014005} {\bibfield  {journal} {\bibinfo  {journal} {Physica Scripta}\ }\textbf {\bibinfo {volume} {T164}},\ \bibinfo {pages} {014005} (\bibinfo {year} {2015})}\BibitemShut {NoStop}%
\bibitem [{\citenamefont {Stern}(2016)}]{stern2016fractional}%
  \BibitemOpen
  \bibfield  {author} {\bibinfo {author} {\bibfnamefont {A.}~\bibnamefont {Stern}},\ }\bibfield  {title} {\bibinfo {title} {Fractional topological insulators: a pedagogical review},\ }\href@noop {} {\bibfield  {journal} {\bibinfo  {journal} {Annual Review of Condensed Matter Physics}\ }\textbf {\bibinfo {volume} {7}},\ \bibinfo {pages} {349} (\bibinfo {year} {2016})}\BibitemShut {NoStop}%
\bibitem [{\citenamefont {Metlitski}\ \emph {et~al.}(2015)\citenamefont {Metlitski}, \citenamefont {Mross}, \citenamefont {Sachdev},\ and\ \citenamefont {Senthil}}]{metlitski2015cooper}%
  \BibitemOpen
  \bibfield  {author} {\bibinfo {author} {\bibfnamefont {M.~A.}\ \bibnamefont {Metlitski}}, \bibinfo {author} {\bibfnamefont {D.~F.}\ \bibnamefont {Mross}}, \bibinfo {author} {\bibfnamefont {S.}~\bibnamefont {Sachdev}},\ and\ \bibinfo {author} {\bibfnamefont {T.}~\bibnamefont {Senthil}},\ }\bibfield  {title} {\bibinfo {title} {Cooper pairing in non-fermi liquids},\ }\href@noop {} {\bibfield  {journal} {\bibinfo  {journal} {Physical Review B}\ }\textbf {\bibinfo {volume} {91}},\ \bibinfo {pages} {115111} (\bibinfo {year} {2015})}\BibitemShut {NoStop}%
\bibitem [{\citenamefont {Haldane}(1988)}]{haldane1988model}%
  \BibitemOpen
  \bibfield  {author} {\bibinfo {author} {\bibfnamefont {F.~D.~M.}\ \bibnamefont {Haldane}},\ }\bibfield  {title} {\bibinfo {title} {Model for a quantum hall effect without landau levels: Condensed-matter realization of the "parity anomaly"},\ }\href {https://doi.org/10.1103/PhysRevLett.61.2015} {\bibfield  {journal} {\bibinfo  {journal} {Phys. Rev. Lett.}\ }\textbf {\bibinfo {volume} {61}},\ \bibinfo {pages} {2015} (\bibinfo {year} {1988})}\BibitemShut {NoStop}%
\bibitem [{\citenamefont {Kane}\ and\ \citenamefont {Mele}(2005)}]{kane2005z2}%
  \BibitemOpen
  \bibfield  {author} {\bibinfo {author} {\bibfnamefont {C.~L.}\ \bibnamefont {Kane}}\ and\ \bibinfo {author} {\bibfnamefont {E.~J.}\ \bibnamefont {Mele}},\ }\bibfield  {title} {\bibinfo {title} {${Z}_{2}$ topological order and the quantum spin hall effect},\ }\href {https://doi.org/10.1103/PhysRevLett.95.146802} {\bibfield  {journal} {\bibinfo  {journal} {Phys. Rev. Lett.}\ }\textbf {\bibinfo {volume} {95}},\ \bibinfo {pages} {146802} (\bibinfo {year} {2005})}\BibitemShut {NoStop}%
\bibitem [{\citenamefont {Wu}\ \emph {et~al.}(2012)\citenamefont {Wu}, \citenamefont {Bernevig},\ and\ \citenamefont {Regnault}}]{wu2012zoology}%
  \BibitemOpen
  \bibfield  {author} {\bibinfo {author} {\bibfnamefont {Y.-L.}\ \bibnamefont {Wu}}, \bibinfo {author} {\bibfnamefont {B.~A.}\ \bibnamefont {Bernevig}},\ and\ \bibinfo {author} {\bibfnamefont {N.}~\bibnamefont {Regnault}},\ }\bibfield  {title} {\bibinfo {title} {Zoology of fractional chern insulators},\ }\bibfield  {journal} {\bibinfo  {journal} {Physical Review B}\ }\textbf {\bibinfo {volume} {85}},\ \href {https://doi.org/10.1103/physrevb.85.075116} {10.1103/physrevb.85.075116} (\bibinfo {year} {2012})\BibitemShut {NoStop}%
\bibitem [{\citenamefont {Repellin}\ \emph {et~al.}(2014)\citenamefont {Repellin}, \citenamefont {Bernevig},\ and\ \citenamefont {Regnault}}]{repellin2014z}%
  \BibitemOpen
  \bibfield  {author} {\bibinfo {author} {\bibfnamefont {C.}~\bibnamefont {Repellin}}, \bibinfo {author} {\bibfnamefont {B.~A.}\ \bibnamefont {Bernevig}},\ and\ \bibinfo {author} {\bibfnamefont {N.}~\bibnamefont {Regnault}},\ }\bibfield  {title} {\bibinfo {title} {Z 2 fractional topological insulators in two dimensions},\ }\href@noop {} {\bibfield  {journal} {\bibinfo  {journal} {Physical Review B}\ }\textbf {\bibinfo {volume} {90}},\ \bibinfo {pages} {245401} (\bibinfo {year} {2014})}\BibitemShut {NoStop}%
\bibitem [{\citenamefont {Scalapino}\ \emph {et~al.}(1993)\citenamefont {Scalapino}, \citenamefont {White},\ and\ \citenamefont {Zhang}}]{scalapino1993insulator}%
  \BibitemOpen
  \bibfield  {author} {\bibinfo {author} {\bibfnamefont {D.~J.}\ \bibnamefont {Scalapino}}, \bibinfo {author} {\bibfnamefont {S.~R.}\ \bibnamefont {White}},\ and\ \bibinfo {author} {\bibfnamefont {S.}~\bibnamefont {Zhang}},\ }\bibfield  {title} {\bibinfo {title} {Insulator, metal, or superconductor: The criteria},\ }\href@noop {} {\bibfield  {journal} {\bibinfo  {journal} {Physical Review B}\ }\textbf {\bibinfo {volume} {47}},\ \bibinfo {pages} {7995} (\bibinfo {year} {1993})}\BibitemShut {NoStop}%
\bibitem [{\citenamefont {Loder}\ \emph {et~al.}(2008)\citenamefont {Loder}, \citenamefont {Kampf},\ and\ \citenamefont {Kopp}}]{loder2008crossover}%
  \BibitemOpen
  \bibfield  {author} {\bibinfo {author} {\bibfnamefont {F.}~\bibnamefont {Loder}}, \bibinfo {author} {\bibfnamefont {A.~P.}\ \bibnamefont {Kampf}},\ and\ \bibinfo {author} {\bibfnamefont {T.}~\bibnamefont {Kopp}},\ }\bibfield  {title} {\bibinfo {title} {Crossover from hc/e to hc/2 e current oscillations in rings of s-wave superconductors},\ }\href@noop {} {\bibfield  {journal} {\bibinfo  {journal} {Physical Review B—Condensed Matter and Materials Physics}\ }\textbf {\bibinfo {volume} {78}},\ \bibinfo {pages} {174526} (\bibinfo {year} {2008})}\BibitemShut {NoStop}%
\bibitem [{\citenamefont {Kane}\ \emph {et~al.}(2002)\citenamefont {Kane}, \citenamefont {Mukhopadhyay},\ and\ \citenamefont {Lubensky}}]{kane2002fractional}%
  \BibitemOpen
  \bibfield  {author} {\bibinfo {author} {\bibfnamefont {C.}~\bibnamefont {Kane}}, \bibinfo {author} {\bibfnamefont {R.}~\bibnamefont {Mukhopadhyay}},\ and\ \bibinfo {author} {\bibfnamefont {T.}~\bibnamefont {Lubensky}},\ }\bibfield  {title} {\bibinfo {title} {Fractional quantum hall effect in an array of quantum wires},\ }\href@noop {} {\bibfield  {journal} {\bibinfo  {journal} {Physical review letters}\ }\textbf {\bibinfo {volume} {88}},\ \bibinfo {pages} {036401} (\bibinfo {year} {2002})}\BibitemShut {NoStop}%
\bibitem [{\citenamefont {Teo}\ and\ \citenamefont {Kane}(2014)}]{teo2014luttinger}%
  \BibitemOpen
  \bibfield  {author} {\bibinfo {author} {\bibfnamefont {J.~C.}\ \bibnamefont {Teo}}\ and\ \bibinfo {author} {\bibfnamefont {C.}~\bibnamefont {Kane}},\ }\bibfield  {title} {\bibinfo {title} {From luttinger liquid to non-abelian quantum hall states},\ }\href@noop {} {\bibfield  {journal} {\bibinfo  {journal} {Physical Review B}\ }\textbf {\bibinfo {volume} {89}},\ \bibinfo {pages} {085101} (\bibinfo {year} {2014})}\BibitemShut {NoStop}%
\bibitem [{\citenamefont {Cr{\'e}pel}\ \emph {et~al.}(2020)\citenamefont {Cr{\'e}pel}, \citenamefont {Estienne},\ and\ \citenamefont {Regnault}}]{crepel2020microscopic}%
  \BibitemOpen
  \bibfield  {author} {\bibinfo {author} {\bibfnamefont {V.}~\bibnamefont {Cr{\'e}pel}}, \bibinfo {author} {\bibfnamefont {B.}~\bibnamefont {Estienne}},\ and\ \bibinfo {author} {\bibfnamefont {N.}~\bibnamefont {Regnault}},\ }\bibfield  {title} {\bibinfo {title} {Microscopic study of the coupled-wire construction and plausible realization in spin-dependent optical lattices},\ }\href@noop {} {\bibfield  {journal} {\bibinfo  {journal} {Physical Review B}\ }\textbf {\bibinfo {volume} {101}},\ \bibinfo {pages} {235158} (\bibinfo {year} {2020})}\BibitemShut {NoStop}%
\bibitem [{\citenamefont {Neupert}\ \emph {et~al.}(2014)\citenamefont {Neupert}, \citenamefont {Chamon}, \citenamefont {Mudry},\ and\ \citenamefont {Thomale}}]{neupert2014wire}%
  \BibitemOpen
  \bibfield  {author} {\bibinfo {author} {\bibfnamefont {T.}~\bibnamefont {Neupert}}, \bibinfo {author} {\bibfnamefont {C.}~\bibnamefont {Chamon}}, \bibinfo {author} {\bibfnamefont {C.}~\bibnamefont {Mudry}},\ and\ \bibinfo {author} {\bibfnamefont {R.}~\bibnamefont {Thomale}},\ }\bibfield  {title} {\bibinfo {title} {Wire deconstructionism of two-dimensional topological phases},\ }\href@noop {} {\bibfield  {journal} {\bibinfo  {journal} {Physical Review B}\ }\textbf {\bibinfo {volume} {90}},\ \bibinfo {pages} {205101} (\bibinfo {year} {2014})}\BibitemShut {NoStop}%
\bibitem [{\citenamefont {Klinovaja}\ and\ \citenamefont {Tserkovnyak}(2014)}]{klinovaja2014quantum}%
  \BibitemOpen
  \bibfield  {author} {\bibinfo {author} {\bibfnamefont {J.}~\bibnamefont {Klinovaja}}\ and\ \bibinfo {author} {\bibfnamefont {Y.}~\bibnamefont {Tserkovnyak}},\ }\bibfield  {title} {\bibinfo {title} {Quantum spin hall effect in strip of stripes model},\ }\href@noop {} {\bibfield  {journal} {\bibinfo  {journal} {Physical Review B}\ }\textbf {\bibinfo {volume} {90}},\ \bibinfo {pages} {115426} (\bibinfo {year} {2014})}\BibitemShut {NoStop}%
\bibitem [{\citenamefont {Santos}\ \emph {et~al.}(2015)\citenamefont {Santos}, \citenamefont {Huang}, \citenamefont {Gefen},\ and\ \citenamefont {Gutman}}]{santos2015fractional}%
  \BibitemOpen
  \bibfield  {author} {\bibinfo {author} {\bibfnamefont {R.~A.}\ \bibnamefont {Santos}}, \bibinfo {author} {\bibfnamefont {C.-W.}\ \bibnamefont {Huang}}, \bibinfo {author} {\bibfnamefont {Y.}~\bibnamefont {Gefen}},\ and\ \bibinfo {author} {\bibfnamefont {D.}~\bibnamefont {Gutman}},\ }\bibfield  {title} {\bibinfo {title} {Fractional topological insulators: From sliding luttinger liquids to chern-simons theory},\ }\href@noop {} {\bibfield  {journal} {\bibinfo  {journal} {Physical Review B}\ }\textbf {\bibinfo {volume} {91}},\ \bibinfo {pages} {205141} (\bibinfo {year} {2015})}\BibitemShut {NoStop}%
\bibitem [{Note1()}]{Note1}%
  \BibitemOpen
  \bibinfo {note} {See Refs.~\cite {kane2002fractional,teo2014luttinger,neupert2014wire,klinovaja2014quantum,santos2015fractional,mukhopadhyay2001sliding} for details and App.~\ref {app_exactpoint} for a short derivation of this effective action.}\BibitemShut {Stop}%
\bibitem [{\citenamefont {Neupert}\ \emph {et~al.}(2011{\natexlab{b}})\citenamefont {Neupert}, \citenamefont {Santos}, \citenamefont {Chamon},\ and\ \citenamefont {Mudry}}]{neupert2011fractional2}%
  \BibitemOpen
  \bibfield  {author} {\bibinfo {author} {\bibfnamefont {T.}~\bibnamefont {Neupert}}, \bibinfo {author} {\bibfnamefont {L.}~\bibnamefont {Santos}}, \bibinfo {author} {\bibfnamefont {C.}~\bibnamefont {Chamon}},\ and\ \bibinfo {author} {\bibfnamefont {C.}~\bibnamefont {Mudry}},\ }\bibfield  {title} {\bibinfo {title} {Fractional quantum hall states at zero magnetic field},\ }\href@noop {} {\bibfield  {journal} {\bibinfo  {journal} {Physical review letters}\ }\textbf {\bibinfo {volume} {106}},\ \bibinfo {pages} {236804} (\bibinfo {year} {2011}{\natexlab{b}})}\BibitemShut {NoStop}%
\bibitem [{\citenamefont {Bernevig}\ and\ \citenamefont {Regnault}(2012)}]{bernevig2012emergent}%
  \BibitemOpen
  \bibfield  {author} {\bibinfo {author} {\bibfnamefont {B.~A.}\ \bibnamefont {Bernevig}}\ and\ \bibinfo {author} {\bibfnamefont {N.}~\bibnamefont {Regnault}},\ }\bibfield  {title} {\bibinfo {title} {Emergent many-body translational symmetries of abelian and non-abelian fractionally filled topological insulators},\ }\href@noop {} {\bibfield  {journal} {\bibinfo  {journal} {Physical Review B}\ }\textbf {\bibinfo {volume} {85}},\ \bibinfo {pages} {075128} (\bibinfo {year} {2012})}\BibitemShut {NoStop}%
\bibitem [{\citenamefont {Carr}\ \emph {et~al.}(2018)\citenamefont {Carr}, \citenamefont {Massatt}, \citenamefont {Torrisi}, \citenamefont {Cazeaux}, \citenamefont {Luskin},\ and\ \citenamefont {Kaxiras}}]{carr2018relaxation}%
  \BibitemOpen
  \bibfield  {author} {\bibinfo {author} {\bibfnamefont {S.}~\bibnamefont {Carr}}, \bibinfo {author} {\bibfnamefont {D.}~\bibnamefont {Massatt}}, \bibinfo {author} {\bibfnamefont {S.~B.}\ \bibnamefont {Torrisi}}, \bibinfo {author} {\bibfnamefont {P.}~\bibnamefont {Cazeaux}}, \bibinfo {author} {\bibfnamefont {M.}~\bibnamefont {Luskin}},\ and\ \bibinfo {author} {\bibfnamefont {E.}~\bibnamefont {Kaxiras}},\ }\bibfield  {title} {\bibinfo {title} {Relaxation and domain formation in incommensurate two-dimensional heterostructures},\ }\href@noop {} {\bibfield  {journal} {\bibinfo  {journal} {Physical Review B}\ }\textbf {\bibinfo {volume} {98}},\ \bibinfo {pages} {224102} (\bibinfo {year} {2018})}\BibitemShut {NoStop}%
\bibitem [{\citenamefont {Cr{\'e}pel}\ and\ \citenamefont {Fu}(2022)}]{crepel2022spin}%
  \BibitemOpen
  \bibfield  {author} {\bibinfo {author} {\bibfnamefont {V.}~\bibnamefont {Cr{\'e}pel}}\ and\ \bibinfo {author} {\bibfnamefont {L.}~\bibnamefont {Fu}},\ }\bibfield  {title} {\bibinfo {title} {Spin-triplet superconductivity from excitonic effect in doped insulators},\ }\href@noop {} {\bibfield  {journal} {\bibinfo  {journal} {Proceedings of the National Academy of Sciences}\ }\textbf {\bibinfo {volume} {119}},\ \bibinfo {pages} {e2117735119} (\bibinfo {year} {2022})}\BibitemShut {NoStop}%
\bibitem [{\citenamefont {Cr{\'e}pel}\ \emph {et~al.}(2022)\citenamefont {Cr{\'e}pel}, \citenamefont {Cea}, \citenamefont {Fu},\ and\ \citenamefont {Guinea}}]{crepel2022unconventional}%
  \BibitemOpen
  \bibfield  {author} {\bibinfo {author} {\bibfnamefont {V.}~\bibnamefont {Cr{\'e}pel}}, \bibinfo {author} {\bibfnamefont {T.}~\bibnamefont {Cea}}, \bibinfo {author} {\bibfnamefont {L.}~\bibnamefont {Fu}},\ and\ \bibinfo {author} {\bibfnamefont {F.}~\bibnamefont {Guinea}},\ }\bibfield  {title} {\bibinfo {title} {Unconventional superconductivity due to interband polarization},\ }\href@noop {} {\bibfield  {journal} {\bibinfo  {journal} {Physical Review B}\ }\textbf {\bibinfo {volume} {105}},\ \bibinfo {pages} {094506} (\bibinfo {year} {2022})}\BibitemShut {NoStop}%
\bibitem [{\citenamefont {Cr\'epel}\ \emph {et~al.}(2023)\citenamefont {Cr\'epel}, \citenamefont {Guerci}, \citenamefont {Cano}, \citenamefont {Pixley},\ and\ \citenamefont {Millis}}]{crepel2023topological}%
  \BibitemOpen
  \bibfield  {author} {\bibinfo {author} {\bibfnamefont {V.}~\bibnamefont {Cr\'epel}}, \bibinfo {author} {\bibfnamefont {D.}~\bibnamefont {Guerci}}, \bibinfo {author} {\bibfnamefont {J.}~\bibnamefont {Cano}}, \bibinfo {author} {\bibfnamefont {J.~H.}\ \bibnamefont {Pixley}},\ and\ \bibinfo {author} {\bibfnamefont {A.}~\bibnamefont {Millis}},\ }\bibfield  {title} {\bibinfo {title} {Topological superconductivity in doped magnetic moir\'e semiconductors},\ }\href {https://doi.org/10.1103/PhysRevLett.131.056001} {\bibfield  {journal} {\bibinfo  {journal} {Phys. Rev. Lett.}\ }\textbf {\bibinfo {volume} {131}},\ \bibinfo {pages} {056001} (\bibinfo {year} {2023})}\BibitemShut {NoStop}%
\bibitem [{\citenamefont {Heersche}\ \emph {et~al.}(2007)\citenamefont {Heersche}, \citenamefont {Jarillo-Herrero}, \citenamefont {Oostinga}, \citenamefont {Vandersypen},\ and\ \citenamefont {Morpurgo}}]{heersche2007bipolar}%
  \BibitemOpen
  \bibfield  {author} {\bibinfo {author} {\bibfnamefont {H.~B.}\ \bibnamefont {Heersche}}, \bibinfo {author} {\bibfnamefont {P.}~\bibnamefont {Jarillo-Herrero}}, \bibinfo {author} {\bibfnamefont {J.~B.}\ \bibnamefont {Oostinga}}, \bibinfo {author} {\bibfnamefont {L.~M.}\ \bibnamefont {Vandersypen}},\ and\ \bibinfo {author} {\bibfnamefont {A.~F.}\ \bibnamefont {Morpurgo}},\ }\bibfield  {title} {\bibinfo {title} {Bipolar supercurrent in graphene},\ }\href@noop {} {\bibfield  {journal} {\bibinfo  {journal} {Nature}\ }\textbf {\bibinfo {volume} {446}},\ \bibinfo {pages} {56} (\bibinfo {year} {2007})}\BibitemShut {NoStop}%
\bibitem [{\citenamefont {Tong}\ \emph {et~al.}(2017)\citenamefont {Tong}, \citenamefont {Yu}, \citenamefont {Zhu}, \citenamefont {Wang}, \citenamefont {Xu},\ and\ \citenamefont {Yao}}]{tong2017topological}%
  \BibitemOpen
  \bibfield  {author} {\bibinfo {author} {\bibfnamefont {Q.}~\bibnamefont {Tong}}, \bibinfo {author} {\bibfnamefont {H.}~\bibnamefont {Yu}}, \bibinfo {author} {\bibfnamefont {Q.}~\bibnamefont {Zhu}}, \bibinfo {author} {\bibfnamefont {Y.}~\bibnamefont {Wang}}, \bibinfo {author} {\bibfnamefont {X.}~\bibnamefont {Xu}},\ and\ \bibinfo {author} {\bibfnamefont {W.}~\bibnamefont {Yao}},\ }\bibfield  {title} {\bibinfo {title} {Topological mosaics in moir{\'e} superlattices of van der waals heterobilayers},\ }\href@noop {} {\bibfield  {journal} {\bibinfo  {journal} {Nature Physics}\ }\textbf {\bibinfo {volume} {13}},\ \bibinfo {pages} {356} (\bibinfo {year} {2017})}\BibitemShut {NoStop}%
\bibitem [{\citenamefont {Cr{\'e}pel}\ \emph {et~al.}(2023)\citenamefont {Cr{\'e}pel}, \citenamefont {Regnault},\ and\ \citenamefont {Queiroz}}]{crepel2023chiral}%
  \BibitemOpen
  \bibfield  {author} {\bibinfo {author} {\bibfnamefont {V.}~\bibnamefont {Cr{\'e}pel}}, \bibinfo {author} {\bibfnamefont {N.}~\bibnamefont {Regnault}},\ and\ \bibinfo {author} {\bibfnamefont {R.}~\bibnamefont {Queiroz}},\ }\bibfield  {title} {\bibinfo {title} {The chiral limits of moir\'e semiconductors: origin of flat bands and topology in twisted transition metal dichalcogenides homobilayers},\ }\href@noop {} {\bibfield  {journal} {\bibinfo  {journal} {arXiv preprint arXiv:2305.10477}\ } (\bibinfo {year} {2023})}\BibitemShut {NoStop}%
\bibitem [{\citenamefont {Cr{\'e}pel}\ \emph {et~al.}(2024)\citenamefont {Cr{\'e}pel}, \citenamefont {Ding}, \citenamefont {Verma}, \citenamefont {Regnault},\ and\ \citenamefont {Queiroz}}]{crepel2024topologically}%
  \BibitemOpen
  \bibfield  {author} {\bibinfo {author} {\bibfnamefont {V.}~\bibnamefont {Cr{\'e}pel}}, \bibinfo {author} {\bibfnamefont {P.}~\bibnamefont {Ding}}, \bibinfo {author} {\bibfnamefont {N.}~\bibnamefont {Verma}}, \bibinfo {author} {\bibfnamefont {N.}~\bibnamefont {Regnault}},\ and\ \bibinfo {author} {\bibfnamefont {R.}~\bibnamefont {Queiroz}},\ }\bibfield  {title} {\bibinfo {title} {Topologically protected flatness in chiral moir$\backslash$'e heterostructures},\ }\href@noop {} {\bibfield  {journal} {\bibinfo  {journal} {arXiv preprint arXiv:2403.19656}\ } (\bibinfo {year} {2024})}\BibitemShut {NoStop}%
\bibitem [{\citenamefont {Bultinck}\ \emph {et~al.}(2020)\citenamefont {Bultinck}, \citenamefont {Khalaf}, \citenamefont {Liu}, \citenamefont {Chatterjee}, \citenamefont {Vishwanath},\ and\ \citenamefont {Zaletel}}]{bultinck2020ground}%
  \BibitemOpen
  \bibfield  {author} {\bibinfo {author} {\bibfnamefont {N.}~\bibnamefont {Bultinck}}, \bibinfo {author} {\bibfnamefont {E.}~\bibnamefont {Khalaf}}, \bibinfo {author} {\bibfnamefont {S.}~\bibnamefont {Liu}}, \bibinfo {author} {\bibfnamefont {S.}~\bibnamefont {Chatterjee}}, \bibinfo {author} {\bibfnamefont {A.}~\bibnamefont {Vishwanath}},\ and\ \bibinfo {author} {\bibfnamefont {M.~P.}\ \bibnamefont {Zaletel}},\ }\bibfield  {title} {\bibinfo {title} {Ground state and hidden symmetry of magic-angle graphene at even integer filling},\ }\href@noop {} {\bibfield  {journal} {\bibinfo  {journal} {Physical Review X}\ }\textbf {\bibinfo {volume} {10}},\ \bibinfo {pages} {031034} (\bibinfo {year} {2020})}\BibitemShut {NoStop}%
\bibitem [{\citenamefont {Cr{\'e}pel}\ \emph {et~al.}(2019{\natexlab{a}})\citenamefont {Cr{\'e}pel}, \citenamefont {Claussen}, \citenamefont {Regnault},\ and\ \citenamefont {Estienne}}]{crepel2019microscopic}%
  \BibitemOpen
  \bibfield  {author} {\bibinfo {author} {\bibfnamefont {V.}~\bibnamefont {Cr{\'e}pel}}, \bibinfo {author} {\bibfnamefont {N.}~\bibnamefont {Claussen}}, \bibinfo {author} {\bibfnamefont {N.}~\bibnamefont {Regnault}},\ and\ \bibinfo {author} {\bibfnamefont {B.}~\bibnamefont {Estienne}},\ }\bibfield  {title} {\bibinfo {title} {Microscopic study of the halperin--laughlin interface through matrix product states},\ }\href@noop {} {\bibfield  {journal} {\bibinfo  {journal} {Nature communications}\ }\textbf {\bibinfo {volume} {10}},\ \bibinfo {pages} {1860} (\bibinfo {year} {2019}{\natexlab{a}})}\BibitemShut {NoStop}%
\bibitem [{\citenamefont {Cr{\'e}pel}\ \emph {et~al.}(2019{\natexlab{b}})\citenamefont {Cr{\'e}pel}, \citenamefont {Claussen}, \citenamefont {Estienne},\ and\ \citenamefont {Regnault}}]{crepel2019model}%
  \BibitemOpen
  \bibfield  {author} {\bibinfo {author} {\bibfnamefont {V.}~\bibnamefont {Cr{\'e}pel}}, \bibinfo {author} {\bibfnamefont {N.}~\bibnamefont {Claussen}}, \bibinfo {author} {\bibfnamefont {B.}~\bibnamefont {Estienne}},\ and\ \bibinfo {author} {\bibfnamefont {N.}~\bibnamefont {Regnault}},\ }\bibfield  {title} {\bibinfo {title} {Model states for a class of chiral topological order interfaces},\ }\href@noop {} {\bibfield  {journal} {\bibinfo  {journal} {Nature communications}\ }\textbf {\bibinfo {volume} {10}},\ \bibinfo {pages} {1861} (\bibinfo {year} {2019}{\natexlab{b}})}\BibitemShut {NoStop}%
\bibitem [{\citenamefont {Cr{\'e}pel}\ \emph {et~al.}(2019{\natexlab{c}})\citenamefont {Cr{\'e}pel}, \citenamefont {Estienne},\ and\ \citenamefont {Regnault}}]{crepel2019variational}%
  \BibitemOpen
  \bibfield  {author} {\bibinfo {author} {\bibfnamefont {V.}~\bibnamefont {Cr{\'e}pel}}, \bibinfo {author} {\bibfnamefont {B.}~\bibnamefont {Estienne}},\ and\ \bibinfo {author} {\bibfnamefont {N.}~\bibnamefont {Regnault}},\ }\bibfield  {title} {\bibinfo {title} {Variational ansatz for an abelian to non-abelian topological phase transition in $\nu$= 1/2+ 1/2 bilayers},\ }\href@noop {} {\bibfield  {journal} {\bibinfo  {journal} {Physical review letters}\ }\textbf {\bibinfo {volume} {123}},\ \bibinfo {pages} {126804} (\bibinfo {year} {2019}{\natexlab{c}})}\BibitemShut {NoStop}%
\bibitem [{\citenamefont {Jaworowski}\ and\ \citenamefont {Nielsen}(2020)}]{jaworowski2020model}%
  \BibitemOpen
  \bibfield  {author} {\bibinfo {author} {\bibfnamefont {B.}~\bibnamefont {Jaworowski}}\ and\ \bibinfo {author} {\bibfnamefont {A.~E.}\ \bibnamefont {Nielsen}},\ }\bibfield  {title} {\bibinfo {title} {Model wave functions for interfaces between lattice laughlin states},\ }\href@noop {} {\bibfield  {journal} {\bibinfo  {journal} {Physical Review B}\ }\textbf {\bibinfo {volume} {101}},\ \bibinfo {pages} {245164} (\bibinfo {year} {2020})}\BibitemShut {NoStop}%
\bibitem [{\citenamefont {Jaworowski}\ and\ \citenamefont {Nielsen}(2021)}]{jaworowski2021model}%
  \BibitemOpen
  \bibfield  {author} {\bibinfo {author} {\bibfnamefont {B.}~\bibnamefont {Jaworowski}}\ and\ \bibinfo {author} {\bibfnamefont {A.~E.}\ \bibnamefont {Nielsen}},\ }\bibfield  {title} {\bibinfo {title} {Model wave functions for an interface between lattice laughlin and moore-read states},\ }\href@noop {} {\bibfield  {journal} {\bibinfo  {journal} {Physical Review B}\ }\textbf {\bibinfo {volume} {103}},\ \bibinfo {pages} {205149} (\bibinfo {year} {2021})}\BibitemShut {NoStop}%
\bibitem [{\citenamefont {Zhu}\ \emph {et~al.}(2020)\citenamefont {Zhu}, \citenamefont {Sheng},\ and\ \citenamefont {Yang}}]{zhu2020topological}%
  \BibitemOpen
  \bibfield  {author} {\bibinfo {author} {\bibfnamefont {W.}~\bibnamefont {Zhu}}, \bibinfo {author} {\bibfnamefont {D.}~\bibnamefont {Sheng}},\ and\ \bibinfo {author} {\bibfnamefont {K.}~\bibnamefont {Yang}},\ }\bibfield  {title} {\bibinfo {title} {Topological interface between pfaffian and anti-pfaffian order in $\nu$= 5/2 quantum hall effect},\ }\href@noop {} {\bibfield  {journal} {\bibinfo  {journal} {Physical Review Letters}\ }\textbf {\bibinfo {volume} {125}},\ \bibinfo {pages} {146802} (\bibinfo {year} {2020})}\BibitemShut {NoStop}%
\bibitem [{\citenamefont {Li}\ \emph {et~al.}(2021{\natexlab{b}})\citenamefont {Li}, \citenamefont {Ma}, \citenamefont {Wang}, \citenamefont {Hu}, \citenamefont {Wang},\ and\ \citenamefont {Yang}}]{li2021dynamics}%
  \BibitemOpen
  \bibfield  {author} {\bibinfo {author} {\bibfnamefont {Q.}~\bibnamefont {Li}}, \bibinfo {author} {\bibfnamefont {K.~K.}\ \bibnamefont {Ma}}, \bibinfo {author} {\bibfnamefont {R.}~\bibnamefont {Wang}}, \bibinfo {author} {\bibfnamefont {Z.-X.}\ \bibnamefont {Hu}}, \bibinfo {author} {\bibfnamefont {H.}~\bibnamefont {Wang}},\ and\ \bibinfo {author} {\bibfnamefont {K.}~\bibnamefont {Yang}},\ }\bibfield  {title} {\bibinfo {title} {Dynamics of quantum hall interfaces},\ }\href@noop {} {\bibfield  {journal} {\bibinfo  {journal} {Physical Review B}\ }\textbf {\bibinfo {volume} {104}},\ \bibinfo {pages} {125303} (\bibinfo {year} {2021}{\natexlab{b}})}\BibitemShut {NoStop}%
\bibitem [{\citenamefont {Haldane}\ and\ \citenamefont {Rezayi}(1988)}]{haldane1988spin}%
  \BibitemOpen
  \bibfield  {author} {\bibinfo {author} {\bibfnamefont {F.}~\bibnamefont {Haldane}}\ and\ \bibinfo {author} {\bibfnamefont {E.}~\bibnamefont {Rezayi}},\ }\bibfield  {title} {\bibinfo {title} {Spin-singlet wave function for the half-integral quantum hall effect},\ }\href@noop {} {\bibfield  {journal} {\bibinfo  {journal} {Physical review letters}\ }\textbf {\bibinfo {volume} {60}},\ \bibinfo {pages} {956} (\bibinfo {year} {1988})}\BibitemShut {NoStop}%
\bibitem [{\citenamefont {Cr{\'e}pel}\ \emph {et~al.}(2019{\natexlab{d}})\citenamefont {Cr{\'e}pel}, \citenamefont {Regnault},\ and\ \citenamefont {Estienne}}]{crepel2019matrix}%
  \BibitemOpen
  \bibfield  {author} {\bibinfo {author} {\bibfnamefont {V.}~\bibnamefont {Cr{\'e}pel}}, \bibinfo {author} {\bibfnamefont {N.}~\bibnamefont {Regnault}},\ and\ \bibinfo {author} {\bibfnamefont {B.}~\bibnamefont {Estienne}},\ }\bibfield  {title} {\bibinfo {title} {Matrix product state description and gaplessness of the haldane-rezayi state},\ }\href@noop {} {\bibfield  {journal} {\bibinfo  {journal} {Physical Review B}\ }\textbf {\bibinfo {volume} {100}},\ \bibinfo {pages} {125128} (\bibinfo {year} {2019}{\natexlab{d}})}\BibitemShut {NoStop}%
\bibitem [{\citenamefont {Repellin}\ \emph {et~al.}(2017)\citenamefont {Repellin}, \citenamefont {Yefsah},\ and\ \citenamefont {Sterdyniak}}]{repellin2017creating}%
  \BibitemOpen
  \bibfield  {author} {\bibinfo {author} {\bibfnamefont {C.}~\bibnamefont {Repellin}}, \bibinfo {author} {\bibfnamefont {T.}~\bibnamefont {Yefsah}},\ and\ \bibinfo {author} {\bibfnamefont {A.}~\bibnamefont {Sterdyniak}},\ }\bibfield  {title} {\bibinfo {title} {Creating a bosonic fractional quantum hall state by pairing fermions},\ }\href@noop {} {\bibfield  {journal} {\bibinfo  {journal} {Physical Review B}\ }\textbf {\bibinfo {volume} {96}},\ \bibinfo {pages} {161111} (\bibinfo {year} {2017})}\BibitemShut {NoStop}%
\bibitem [{\citenamefont {Furukawa}\ and\ \citenamefont {Ueda}(2014)}]{furukawa2014global}%
  \BibitemOpen
  \bibfield  {author} {\bibinfo {author} {\bibfnamefont {S.}~\bibnamefont {Furukawa}}\ and\ \bibinfo {author} {\bibfnamefont {M.}~\bibnamefont {Ueda}},\ }\bibfield  {title} {\bibinfo {title} {Global phase diagram of two-component bose gases in antiparallel magnetic fields},\ }\href@noop {} {\bibfield  {journal} {\bibinfo  {journal} {Physical Review A}\ }\textbf {\bibinfo {volume} {90}},\ \bibinfo {pages} {033602} (\bibinfo {year} {2014})}\BibitemShut {NoStop}%
\bibitem [{\citenamefont {Hansson}\ \emph {et~al.}(2017)\citenamefont {Hansson}, \citenamefont {Hermanns}, \citenamefont {Simon},\ and\ \citenamefont {Viefers}}]{hansson2017quantum}%
  \BibitemOpen
  \bibfield  {author} {\bibinfo {author} {\bibfnamefont {T.~H.}\ \bibnamefont {Hansson}}, \bibinfo {author} {\bibfnamefont {M.}~\bibnamefont {Hermanns}}, \bibinfo {author} {\bibfnamefont {S.~H.}\ \bibnamefont {Simon}},\ and\ \bibinfo {author} {\bibfnamefont {S.~F.}\ \bibnamefont {Viefers}},\ }\bibfield  {title} {\bibinfo {title} {Quantum hall physics: Hierarchies and conformal field theory techniques},\ }\href@noop {} {\bibfield  {journal} {\bibinfo  {journal} {Reviews of Modern Physics}\ }\textbf {\bibinfo {volume} {89}},\ \bibinfo {pages} {025005} (\bibinfo {year} {2017})}\BibitemShut {NoStop}%
\bibitem [{\citenamefont {Cr{\'e}pel}\ \emph {et~al.}(2018)\citenamefont {Cr{\'e}pel}, \citenamefont {Estienne}, \citenamefont {Bernevig}, \citenamefont {Lecheminant},\ and\ \citenamefont {Regnault}}]{crepel2018matrix}%
  \BibitemOpen
  \bibfield  {author} {\bibinfo {author} {\bibfnamefont {V.}~\bibnamefont {Cr{\'e}pel}}, \bibinfo {author} {\bibfnamefont {B.}~\bibnamefont {Estienne}}, \bibinfo {author} {\bibfnamefont {B.~A.}\ \bibnamefont {Bernevig}}, \bibinfo {author} {\bibfnamefont {P.}~\bibnamefont {Lecheminant}},\ and\ \bibinfo {author} {\bibfnamefont {N.}~\bibnamefont {Regnault}},\ }\bibfield  {title} {\bibinfo {title} {Matrix product state description of halperin states},\ }\href@noop {} {\bibfield  {journal} {\bibinfo  {journal} {Physical Review B}\ }\textbf {\bibinfo {volume} {97}},\ \bibinfo {pages} {165136} (\bibinfo {year} {2018})}\BibitemShut {NoStop}%
\bibitem [{\citenamefont {Mukhopadhyay}\ \emph {et~al.}(2001)\citenamefont {Mukhopadhyay}, \citenamefont {Kane},\ and\ \citenamefont {Lubensky}}]{mukhopadhyay2001sliding}%
  \BibitemOpen
  \bibfield  {author} {\bibinfo {author} {\bibfnamefont {R.}~\bibnamefont {Mukhopadhyay}}, \bibinfo {author} {\bibfnamefont {C.}~\bibnamefont {Kane}},\ and\ \bibinfo {author} {\bibfnamefont {T.}~\bibnamefont {Lubensky}},\ }\bibfield  {title} {\bibinfo {title} {Sliding luttinger liquid phases},\ }\href@noop {} {\bibfield  {journal} {\bibinfo  {journal} {Physical Review B}\ }\textbf {\bibinfo {volume} {64}},\ \bibinfo {pages} {045120} (\bibinfo {year} {2001})}\BibitemShut {NoStop}%
\bibitem [{\citenamefont {Thouless}(1989)}]{thouless1989level}%
  \BibitemOpen
  \bibfield  {author} {\bibinfo {author} {\bibfnamefont {D.}~\bibnamefont {Thouless}},\ }\bibfield  {title} {\bibinfo {title} {Level crossing and the fractional quantum hall effect},\ }\href@noop {} {\bibfield  {journal} {\bibinfo  {journal} {Physical Review B}\ }\textbf {\bibinfo {volume} {40}},\ \bibinfo {pages} {12034} (\bibinfo {year} {1989})}\BibitemShut {NoStop}%
\bibitem [{\citenamefont {Giamarchi}(2003)}]{giamarchi2003quantum}%
  \BibitemOpen
  \bibfield  {author} {\bibinfo {author} {\bibfnamefont {T.}~\bibnamefont {Giamarchi}},\ }\href@noop {} {\emph {\bibinfo {title} {Quantum physics in one dimension}}},\ Vol.\ \bibinfo {volume} {121}\ (\bibinfo  {publisher} {Clarendon press},\ \bibinfo {year} {2003})\BibitemShut {NoStop}%
\bibitem [{\citenamefont {Mukherjee}\ and\ \citenamefont {Park}(2019)}]{mukherjee2019spin}%
  \BibitemOpen
  \bibfield  {author} {\bibinfo {author} {\bibfnamefont {S.}~\bibnamefont {Mukherjee}}\ and\ \bibinfo {author} {\bibfnamefont {K.}~\bibnamefont {Park}},\ }\bibfield  {title} {\bibinfo {title} {Spin separation in the half-filled fractional topological insulator},\ }\bibfield  {journal} {\bibinfo  {journal} {Physical Review B}\ }\textbf {\bibinfo {volume} {99}},\ \href {https://doi.org/10.1103/physrevb.99.115131} {10.1103/physrevb.99.115131} (\bibinfo {year} {2019})\BibitemShut {NoStop}%
\end{thebibliography}%

\iftrue

\appendix 

\newpage

\section{Layer polarization} \label{app_layerpolarization}

In Fig.~\ref{fig_layerpolarization}, we show the layer polarization $P_z$ of the ground state obtained by exact diagonalization of Eq.~\ref{eq_fullmodel} in all the $(\nu_+,\nu_-)$ sectors with $\nu_+ \geq \nu_-$; the energies in the $(x,y)$ and  $(y,x)$ sectors are equal due to the $\tilde{\mathcal{T}}$-symmetry of the model. The balanced bilayer is always energetically favored for the range of the parameters and system sizes considered.

\begin{figure}
\centering
\includegraphics[width=\columnwidth]{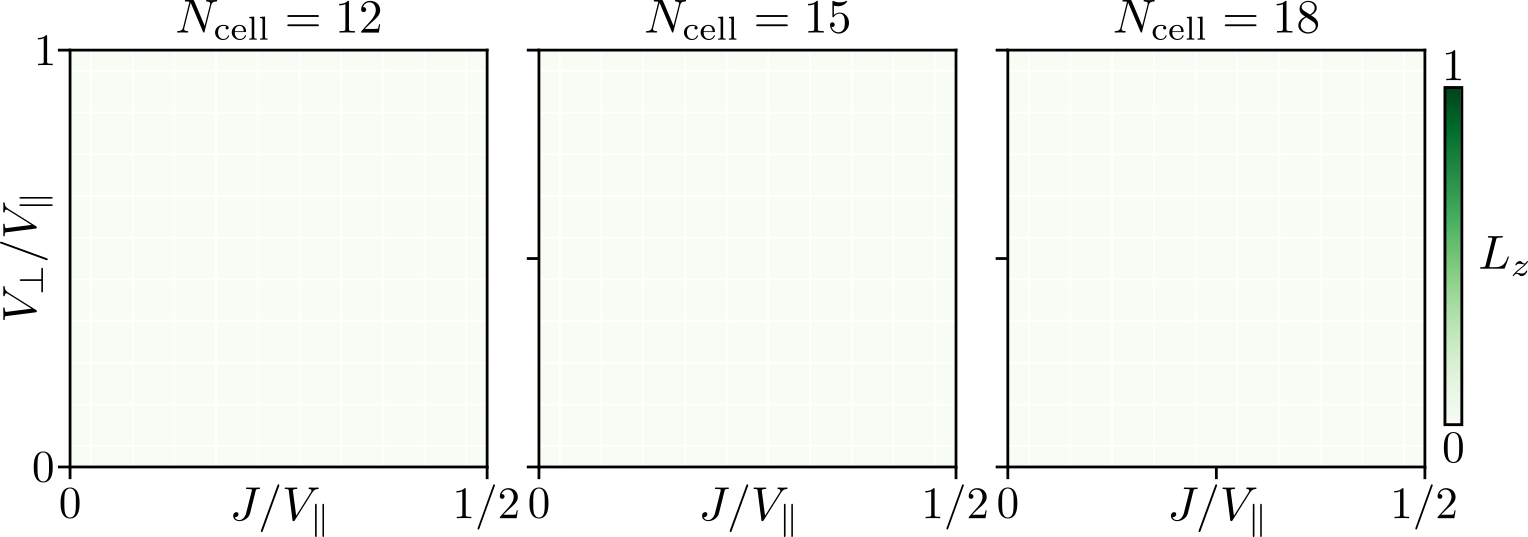}
\caption{Layer polarization $L_z$ of the ground state of Eq.~\ref{eq_fullmodel} for different system sizes $N_{\rm cell}$. The balanced bilayer is always energetically favored for the range of the parameters considered.}
\label{fig_layerpolarization}
\end{figure}

\section{Spread-to-gap ratios} \label{app_phaseboundaries}

In Fig.~\ref{fig_barespread}, we plot the bare numerical data for the spread $(E_n-E_0)$ and spread-to-gap ratio $(E_n-E_0)/(E_9 - E_0)$ for $n=1,\cdots ,8$ computed on the $N_{\rm cell} = 18$ cluster. Here, $E_n$ denotes the $n$-th lowest eigenvalue of the Hamiltonian. In that figure, an $N$-fold degenerate ground state manifold will appear in light and dark red for $n<N$ and $n\geq N$, respectively. We find very good agreement with the approximate phase boundaries estimated in Fig.~\ref{fig_GSdegeneracy} of the main text, which are overlayed with the data in Fig.~\ref{fig_barespread}. 

\begin{figure}
\centering
\includegraphics[width=0.69\columnwidth]{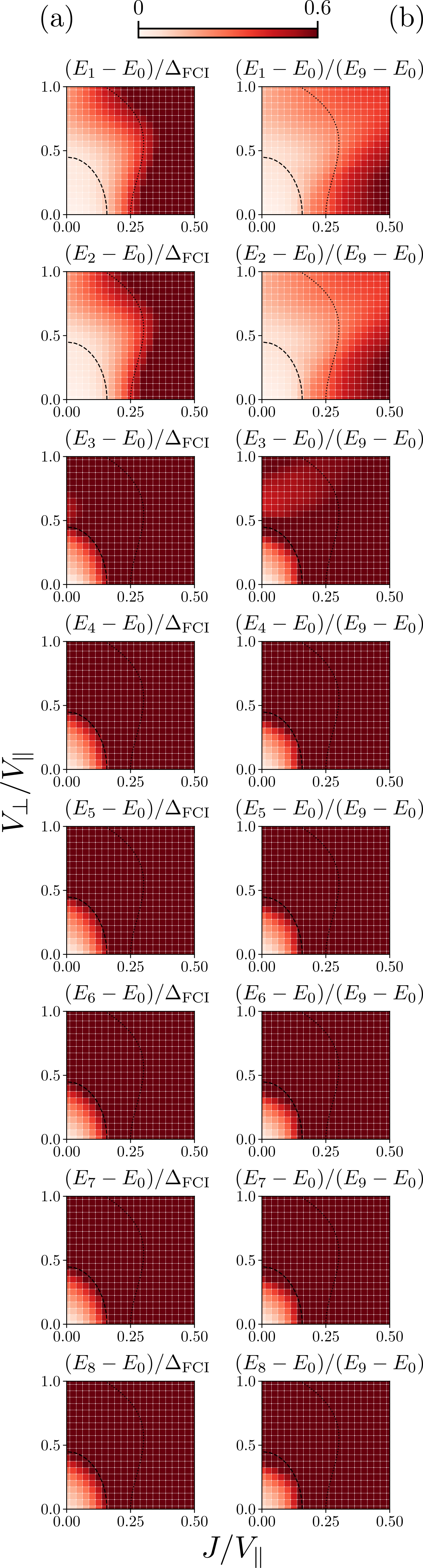}
\caption{Spread $(E_n-E_0)$ (a) and spread-to-gap ratio $(E_n-E_0)/(E_9 - E_0)$ (b) obtained for $n=1,\cdots ,8$ (rows) on the $N_{\rm cell} = 18$ cluster.}
\label{fig_barespread}
\end{figure}

\section{Layer polarized FCI} \label{app_proofFCI}

We here perform a careful finite-size scaling of the gap of the FCI obtained in the layer-polarized limit $\nu_+ = 1/3$ and $\nu_-=0$~\cite{wu2012zoology}, from which the FTI discussed in the main text derives. This state was identified numerically using a tight-binding model similar to Eq.~\ref{eq_fullmodel} in Ref.~\cite{crepel2023anomalous}, and we only highlight its existence for our specific choice of parameters by displaying the many-body spectra and flux-threading behaviors for the three system sizes studied in the main text (Eq.~\ref{eq_tiltedclusters}). As depicted in Fig.~\ref{fig_proofFCI}, the many-body momenta of the three degenerate ground state satisfy to the FCI counting rule~\cite{bernevig2012emergent}, and they undergo a cyclic permutation under $2\pi$-flux insertion~\cite{thouless1989level}; which together present strong evidence for the FCI nature of the ground state manifold. 

\begin{figure}
\centering
\includegraphics[width=\columnwidth]{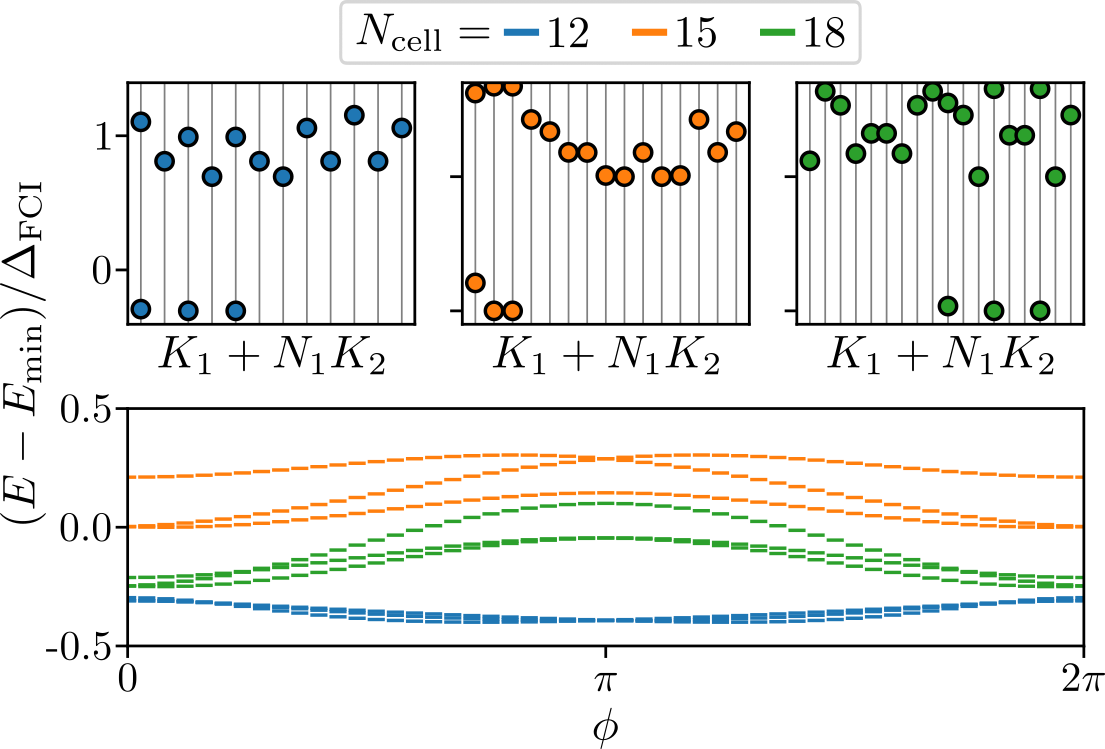}
\caption{Momentum resolved many-body spectra (top row) and spectral flow of the three lowest lying states (bottom row) characterizing the layer-polarized FCI ($\nu_+ = 1/3$ and $\nu_-=0$) for the three finite-size lattices considered in the main text (different colors). For the spectral flow, we shifted the data corresponding to the different sizes by a constant amount to better discern the various curves. As mentioned in Sec.~\ref{sec_thermolimit}, the three ground states of forming the 1/3 Laughlin FCI appear in distinct momentum sectors.}
\label{fig_proofFCI}
\end{figure}

\begin{figure}
\centering
\includegraphics[width=\columnwidth]{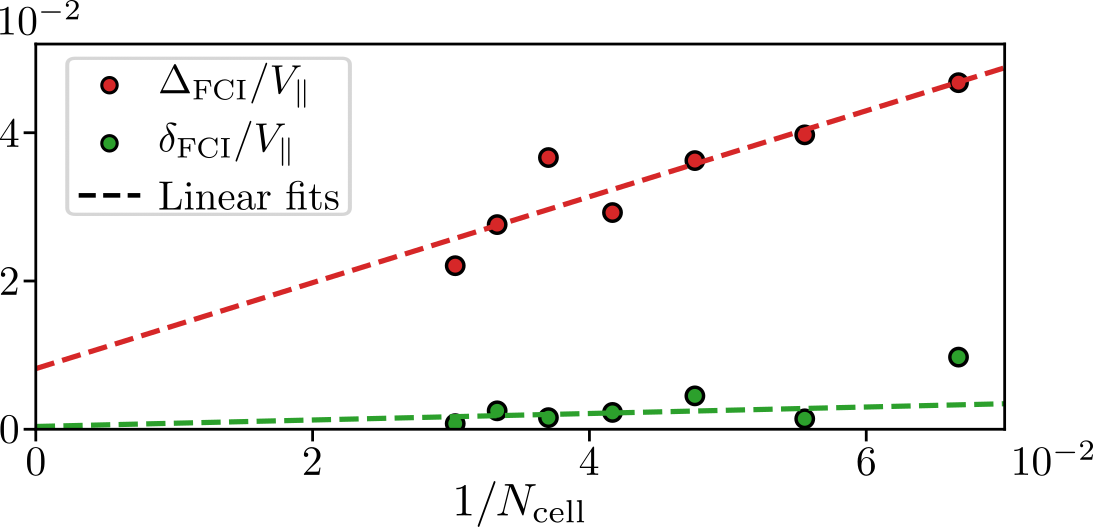}
\caption{Gap (red dots) and spread (green dots) of the FCI obtained in the layer-polarized limit $\nu_+ = 1/3$ and $\nu_-=0$ as a function of the inverse system size $1/N_{\rm cell}$. The aspect ratios are kept close to $\sqrt{3}/2$ but are not all equal. Extrapolation to the thermodynamic limit are highlighted with dashed lines.}
\label{fig_extrapolatedgap}
\end{figure}

The layer polarization allows to access larger system sizes (up to 36 unit cells), for which we also use tilted cluster in order to keep aspect ratios close to $\sqrt{3}/2$ using the tilts
\begin{equation}
\begin{array}{c||c|c|c|c|c|c|c|c}
N_s & N^{(1)} & N^{(2)} & n_1^{(1)} & n_1^{(2)} & n_2^{(1)} & n_2^{(2)} & \alpha & {\rm aspect} \\ \hline \hline
21 & 7 & 3 & 1 &-6 & 4 & -3 & 1 & 0.866  \\
24 &24 & 1 & 4 &-4 & 1 & 5 & 1-5 & 0.990  \\
27 &27 & 1 & 4 & 1 & 1 & 7 & 4 & 0.898  \\
30 & 5 & 6 & 5 &0 & 1 &6 & 1 & 0.838  \\
33 &33 & 1 & 1 & -8 & 5 &-7 & 4 & 0.922 \\
36 &6 & 6 & 6 & 0 & 0 &6 & 0 & 0.866 \\
\end{array} ,
\end{equation}
where we used the notations of Eq.~\ref{eq_tiltedclusters}. For all accessible system sizes, we compute the FCI gap $\Delta_{\rm FCI}$ and spread $\delta_{\rm FCI}$, shown in Fig.~\ref{fig_extrapolatedgap}. We extrapolate them to $1/N_{\rm cell} \to 0$ using a linear fit on the $N_{\rm cell}\geq 15$ points, leading to the following estimate of their thermodynamic values 
\begin{equation}
\Delta_{\rm FCI}^{(N_{\rm cell} \to \infty)} \simeq 8 \cdot 10^{-3} V_\parallel , 
\end{equation}
and $\delta_{\rm FCI}^{(N_{\rm cell} \to \infty)} \simeq 0$ within error bars of the fit. While small in unit of $V_\parallel$, the gap of the FCI remains finite and much larger than its spread, ensuring that the phases observed in our finite clusters still exist in the thermodynamic limit. Note that the FCI gap $\Delta_{\rm FCI}$ used as units of energy in the main text corresponds to the one obtained in finite size, and not the extrapolated one.

\section{Choice of coupled-wire model coefficients for 1/3+1/3 FTI} \label{app_exactpoint}

In this appendix, we go through the important steps of the coupled-wire construction of the Laughlin 1/3 topological order from Refs.~\cite{kane2002fractional,teo2014luttinger} using a special exactly-solvable model that directly leads to Eq.~\ref{eq_coupledwirebeforeattration} in the main text. 

We start from a bosonized description of the decoupled array of quantum wires, in which each wire $w$ is described by four bosonic fields $\Tilde{\varphi}_{w,\ell}(x)$ and $\Tilde{\theta}_{w,\ell}(x)$ with $x$ the position along the compact dimension of the wire and $\ell = \pm$ a ``layer'' index analogous to the one in Eq.~\ref{eq_Haldanebilayer}. Physically, $\Tilde{\varphi}$ and $(\partial_x \Tilde{\theta})$ fields respectively represent the phase and density fluctuations of the fermionic fields near the Fermi energy of the wires~\cite{giamarchi2003quantum}. They satisfy the commutation relations $[\partial_x \Tilde{\theta}_{w,\ell}, \Tilde{\varphi}_{w',\ell'}] = i \pi \ell \delta_{\ell,\ell'} \delta_{w,w'}$, where the factor of $\ell$ comes from the fact that the two layers are related by an anti-unitary effective time-reversal symmetry ($\Tilde{\mathcal{T}}$ int he main text), which acts as a complex conjugation and therefore flips the sign of $\Tilde{\varphi}$~\cite{neupert2014wire}. In terms of these variables, the Hamiltonian of the quantum wire array takes the form of coupled sliding Luttinger liquids~\cite{teo2014luttinger}
\begin{equation}
\mathcal{H}_{\rm SLL}  =  \sum_{w, w'} \int {\rm d} x \,\partial_x \begin{bmatrix} \Tilde{\varphi}_{w,-} \\ \Tilde{\varphi}_{w,+} \\ \Tilde{\theta}_{w,-} \\ \Tilde{\theta}_{w,+} \end{bmatrix}^T \Tilde{M}_{w,w'} \begin{bmatrix} \Tilde{\varphi}_{w',-} \\ \Tilde{\varphi}_{w',+} \\ \Tilde{\theta}_{w',-} \\ \Tilde{\theta}_{w',+} \end{bmatrix} , 
\end{equation}
where $\Tilde{M}_{w,w}$ contains both kinetic and intra-wire density-density interactions, while all other $\Tilde{M}_{w,w'\neq w}$ describe density-density interactions between wires. 

The link variables introduced to capture the Laughlin $1/m$ topological order in both layers read~\cite{kane2002fractional}
\begin{align}
\varphi_{j,\ell} & = \frac{\Tilde{\varphi}_{w,\ell} + \Tilde{\varphi}_{w+1,\ell}}{2} + \ell m \frac{\Tilde{\theta}_{w,\ell} - \Tilde{\theta}_{w+1,\ell}}{2} , \\
\theta_{j,\ell} & = \frac{\Tilde{\varphi}_{w,\ell} - \Tilde{\varphi}_{w+1,\ell}}{2} + \ell m \frac{\Tilde{\theta}_{w,\ell} + \Tilde{\theta}_{w+1,\ell}}{2} 
\end{align}
with $j = w+1/2$ a link index. Their commutation relation is $[\partial_x \theta_{j,\ell}, \varphi_{j',\ell'}] = m i \pi \ell \delta_{\ell,\ell'} \delta_{j,j'}$, and they correspond to the fields introduced in Eq.~\ref{eq_coupledwirebeforeattration} of the main text when $m=3$.

In terms of these new variables, the form of the sliding Luttinger liquid Hamiltonian remains same albeit with new matrices $M_{j,j'}$. It is always possible to choose the interaction terms in the original theory, \textit{i.e.} the $\Tilde{M}$, such that the $M$ matrices become local in the link indices $M_{j,j'} = M \delta_{j,j'}$; the recipe for doing so is given in Ref.~\cite{santos2015fractional}. In absence of any attrractive interactions, we choose the diagonal for $M = \diag (K_0, K_0, K_0^{-1}, K_0^{-1})$ corresponding to a time-reversal invariant coventional Luttinger liquid with parameter $K_0$ on each links of the wire array. The phenomenological addition of density-density attractive interaction between the two layers, as described in Sec.~\ref{subsec_coupledwire}, changes this form to 
\begin{equation} \label{eq_Mmatrixcoupledwire}
M = \begin{bmatrix} K_0 & . & . & . \\ . & K_0 & . & . \\ . & . & K_0^{-1} & - J^{\rm eff} \\  . & . & - J^{\rm eff} &  K_0^{-1} \end{bmatrix} . 
\end{equation}

The final ingredient of the coupled-wire construction is the sine-Gordon mass term provided by correlated single particle tunneling between wires, which we include in the standard way~\cite{kane2002fractional,teo2014luttinger}. It was shown in Ref.~\cite{santos2015fractional} that this coupled wire construction continued to describe a FTI order adiabatically connected to two copies of Laughlin 1/$m$ order with opposite chirality as long as the renormalized Luttinger parameters resulting from Eq.~\ref{eq_Mmatrixcoupledwire} remained finite (in other words, as long as the bosonic theory remains adiabatically connected to the one obtained in absence of attractive interactions $J^{\rm eff}=0$). 

We therefore compute the new Luttinger parameters corresponding to the theory Eq.~\ref{eq_coupledwirebeforeattration} in the main text before turning on the mass terms (\textit{i.e.} $g=0$). The action following from the Hamiltonian $M$ and reproducing the commutation relations of the link bosonic fields is 
\begin{widetext}
\begin{equation}
\mathcal{S}_j = \int {\rm d}x {\rm d}\tau \sum_{\ell}  \left[ \frac{i\ell}{m \pi}   (\partial_\tau \varphi_{j,\ell})(\partial_x \theta_{j,\ell}) - K_0 (\partial_x \varphi_{j,\ell})^2 - K_0^{-1} (\partial_x \theta_{j,\ell})^2 \right] - 2 J^{\rm eff} (\partial_x \theta_{j,+}) (\partial_x \theta_{j,-}) . 
\end{equation}
\end{widetext}
Integrating out the phase variables $\varphi_{j,\ell}$, and introducing the spin and charge densities
\begin{equation}
\Lambda_{j,c} = \frac{\theta_{j,+}+\theta_{j,-}}{\sqrt{2}} , \quad \Lambda_{j,s} = \frac{\theta_{j,+}-\theta_{j,-}}{\sqrt{2}} , 
\end{equation}
yields the bosonic action 
\begin{equation}
\mathcal{S}_j = \int \frac{{\rm d}x {\rm d}\tau }{2\pi K_0} \sum_{\alpha= c/s} (\partial_\tau \Lambda_{j,\alpha})^2 + \frac{K_0^2}{K_\alpha^2} (\partial_x \Lambda_{j,\alpha})^2, 
\end{equation}
leanding to the results quoted in the main text for the Luttinger parameters 
\begin{align}
 \frac{K_0}{K_c} = m \sqrt{1 - K_0 J^{\rm eff}} , \quad  \frac{K_0}{K_s} = m \sqrt{1 + K_0 J^{\rm eff}} . 
\end{align}

\section{Absence of three-fold phase} \label{app_rankoneperturbation}

\subsection{Perturbation theory}

We here argue using perturbation theory that the intermediate three-fold degenerate phase from Fig.~\ref{fig_GSdegeneracy} does not exist in the thermodynamic limit, where all states of the FCI states $\ket{{\rm FCI}_{j=1,2,3}}$ appear at the same many-body momentum~\cite{bernevig2012emergent}. In absence of inter-layer coupling, the corresponding FTI states are simply given by the products $\ket{\psi_{ij}} = \ket{\overline{\rm FCI}_i} \otimes \ket{{\rm FCI}_j}$. Consider a small and generic inter-layer interaction in momentum space $H_{\rm inter} = \frac{1}{N_{\rm cell}} \sum_q v(q) \rho_{q,\uparrow} \rho_{-q,\downarrow}$. Its perturbative effect on the FTI states can be determined by considering the matrix elements 
\begin{align} 
&M_{im}^{jn} = \braOket{\psi_{ij}}{H_{\rm inter}}{\psi_{mn}} \\
& = \frac{1}{N_{\rm cell}} \sum_q v(q) \braOket{\overline{\rm FCI}_i}{\rho_{-q,\downarrow}}{\overline{\rm FCI}_m} \braOket{{\rm FCI}_j}{\rho_{q,\uparrow}}{{\rm FCI}_n} \notag \\ 
& = \frac{v(0)}{N_{\rm cell}} X_{im}^* X_{jn} ,  \quad X_{ab} = \braOket{{\rm FCI}_a}{\rho_{0,\uparrow}}{{\rm FCI}_b} , \notag
\end{align}
where we have used the fact that the FCI states all appear at the same momentum in the second line to restrict the sum to $q=0$. In this form, it is clear that the matrix describing the perturbative effect of inter-layer interactions within the FTI manifold has rank one and has a single non-zero eigenvalue. As a result, the interlayer interaction singles out one of the FTI states from the eight others, and we do not expect a three-fold degenerate phase to appear. Note that when the FCI/FTI states are located at three different momenta, a similar calculation shows that one state is perturbatively singled out by inter-layer interaction in each momentum sector, which leads to the three-fold degenerate phase observed in Fig.~\ref{fig_GSdegeneracy}. This behavior can be clearly seen in Fig.~\ref{fig_edexamples} for small value of the interaction $J < 0.2$ (where the interaction can be treated perturbatively).

Note that this argument does not contradict the exact degeneracy between the FTI ground states in presence of small perturbations. Indeed, the perturbative splitting identified above involves a single momentum exchange component $q=0$ of the interaction, whose amplitude scales with the inverse of the number of unit cells $N_{\rm cell}$. Our derivation should henceforth be understood as follows. For large system close to thermodynamic limit, a single FTI state is single out from the others by interlayer interaction by an amount $\propto N_{\rm cell}^{-1}$ that strictly vanish when $N_{\rm cell} \to \infty$.

\subsection{Numerical example}

The arguments provided in Sec.~\ref{sec_thermolimit}-B and above suggest that, in finite-size, the presence or absence of a three-fold degenerate region between the FTI and SC phases depends on the geometry chosen for the calculations. Unfortunately, for the parameters chosen in Eq.~\ref{eq_fullmodel}, we were not able to find a finite size cluster containing 18 unit cells and featuring an approximately nine-fold degenerate ground state for $J=V_\perp=0$. 

As explained in the main text, Eq.~\ref{eq_fullmodel} is obtained from the three-orbital model derived in Ref.~\cite{crepel2024bridging}, in which two orbitals forming a honeycomb lattice and a third orbital living at the center of the hexagons, after adiabatic elimination of hexagon-centered orbital due to its larger on-site potential energy. 
Taking the original model of Ref.~\cite{crepel2024bridging} allows to obtained, in some regimes of parameters, a more uniform Berry curvature, that allows to stabilize the FTI on smaller finite size clusters. In particular, we were able to find an FTI with all nine nearly degenerate ground state in the same momentum sector on a $6\times 3$ lattice using the parameters $t_{\rm TH}^{(1)} = 5 t_{HH}^{(1)} = 10 t_{HH}^{(3)} = \delta/ 2.5$ (the notations follow Ref.~\cite{crepel2024bridging}) and an equal intra-specie honeycomb-honeycomb and triangular-honeycomb interaction strength $V_\perp$.

\begin{figure}
\centering
\includegraphics[width=\columnwidth]{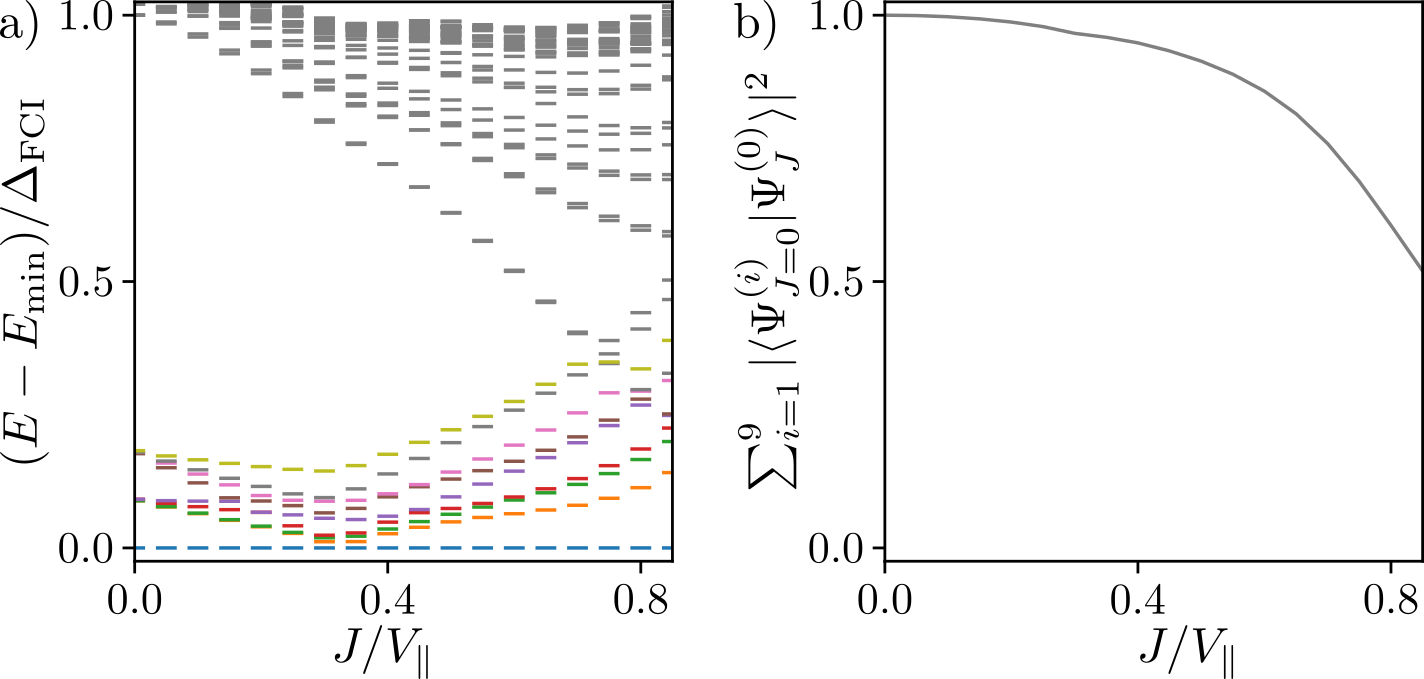}
\caption{a) Same as Fig.~\ref{fig_threestates}a for the model of Ref.~\cite{crepel2024bridging} with $t_{\rm TH}^{(1)} = 5 t_{HH}^{(1)} = 10 t_{HH}^{(3)} = \delta/ 2.5$ on a $6\times 3$ lattice, for which the nine FTI degenerate ground states (colored for the sake of visibility) occupy the same momentum sector. b) Sum of the square overlap between $\ket{\Psi_J^{(0)}}$, the absolute lowest energy many-body ground state, and $\ket{\Psi_{J=0}^{(i)}}$ with $i=1,\cdots,9$ the nine FTI states at $J=0$.}
\label{fig_noitd}
\end{figure}

Introducing an on-site attractive interaction $J$, we observe a transition from a approximately nine-fold degenerate phase to a non-degenerate phase, as an be seen in Fig.~\ref{fig_noitd}a. In between, no sign of an ITD can be observed, substantiating our perturbative argument and the interpretation of the ITD as a finite-size artefact. To makes sure that the non-degenerate ground state is not part of an FTI phase with very strong spread, we computed its overlap with the nine degenerate FTI at $J=0$, which is shown in Fig.~\ref{fig_noitd}b. The collapse of the overlap around $J \simeq 0.5 V_\parallel$ suggests that the two phases are distinct in the thermodynamic limit. This strengthen the claim made in the main text that the transition between the FTI and the superconductor does not involve any intermediate phase.

\section{Phase separation for $V_\perp > V_\parallel$} \label{app_phaseseparation}

Fig.~\ref{fig_phaseseparation} shows the exact diagonalization spectrum obtained as a function of $V_\perp$ along the $J=0$ line of the phase diagram beyond the physically relevant regimes identified in Sec.~\ref{sec_MicroscopicModel}. For $V_\perp < 0.8 V_\parallel$, this is the data shown in Fig.~\ref{fig_threestates}b. When $V_\perp \simeq 1.4-1.5 V_\parallel$, we observe a collapse of the many-body continuum. The phase stabilized beyond this collapse ($V_\perp > 1.5 V_\parallel$) has one ground state in each of the many-body momentum sectors. We can therefore understand it as a large and immobile cluster of particles in real-space that can be translated in the finite-size lattice at small energy cost; a clear signature of phase separation. A similar behavior has been observed in Refs.~\cite{mukherjee2019spin,chen2012interaction}.

\begin{figure}
\centering
\includegraphics[width=\columnwidth]{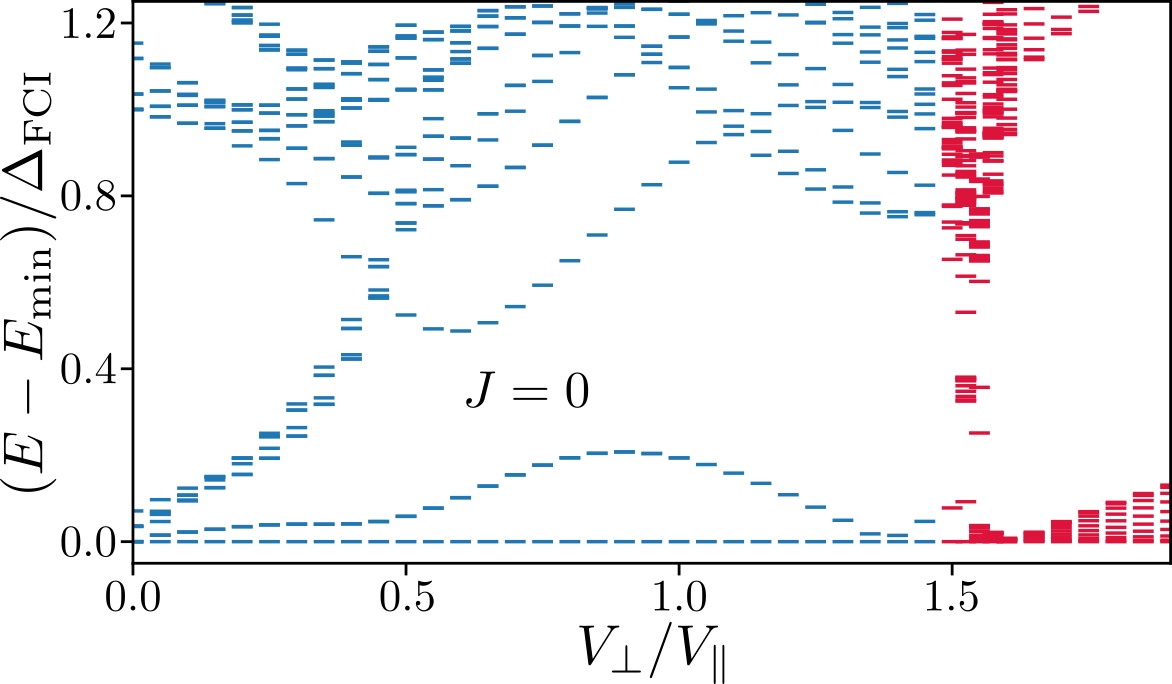}
\caption{Same as Fig.~\ref{fig_threestates}b for a wider range of $V_\perp / V_\parallel$. Phase separation is observed for $V_\perp \gtrsim 1.5 V_\parallel$, probed here by an extensive degeneracy of the ground state. 
This is the reason for the slow increase of the doublet in Fig.~\ref{fig_threestates}b. A similar behavior has been observed in Refs.~\cite{mukherjee2019spin,chen2012interaction}.
Near the transition point $V_\perp = 1.5 V_\parallel$, we used a finer discretization in $V_\perp$ and diagonalized for a larger number of low-lying states to better observe the collapse, resulting in a denser spectrum (shown in red).}
\label{fig_phaseseparation}
\end{figure}

\fi
\end{document}